\documentclass[%
 aip,
 amsmath,amssymb,
]{revtex4-1}
\usepackage{graphicx}
\usepackage{dcolumn}
\usepackage{ulem}
\usepackage{bm}
\usepackage{subfigure}
\usepackage[utf8]{inputenc}
\usepackage[T1]{fontenc}
\usepackage{mathptmx}
\usepackage{xcolor}
\usepackage{soul}
\usepackage{float}
\begin{document}


\title[]{Transient shear banding during startup flow: Insights from nonlinear simulations}

\author{Shweta Sharma}
\author{Yogesh M.~Joshi}%
 \email{joshi@iitk.ac.in}
 \author{V.~Shankar}
 \email{vshankar@iitk.ac.in}
\affiliation{Department of Chemical Engineering, Indian Institute of Technology Kanpur, Kanpur 208016, India
}%


\date{\today}

\begin{abstract}

We study the dynamics of shear startup of the Johnson-Segalman and non-stretching Rolie-Poly models using nonlinear simulations. Signatures of transient and steady-state shear banding are quantified by monitoring the `degree of banding' defined as the difference between maximum and minimum shear rates during shear startup. We consider cases where the startup is from zero shear rate to shear rates in both the monotonic and nonmonotonic regions of the constitutive curve. For the Johnson-Segalman model, which exhibits a shear stress overshoot during startup, our nonlinear simulations show that transient shear banding is absent regardless of whether the start-up shear rate is in the monotonic or nonmonotonic regions of the constitutive curve. In the latter case, while there is clearly an inhomogeneity \textit{en route} to the banded state, the magnitude of the degree of banding is not substantially large compared to that of the eventual banded state. Marked inhomogeneity in the velocity profile is predicted for the non-stretching Rolie-Poly model only if the solvent to solution viscosity ratio is smaller than $O(10^{-3})$, but its occurrence does not appear to have any correlation with the stress overshoot during startup. These inhomogeneities are also very sensitive to initial amplitude of perturbations and the magnitude of Reynolds number. We also compare the results obtained within the realm of linearized dynamics in our previous study (Sharma \textit{et al.}, Journal of Rheology 65, 1391 (2021)) with the present nonlinear results and show that nonlinearities have a stabilizing effect and mitigate the divergence of perturbations (as predicted within the linearized dynamics) during shear startup. We argue that the neglect of inertia in the nonlinear simulations is not self-consistent if the solvent to solution viscosity ratio is very small, and that inertial effects need to be included in order to obtain physically realistic results. Our nonlinear simulations show that the transient evolution during shear startup is quite sensitive to the Reynolds number when the solvent viscosity parameter is much smaller than unity for non-stretching Rolie-Poly model. However, the results of the Johnson-Segalman model are very robust for solvent to solution viscosity greater than $O(10^{-3})$ and do not reveal any transient shear banding during shear startup. 

\end{abstract}

\maketitle
\section{Introduction} \label{section_intro}

Viscoelastic fluids undergoing shear and extensional flows are often susceptible to instabilities of various kinds [\onlinecite{denn2001extrusion,joshi2000slipping,petrie1976instabilities,howells1962flow,joshi2003planar,joshi2004rupture,Shaqfeh_1996}]. Shear banding is one such example of an instability exhibited by viscoelastic materials [\onlinecite{olmsted2008perspectives,divoux2016shear,germann2019shear,varchanis2022evaluation}]. Shear banding is best understood in the context of a planar Couette flow of a viscoelastic fluid with the applied shear rate in the underlying nonmonotonic region of the constitutive curve. In this situation, the linear velocity profile becomes unstable and transforms into bands of two distinct shear rates. The shear rates in the two bands lie in the stable branch of the constitutive curve.

Shear banding seen during shear startup, when the flow is transitioning to its eventual steady state, is referred to as transient shear banding. Shear banding observed after the flow has reached steady state is referred to as steady-state shear banding. Transient shear banding can, in principle, occur in both the scenarios, regardless of whether the steady state is banded or homogeneous. If the steady state is banded, however, transient shear banding must be distinguished from the inevitable inhomogeneities that will ensue \textit{en route} to the banded state by measuring the extent of the inhomogeneity. In recent studies, transient shear banding is quantified by measuring the difference between maximum and minimum shear rates, which has been dubbed as the `degree of banding'  [\onlinecite{adams2011transient,moorcroft14}]. If the steady state is homogeneous, then the evolution of degree of banding with time will transiently increase to attain a peak and then decrease to zero. If the steady state is banded, then the presence of transient shear banding can be ascertained only if degree of banding shows a transient peak of substantially high magnitude, before attaining its eventual value at steady state. The evolution of second derivative of velocity in the direction of momentum diffusion has also been utilised in literature to find out the presence of transient shear banding [\onlinecite{jain2018role,kushwaha2022dynamics}]. Also, the width of the interface between two bands must be much smaller than the width of the bands; hence, transient flow inhomogeneities during shear startup may not always be designated as transient shear banding.

The stability of steady state in a shear startup flow was first studied by Yerushalmi et. al. [\onlinecite{yerushalmi1970stability}] who proposed a model independent or fluid universal criterion for existence of steady state shear banding in shear startup flow. In their manuscript, the authors stated that ``It is shown that all values of the steady shear rate where the flow curve exhibits a zero or negative slope the flow is unstable." This criterion has been validated by many researchers in the literature using experiments and various constitutive models with a nonmonotonic region where shear stress decreases as a function of shear rate [\onlinecite{helgeson2009relating,johnson1977model,doi1988theory,cates1987reptation,vasquez2007network,giesekus1982simple,likhtman2003simple,ianniruberto2017shear,briole2021shear,rassolov2022role}]. 

A fluid universal or model independent criterion for existence of transient shear banding during shear startup flow has been proposed in literature by Moorcroft and Fielding [\onlinecite{moorcroft2013criteria,moorcroft14}]. According to this criterion, in a shear startup flow, if shear stress shows an overshoot then the time duration in which shear stress shows a negative slope as a function of time or strain, the shear startup flow becomes unstable and should form bands of different shear rates transiently. The authors utilised the modal stability analysis method and calculated the transient eigenvalue, also known as `frozen time' analysis, to derive the criterion. In our previous study [\onlinecite{sharma2021onset}], we discussed in detail the restrictive nature of the criterion based on the assumptions utilised to obtain the simplified criterion. These assumptions are (i) shear startup is at a high shear rate, (ii) frozen time analysis can be used as a tool to obtain the signature of transient shear banding, (iii) Hopf bifurcation is absent, and (iv) only one real part of transient eigenvalue is positive at any time. 

Moorcroft and Fielding [\onlinecite{moorcroft14}] examined the shear startup of non-stretching Rolie-Poly and Giesekus models using frozen time analysis, linearized evolution of perturbations and nonlinear simulations to check the validity of the criterion. They found that the results obtained using the non-stretching Rolie-Poly model validated the criterion and transient shear banding was observed in presence of stress overshoot, however, for the Giesekus model, transient shear banding was not observed even in presence of stress overshoot. The authors argued that since transient shear banding is an elastic instability, the models that can show stress overshoot at same magnitude of strain can qualify to show transient shear banding. Fielding [\onlinecite{fielding2016triggers}] subsequently analysed the transient shear banding criterion for shear startup flow and and noted that while the criterion is useful, it may not be universal.

The transient shear banding criterion for shear startup flow [\onlinecite{moorcroft14}] has been tested experimentally and theoretically in the literature and it has been found to be consistent with some results but not universally applicable [\onlinecite{cao2012shear,mohagheghi2016elucidating,zhou2008modeling,zhou2012multiple,zhou2014wormlike,sharma2021onset,ianniruberto2017shear,briole2021shear,burroughs2023flow,benzi2021continuum,tapadia2006direct,ravindranath2008banding,wang2011homogeneous,boukany2008use,boukany2009exploring,boukany2009shear,hu2007constitutive,hu2008comparison,li2013flow,wang2014letter,ianniruberto2017shear,2018PhDT.......101P}] as discussed in Section II of our previous study [\onlinecite{sharma2021onset}]. Transient and steady state shear banding has been observed in shear startup of entangled polymeric solutions in literature and researchers have proposed flow concentration coupling to be the underlying reason [\onlinecite{fielding2003kinetics,fielding2003flow,fielding2003early,cromer2013shear,cromer2014study,burroughs2023flow,peterson2016shear}]. Transient shear banding in shear startup flow of wormlike micellar systems has also been explored in the literature but only when the startup shear rate is such that the steady state is also banded or using geometries that has quite high shear stress inhomogeneity [\onlinecite{briole2021shear,rassolov2022role,hu2008comparison,boukany2008use,rassolov2020effects,rassolov2022kinetics,mohammadigoushki2019transient}]. To the best of our knowledge, transient shear banding has not been reported in shear startup flow of wormlike micellar solutions yet, wherein the steady state is homogeneous and shear stress is almost homogeneous between the plates. Since there can be multiple factors governing the shear banding in entangled polymeric solutions, in this study, we focus on understanding the possibility of transient shear banding in wormlike micellar solutions. We also focus on shear startup of viscoelastic constitutive models that can predict a monotonic and nonmonotonic constitutive curve and have no coupling with concentration (Rolie-Poly model) [\onlinecite{giesekus1982simple,likhtman2003simple,zhou2008modeling,cates1987reptation}] and can fit experimental results of wormlike micellar solutions (Johnson-Segalman model [\onlinecite{johnson1977model}]).

In our earlier study [\onlinecite{sharma2021onset}], we had analyzed the transient shear banding criterion for shear startup flow using Johnson-Segalman, non-stretching Rolie-Poly, and Giesekus models using frozen time analysis and fundamental matrix method. In the frozen time analysis, the eigenvalue is calculated at each time instant during shear startup, treating the flow to be quasi steady. This approach is valid only if the transient growth rate is significantly higher than the rate of change of base-state evolution. In the fundamental matrix method [\onlinecite{alam1997influence,schmid1994optimal,schmid1994transient,strang1997linear}], the evolution of linearized perturbations can be obtained along with the evolution of base state. This method also provides maximum amplitude of perturbations that is independent of initial conditions and consequently renders more robust results as compared to frozen time analysis. We used these methods and showed that the assumptions made to derive the criterion for transient shear banding [\onlinecite{moorcroft14}] are restrictive. 
We showed that there is no link between positive transient eigenvalue and stress overshoot. We also reported that the transiently positive eigenvalue does not lead to growth of linearized perturbations for the Johnson-Segalman model. 

For some cases of shear startup of non-stretching Rolie-Poly model, we found agreement between stress overshoot, transiently positive eigenvalue and growth of perturbations as reported earlier by Moorcroft and Fielding [\onlinecite{moorcroft14}]. Using the Giesekus model, we showed that a transiently positive eigenvalue is not necessary for transient growth of linearized perturbations as the transient eigenvalue is not positive if the steady state is stable. We attributed the contrasting observations of Johnson-Segalman and non-stretching Rolie-Poly models to the difference in orders of magnitude of ratio of solvent to solution viscosity which is $0.115$ for the Johnson-Segalman model and $10^{-5}$ for non-stretching Rolie-Poly models. For high shear rates, we showed that the maximum transient eigenvalue diverges on decreasing the ratio of solvent to solution viscosity. We also argued that if transient growth rate becomes significantly high then inertial effects can no longer be ignored. We assumed the base state evolution to be inertialess and the evolution of perturbations in presence of inertia and found that transient eigenvalue does not diverge and saturates to a constant eigenvalue. The growth of linearized perturbations also reduced significantly in the presence of inertia. We concluded that there may not be any transient shear banding if shear startup flow is solved for higher values of solvent to solution viscosity. If the ratio of solvent to solution viscosity is less than $O(10^{-3})$, then inclusion of inertia may regularise the evolution of eigenvalue and perturbations. However, this conclusion is based on an approximate solution in which inertia is considered only during evolution of perturbations and not during base state. Furthermore, the growth of linearized perturbations also does not necessarily guarantee the presence of transient shear banding since the nonlinear terms are ignored. Therefore, it is of interest to examine whether transient shear banding is observed after considering nonlinear terms. We also determine the effect of inertia on transient dynamics of shear startup flow and whether inertia regularises the effect of low ratio of solvent to solution viscosity.

In this work, we study the shear startup flow of the Johnson-Segalman and non-stretching Rolie-Poly models using full nonlinear simulations by imposing perturbations at the beginning of the shear startup. We discuss both the models and governing equations in Sec.\,\ref{section_model}. We also discuss the issue of realistic initial amplitude of perturbations to study transient shear banding in Sec.\,\ref{section_model}. We explore the effect of initial amplitude of perturbation on transient dynamics of shear startup flow in presence and absence of inertia in Sec.\,\ref{section_A_effect}. We also study the effect of decreasing ratio of solvent to solution viscosity and inclusion of inertia in Sec.\,\ref{section_Re_effect}. We discuss the salient conclusions of this study in Sec.\,\ref{section_conclusion}.

\section{Models and governing equations} \label{section_model}

We use the Johnson-Segalman and non-stretching Rolie-Poly models to analyse shear startup of viscoelastic fluids. We consider an incompressible viscoelastic fluid between two parallel plates that are infinite in the $x^*$ and $z^*$ directions and confined in the $y$-direction, with a gap $H$ between the plates. Both plates are at rest for time $t^*<0$, but for time $t^*\geq0$, the top plate moves at a steady velocity $U$ in the $x^*$ direction. The variables in this manuscript that include a superscript $*$ are dimensional, while those without one are dimensionless. In addition, we assume that the no-slip boundary condition holds at both the plates. In this case, the continuity equation is satisfied automatically and the simplified Cauchy momentum equation can be written as follows:
\begin{equation}\label{inertialess}
   \nabla \cdot \underset{\approx}{\Sigma}^*=\rho\frac{\partial \underset{\sim}{u}^*}{\partial t^*},
\end{equation}
where, $\rho$ is the density of the fluid, $\underset{\sim}{u^*}$ is the velocity field, and \({\underset{\approx}{\Sigma }^{*}}\) is the total stress tensor. We consider the total stress to be equal to the sum of viscoelastic stress and Newtonian solvent stress contributions:
\begin{equation}\label{totalstress}
    {\underset{\approx}{\Sigma }^{*}}={\underset{\approx}{\sigma }^{*}}+2{{\bar{\eta}}_{s}}{\underset{\approx}{\dot{\gamma }}^{*},}
\end{equation}
where, $\underset{\approx}{\sigma }^{*}$ is the viscoelastic stress, ${\bar{\eta}}_{s}$ is the viscosity of solvent in a polymeric or wormlike micellar solution, and ${\underset{\approx}{\dot{\gamma }}}^{*}= \frac{1}{2}(\nabla{{\underset{\sim}{u}^{*}}} + (\nabla{{\underset{\sim}{u}^{*}}})^{T}) $ is the shear rate tensor. We use the Johnson-Segalman [\onlinecite{johnson1977model}] and the Rolie-Poly [\onlinecite{likhtman2003simple}] models for the viscoleastic stress $(\underset{\approx}{\sigma }^{*})$. Both models are augmented by an additional stress diffusion term for unique stress selection in the non-monotonic region of the flow curve [\onlinecite{olmsted2000johnson,lu2000effects}]. The models are made dimensionless as follows: $\underset{\approx}{\sigma }=\displaystyle \frac{\underset{\approx}{\sigma }^{*}}{\left((\bar{\eta}_s+\bar{\eta}_p)/\tau\right)}$, $t=t^*/\tau$, $\underset{\sim}{u}=\underset{\sim}{u}^*/U$. In the above expressions, $\bar{\eta}_p$ is the contribution of polymer to the zero shear viscosity of the solution and $\tau$ is the relaxation time of the solution. In the case of Rolie-Poly model, $\tau$ = $\tau_D$, which represents the reptation time, and for Johnson-Segalman model $\tau$ = $\lambda$ which is the longest relaxation time. The relevant dimensionless groups are the Weissenberg number, $Wi = \displaystyle \frac{\tau U}{H}$ which represents non-dimensional shear rate, Reynolds number $Re=\displaystyle \frac{\rho U H}{\bar{\eta}_s+\bar{\eta}_p}$, and the ratio of solvent viscosity to the zero shear viscosity of the solution, $\eta_s = \displaystyle \frac{\bar{\eta}_s}{\bar{\eta}_s+\bar{\eta}_p}$. The dimensionless form of Cauchy momentum equation is given below:

\begin{equation}\label{dimensionless_cauchy}
    Re\frac{\partial u}{\partial t}=\frac{\partial\sigma_{xy}}{\partial y}+\eta_sWi\frac{\partial^2 u}{\partial y^2}
\end{equation}

\textit{Johnson-Segalman model} - The Johnson-Segalman (hereafter JS) model [\onlinecite{johnson1977model}] was developed to introduce non-affine motion in the upper-convected Maxwell (UCM) model by replacing the upper-convected derivative with the Gordon-Schowalter derivative [\onlinecite{birddynamics,gordon1972anisotropic,johnson1977model}]. The degree of non-affine motion is governed by the slip parameter $\xi$. If $\xi=0$, the JS model constitutive equations yields the UCM model and if $\xi=2$, it reduces to the Lower Convective Maxwell (LCM) model [\onlinecite{birddynamics}]. The constitutive equation of the JS model is given by the following equation
\begin{equation}
 \underset{\approx}\sigma^*+\lambda \overset{\square}{\underset{\approx}{\sigma}^*}  =2{\bar{\eta}}_p \underset{\approx}{\dot{\gamma}^*}
\end{equation}
and the Gordon-Schowalter derivative $\overset{\square}{\underset{\approx}{\sigma}}$ [\onlinecite{birddynamics}] is
\begin{equation}
   \overset{\square}{\underset{\approx}{\sigma^*}}= \left( 1-\frac{\xi}{2} \right){\overset{\triangledown}{\underset{\approx}{\sigma^*}}}+\left( \frac{\xi}{2} \right){\overset{\triangle}{\underset{\approx}{\sigma^*}}}
\end{equation}
where, $\lambda$ is the longest relaxation time of the polymeric solution, $\bar{\eta}_p$ is the contribution of polymer to the zero shear viscosity of the solution, and $\xi$ is fixed as $0.01$ throughout the results shown in this manuscript. The upper-convected and the lower-convected derivatives are ${\overset{\triangledown}{\underset{\approx}{\sigma}}}$ and ${\overset{\triangle}{\underset{\approx}{\sigma}}}$, respectively [\onlinecite{LARSON1988129}]. 
The simplified non-dimensional component-wise equations of the model for simple shear flow are
\begin{equation}\label{jsxy}
    \frac{\partial {{\sigma }_{xy}}}{\partial t} = \left( -\left( \frac{\xi }{2} \right){{\sigma }_{xx}}+\left( 1-\frac{\xi }{2} \right){{\sigma }_{yy}}+\left( 1-{\eta}_{s} \right) \right)\dot{\gamma }Wi-{{\sigma }_{xy}}+D\frac{{{\partial }^{2}}\sigma_{xy}^{{}}}{\partial {{y}^{2}}}
\end{equation}

\begin{equation}\label{jsxx}
\frac{\partial {{\sigma }_{xx}}}{\partial t} = 2\left( \left( 1-\frac{\xi }{2} \right){{\sigma }_{xy}} \right)\dot{\gamma }Wi-{{\sigma }_{xx}}+D\frac{{{\partial }^{2}}\sigma_{xx}^{{}}}{\partial {{y}^{2}}}
\end{equation}

\begin{equation}\label{jsyy}
\frac{\partial {{\sigma }_{yy}}}{\partial t} = -\left( \xi {{\sigma }_{xy}} \right)\dot{\gamma }Wi-{{\sigma }_{yy}}+D\frac{{{\partial }^{2}}\sigma_{yy}^{{}}}{\partial {{y}^{2}}}
\end{equation}



\textit{Rolie-Poly model} - The Rolie-Poly model is a single mode molecular model for entangled polymer melts with its name derived from ROuse LInear Entangled POLYmers model. This model was developed by Likhtman and Graham by simplification of the Doi-Edwards tube model [\onlinecite{doi1988theory,likhtman2003simple}]. This model also accounts for most of the molecular processes, including reptation, convective constraint release (CCR), chain stretch, retraction, and contour length fluctuations. The constitutive equation of the model can be expressed as follows 

\begin{equation} \label{rpmodel}
    \left( \underset{\approx}{\sigma}^*-\underset{\approx}{I} \right)+{\tau _D}\overset{\nabla }{\mathop{\underset{\approx}{\sigma}^*}}=-2\frac{{{\tau }_{D}}}{{{\tau }_{R}}}\left( 1-\sqrt{\frac{3}{tr\left( \underset{\approx}{\sigma}^* \right)}} \right)\left( \underset{\approx}{\sigma}^*+\beta \left( \underset{\approx}{\sigma}^*-\underset{\approx}{I}  \right){{\left(\sqrt{ \frac{3}{tr\left( \underset{\approx}{\sigma}^* \right)}} \right)}^{-2\delta }} \right)
\end{equation}
where, \( {\beta} \)  determines the effectiveness of the convective constraint release mechanism, \( {\delta} =\frac{1}{2}\) following [\onlinecite{likhtman2003simple,holroyd2017analytic}] which fixes the strength of convective constraint release. Here, $\underset{\approx}{I}$ is the identity tensor. Also, \({\tau}_{D} \) determines the contribution of reptation to relaxation mechanism, ${\tau }_{R}$ shows the contribution of chain stretching to relaxation mechanism. The number of entanglements is represented by \(Z\) \( \left(Z=\displaystyle \frac{{\tau }_{D}}{{3\tau }_{R}}\right)\), which consequently fixes the two relaxation times. 

Likhtman and Graham [\onlinecite{likhtman2003simple}] further simplified the model by considering  \({\tau }_{R}\rightarrow 0\) and \(tr\left(\underset{\approx}{\sigma}^*\right)=3+\Delta\) (here \(\Delta = 0\) because \({\tau }_{R}\rightarrow 0\)). The simplified model is known as non-stretching Rolie-Poly model referred hereafter as the nRP model whose constitutive equation is given by:

\begin{equation} \label{nrpmodel}
\left( \underset{\approx}{\sigma}^*-\underset{\approx}{I} \right)+{\tau _D}\overset{\nabla }{\mathop{\underset{\approx}{\sigma}^*}}=-\frac{2}{3}{\tau }_{D}\left(tr\left(\underset{\sim}{\nabla}^* \underset{\sim}{u}^*\cdot\underset{\approx}{\sigma}^*\right)\right)\left( \underset{\approx}{\sigma}^*+\beta \left( \underset{\approx}{\sigma}^*-\underset{\approx}{I}  \right)\right)
\end{equation}
The component-wise equations of the nRP model in nondimensional form for simple shear flow are as follows: 

\begin{equation}\label{nrpxy}
    \frac{\partial {{\sigma}_{xy}}}{\partial t}=Wi\dot{\gamma }\left( {{\sigma}_{yy}}-\frac{2}{3}\sigma_{xy}^{2}\left( 1+\beta  \right) \right)-{{\sigma}_{xy}}+D\frac{{{\partial }^{2}}\sigma_{xy}^{{}}}{\partial {{y}^{2}}}
\end{equation}

\begin{equation}\label{nrpxx}
    \frac{\partial {{\sigma}_{yy}}}{\partial t}=\frac{2}{3}Wi\dot{\gamma }\left( \beta {{\sigma}_{xy}}-{{\sigma}_{yy}}\sigma_{xy}^{{}}\left( 1+\beta  \right) \right)-\left( {{\sigma}_{yy}}-1 \right)+D\frac{{{\partial }^{2}}\sigma_{yy}^{{}}}{\partial {{y}^{2}}}
\end{equation}
\\
 The results presented in this manuscript using the nRP model are obtained using $\beta=0.6$ to solve shear startup flow with shear rate in the nonmonotonic region of the constitutive curve and $\beta=1$ is used for shear rate in monotonic region of the constitutive curve. 
We carry out full non-linear simulations for simple shear flow using JS and nRP model and solve Eqs.~\ref{inertialess}, Eqs.~\ref{jsxy}-\ref{jsyy} and Eqs.~\ref{nrpxy}-\ref{nrpxx}. In this study, we impose a perturbation as initial condition for $\underset{\sim}{u}$ and $\underset{\approx}{\sigma}$ which is expressed as follows:
\begin{equation}\label{velocitynp}
    \underset{\sim}{u}=\underset{\sim}{u}^0+A\sin(n \pi y)
\end{equation} 
\begin{equation}\label{stressnp}
    \underset{\approx}{\sigma}={\underset{\approx}{\sigma}^{0}} + A\cos(n\pi y)
\end{equation}
where, $A$ is the amplitude of sine or cosine wave and $n$ fixes its wavelength. All stress components and shear rate perturbation are chosen as cosines, so that the velocity perturbation can be written in terms of sines, and thus the no-slip boundary condition gets satisfied at $y = 0$ and $1$. In this manuscript, $n$ (in Eqs.\,\ref{velocitynp} and \ref{stressnp}) is fixed as $1$ throughout the manuscript, as it yields the most unstable mode and $A$ is considered as $10^{-1}$, $10^{-4}$, and $10^{-7}$. 

\subsection{A note on the magnitude of shear rate initial condition}

When the JS and nRP models are solved for shear startup in the absence of inertia, then it is not possible to specify the initial amplitude of shear rate because the governing Cauchy momentum equation is devoid of any acceleration in the inertialess limit. 
In this limit, the initial value of the shear rate is actually dictated by the initial value of $\sigma_{xy}$ as given by the Cauchy momentum equation (Eq.\,\ref{dimensionless_cauchy}) in the absence of inertia: 
$\dot{\gamma}(t = 0) = \sigma_{xy}(t = 0)/(\eta_s Wi)$. 
As a consequence, for $\eta_s \ll 1$,  the initial shear rate perturbation is $O(\eta_s^{-1}) \sigma_{xy}(t = 0)$, i.e., it is much larger than the magnitude of the stress perturbation. Thus, for $\eta_s \ll 1$, if we impose an $O(1)$ stress perturbation at $t = 0$, the shear rate perturbations become much larger than unity. Note that the step change in the shear rate, in the shear start-up protocol, is itself from 0 to 1. However, if we use a $\sigma_{xy} \sim O(1)$ at $t = 0$, this would yield a very high value of the shear rate at $t = 0$. It is unrealistic to impose a large (i.e. $\gg O(1))$ shear rate perturbation as an initial condition in the nonlinear simulations, since the maximum imposed shear rate in the base state itself is $O(1)$. Note that the initial conditions that we use in our numerical simulations are intended to mimic inevitable (spontaneous) perturbations that would be present in any laboratory experiment. Thus, in order for the initial conditions in the numerical simulations to be physically realistic, one should not have the shear rate perturbation magnitudes to be much larger than the base-state shear rate itself, which is unity in our case. In order to address this issue, especially in the limit of small $\eta_s$, it is necessary to prescribe the initial shear stress perturbation amplitudes in a manner that the initial shear rate perturbation is at most $O(10^{-1})$ (i.e.\,10\% of the imposed shear rate jump in the startup flow).

The non-linear simulations are carried out using COMSOL 5.0\textsuperscript{\textregistered}. The inbuilt partial differential equation solver of COMSOL utilizes the finite element method. 
The domain (0,1) is discretized into 6452 points 
(also, the results have been verified for mesh size and time stepping). The transient evolution of nRP model for shear start-up to shear rate in the non-monotonic region of the constitutive curve is sensitive to the change in number of domain elements by an order of magnitude (owing to the linearly unstable velocity profile). However, the overall conclusion of the study remains same for all these results. The results from the JS model and the nRP model are same for 6000-10000 domain elements. The relative and absolute tolerance is kept at $10^{-5}$ and $10^{-6}$, respectively.


\section{Results and discussion}

In our earlier work [\onlinecite{sharma2021onset}], we demonstrated that if there is a stress overshoot during shear startup, this does not guarantee a positive eigenvalue within the frozen-time linear stability analysis, nor does it result in the growth of perturbations within a more accurate analysis of linearized dynamics using the fundamental matrix approach. (Some of the caveats of the frozen time analysis has also been mentioned in Ref.\,[\onlinecite{moorcroft14}].) The former, in particular, has been used as a signature of transient instability in the literature [\onlinecite{moorcroft14,adams2011transient}]. While results from linearized dynamics are suggestive of the time evolution when the perturbations are infinitesimal, the ultimate answer to the question of whether transient shear banding is present or not, must come from fully-nonlinear solutions of shear startup, which is the objective of the present study. Here, we numerically solve the partial differential equations mentioned in Sec.\,\ref{section_model}, and determine the relevance of stress overshoot, positive eigenvalue within a frozen-time stability analysis, and the growth and decay of linearized perturbations on transient shear banding.
We analyze the shear startup of JS and nRP models with the shear rates in both monotonic and nonmonotonic regions of the constitutive curves. As discussed in Sec.\,\ref{section_intro}, shear banding can be identified and quantified more reliably by peak in second derivative of velocity [\onlinecite{kushwaha2022dynamics,jain2018role}]. However, in this work, to quantify steady state and transient shear banding in each case, we calculate the degree of banding which is defined as the difference between maximum and minimum shear rates of the flow of fluid between parallel plates [\onlinecite{moorcroft14,adams2011transient}]. If the degree of shear banding is zero, then the flow is homogeneous otherwise it is inhomogeneous and its magnitude determines the extent of inhomogeneity in the flow. 
 We also show the corresponding velocity profiles as a function of time to correlate them with degree of banding. As noted in the preceding section, we specifically address the question of realistic amplitudes of the initial imposed stress perturbation under the creeping-flow assumption. In the subsequent subsections (Sec.\,\ref{section_eta_s_effect}-\ref{section_Re_effect}), we examine the influence of flatness of constitutive curve (or solvent viscosity) and inertia.

\subsection{Effect of initial amplitude of perturbation} \label{section_A_effect}

We first discuss the effect of initial amplitude of perturbation $(A)$ by considering both \linebreak $Re=0$ and $Re \neq 0$. By computing the transient dynamics at varying magnitudes of $A$, we examine the extent of validity of results of linearized dynamics, which are strictly valid in the limit \linebreak $A\ll1$. For $Re = 0$, the initial shear rate perturbation amplitude cannot be specified directly and is dictated by the magnitude of the initial stress perturbation $\sigma_{xy}(t = 0)$ from the Cauchy momentum equation~(Eq.\,\ref{dimensionless_cauchy}). However, for $Re\neq0$, initial shear rate perturbation can be specified directly as mentioned in preceding section. In the following subsections, we also compare the shear stress evolution for different $A$ which is obtained using full nonlinear simulations. For comparison, we also show shear stress evolution for forced homogeneous flow which is obtained by solving the JS and nRP models with the assumption of linear velocity profile at all times, $Re=0$ and no stress diffusion.
\begin{figure}
   \centering
    \subfigure[]{
    \includegraphics[scale=0.25]{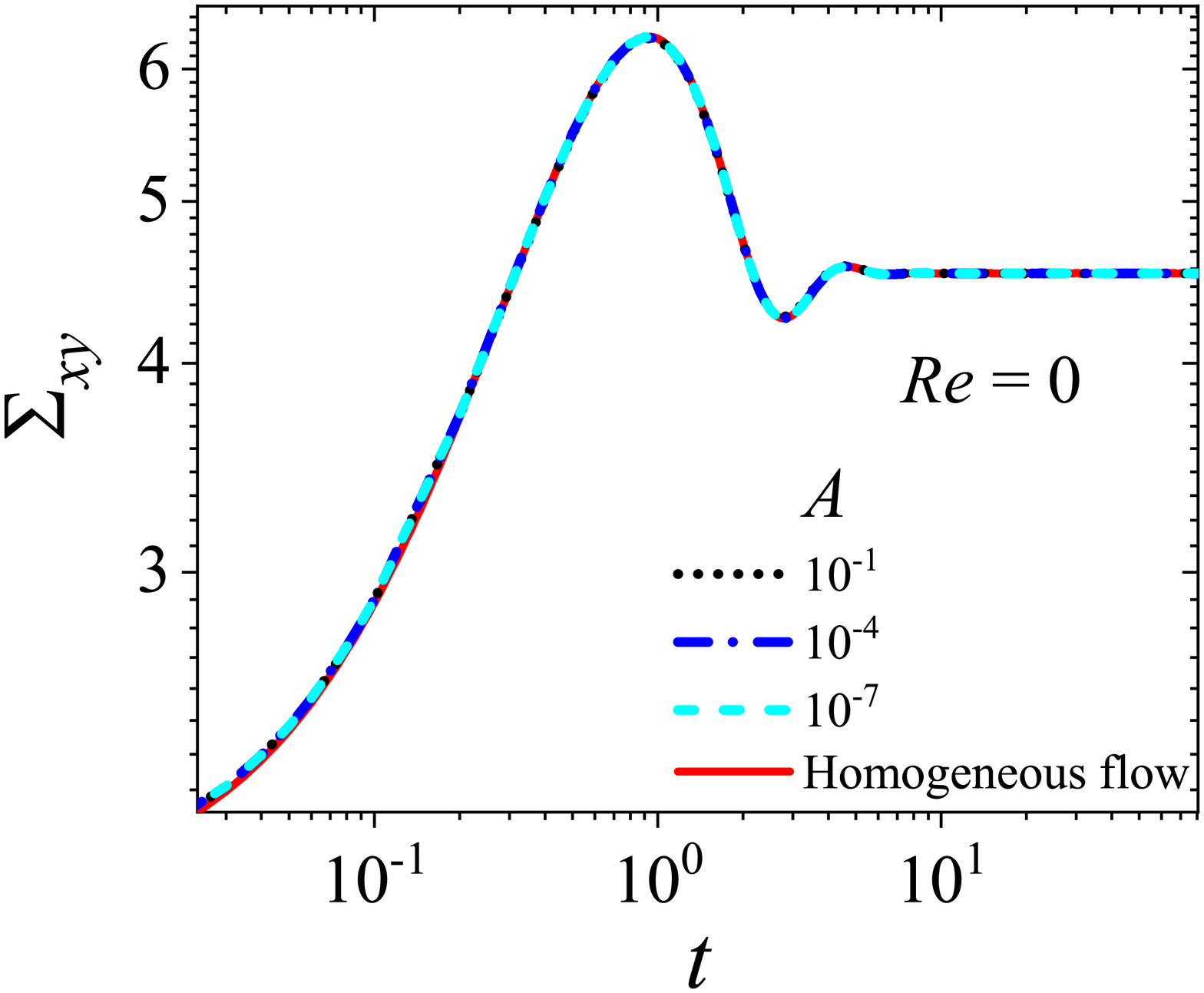}
    \label{js_A_mono_stress_Re_zero}
    }
    \subfigure[]{
    \includegraphics[scale=0.25]{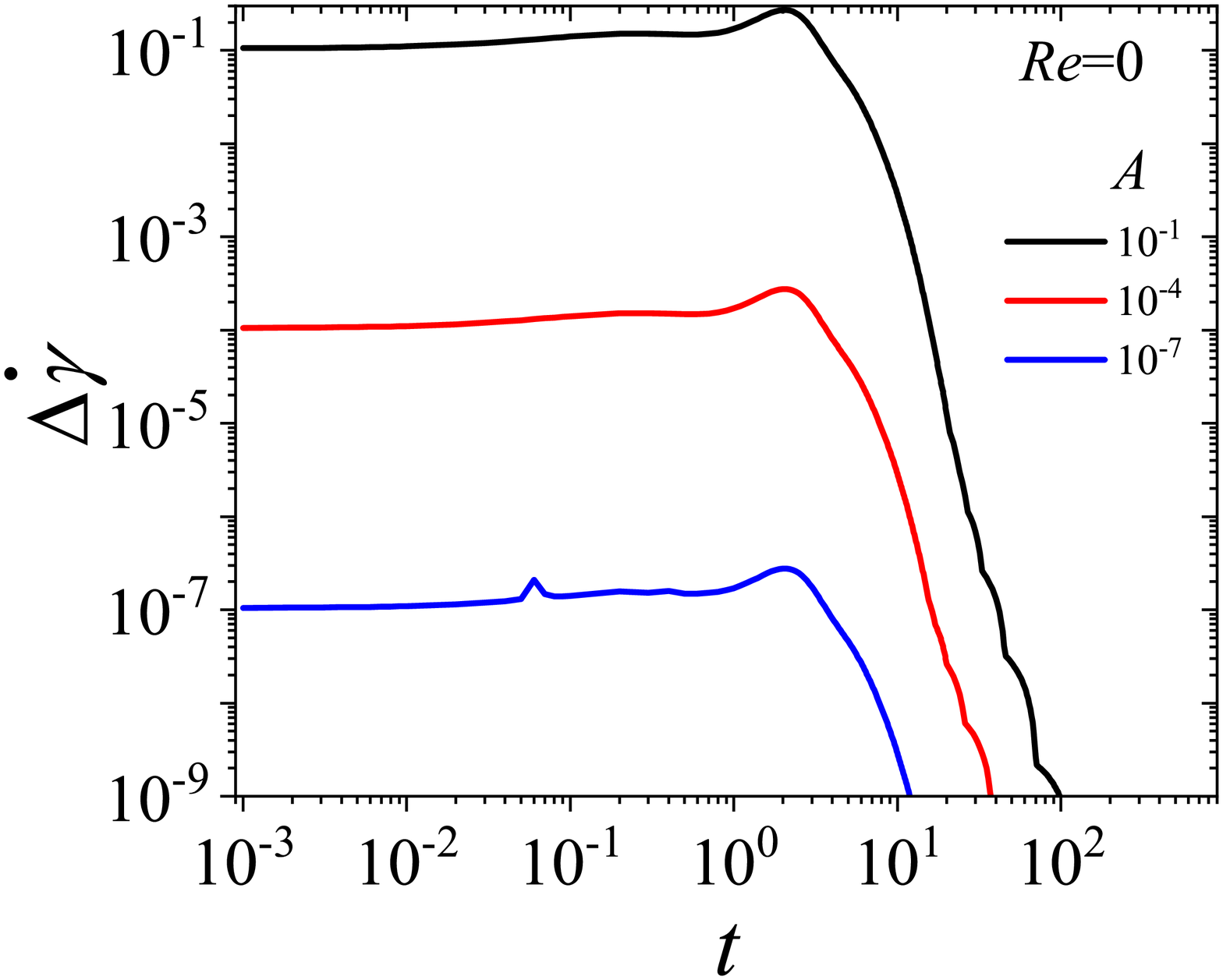}
    \label{js_A_mono_dob_Re_zero}
    }
    \caption{Effect of amplitude of perturbation $(A)$ for shear startup of the JS model for $Re=0$. Here, $Wi=12$, and $\eta_s=0.16$, which corresponds to a shear rate in the monotonic region of the constitutive curve. (a) Shear stress evolution is plotted for $A=$ $10^{-1}$, $10^{-4}$, $10^{-7}$ and for a forced homogeneous flow. (b) The variation of degree of banding $(\Delta\dot{\gamma}=\dot{\gamma}_{max}-\dot{\gamma}_{min})$ with time is shown for different values of $A=$ $10^{-1}$, $10^{-4}$, and $10^{-7}$ for $Re=0$. }
    \label{fig:JS_effect_of_A_monotonic}
\end{figure}

\subsubsection{$Re=0$} \label{section_Re_zero}

\paragraph{JS model} \label{section_Re_zero_js}

\begin{figure} 
   \centering
    \subfigure[]{
    \includegraphics[scale=0.25]{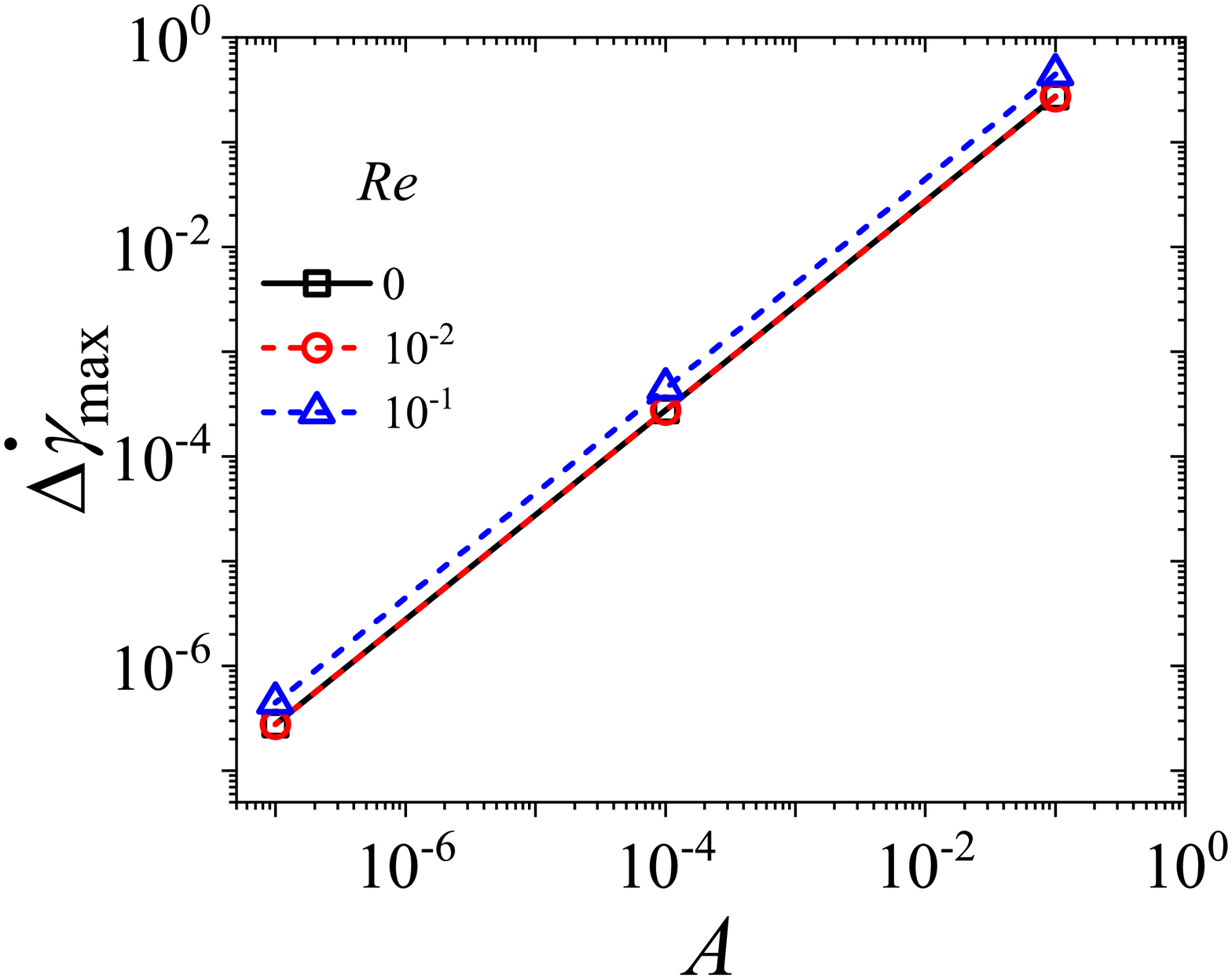}
    \label{js_dobmax_mono}
    }
    \subfigure[]{
    \includegraphics[scale=0.25]{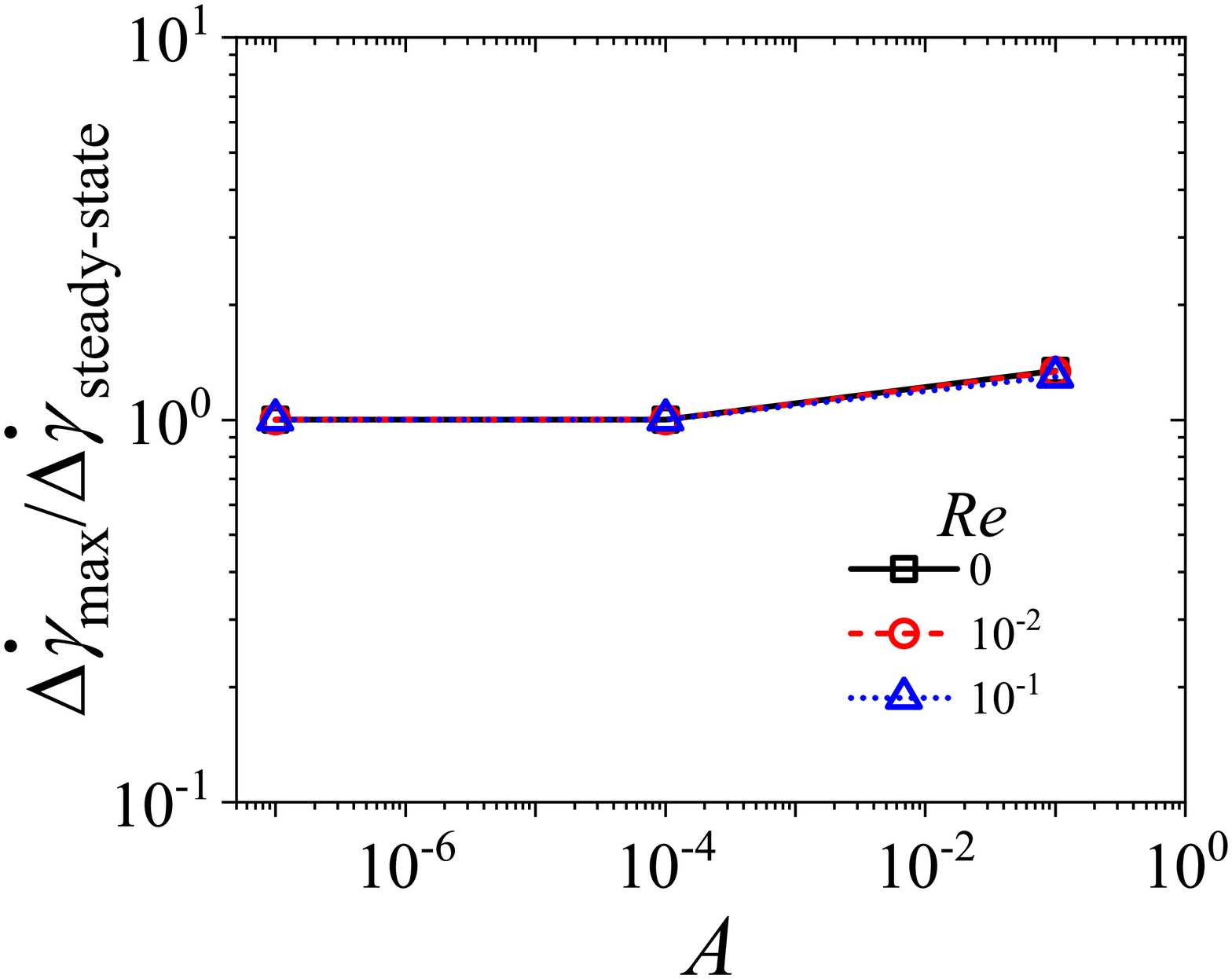}
    \label{js_dobmax_nonmono}
    }
    \caption{(a) The maximum of degree of banding $(\Delta\dot{\gamma}_{max})$ is plotted for shear startup of JS model if the shear rate is in the monotonic $(Wi=12,\eta_s=0.16)$ and (b) the ratio of maximum of degree of banding $(\Delta\dot{\gamma}_{max})$ with degree of banding at steady state $(\Delta\dot{\gamma}_{steady-state})$ is plotted if the shear rate is in the nonmonotonic $(Wi=12,\eta_s=0.05)$ region of the constitutive curve for $A=$ $10^{-1}$, $10^{-4}$, $10^{-7}$ and $Re=$ $0$, $10^{-2}$, $10^{-1}$.}
    \label{fig:JS_dob_max_vs_A}
\end{figure}

We first study shear startup of the JS model in the creeping-flow limit, and for shear rate in the monotonic region of the constitutive curve. In this case, we fix $\eta_s=0.16$ and $Wi=12$ so that the shear rate is in a flatter and monotonic region of the constitutive curve. The value of $\eta_s$ is much higher as compared to $\eta_s=10^{-4}$ used below for the nRP model. This is because in the case of JS model $\eta_s$ cannot be decreased below $1/9$ as for $\eta_s<1/9$ constitutive curve is nonmonotonic as also shown in Fig.\,1 of our previous study [\onlinecite{sharma2021onset}].  We use $A=10^{-1}$, $10^{-4}$, and $10^{-7}$ for the initial amplitude of perturbations. For $Re=0$, the initial shear rate perturbation is of the order of $10^{-1}$, $10^{-4}$, and $10^{-7}$ for $Wi=12$ and $\eta_s=0.16$. Figure\,\ref{fig:JS_effect_of_A_monotonic} shows shear stress and degree of banding $(\Delta\dot{\gamma}=\dot{\gamma}_{max}-\dot{\gamma}_{min})$ as a function of time. The corresponding velocity profiles are shown in Fig.\,S1 of the supplementary information. During shear startup, the shear stress increases, shows an overshoot and then attains a steady state in all the three cases. We find that shear stress for all three values of $A$ overlaps for all times and $\Delta\dot{\gamma}$ is initially of the order of $A$ before finally decaying to zero. These results show that if imposition of perturbation of initial amplitude $A$ to shear startup flow engenders a response which is commensurate with the forcing, this cannot be interpreted as a signature of a transient (elastic) instability. In such situation, the system is merely adjusting itself to the perturbation before reaching the steady state. Therefore, on this count, Fig.\,\ref{js_A_mono_dob_Re_zero} clearly shows that there is no instability in the JS model from nonlinear simulations. We also find that the maximum degree of banding $(\Delta\dot{\gamma}_{max})$ to be of the order of $A$ as shown in Fig.\,\ref{js_dobmax_mono}. It can also be seen from Fig.\,\ref{js_dobmax_mono} that the maximum value of the degree of banding $(\Delta\dot{\gamma}_{max})$ increases in a linear manner with $A$. Similarly, the velocity profiles plotted in Fig.\,S1 of the supplementary information show no transient shear banding or any significant deviation from linear velocity profile for all the explored values of $A$ at $Re=0$.
\begin{figure}
   \centering
     \subfigure[]{
    \includegraphics[scale=0.25]{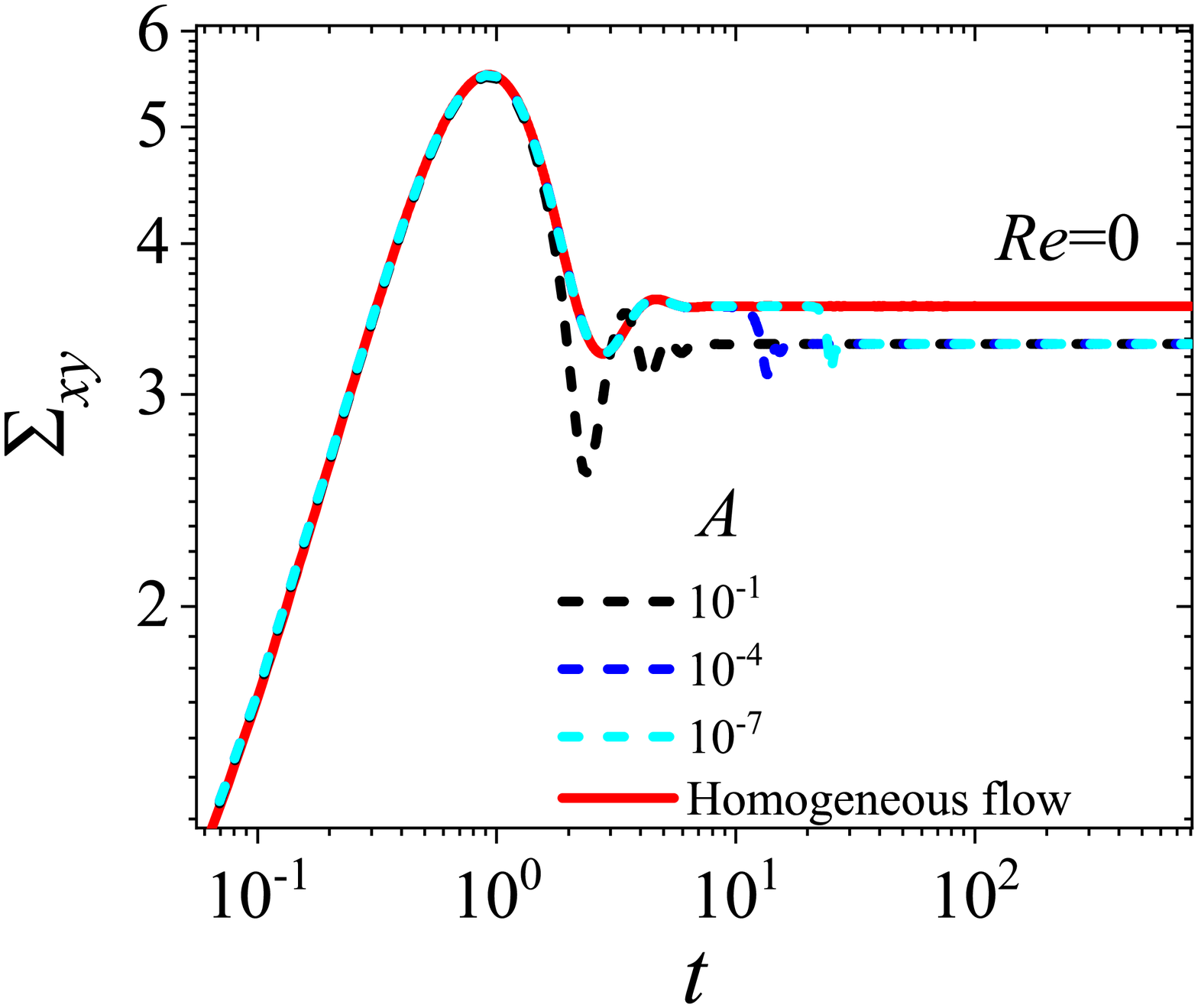}
    \label{js_A_nonmono_stress_Re_zero}
    }
     \subfigure[]{
    \includegraphics[scale=0.25]{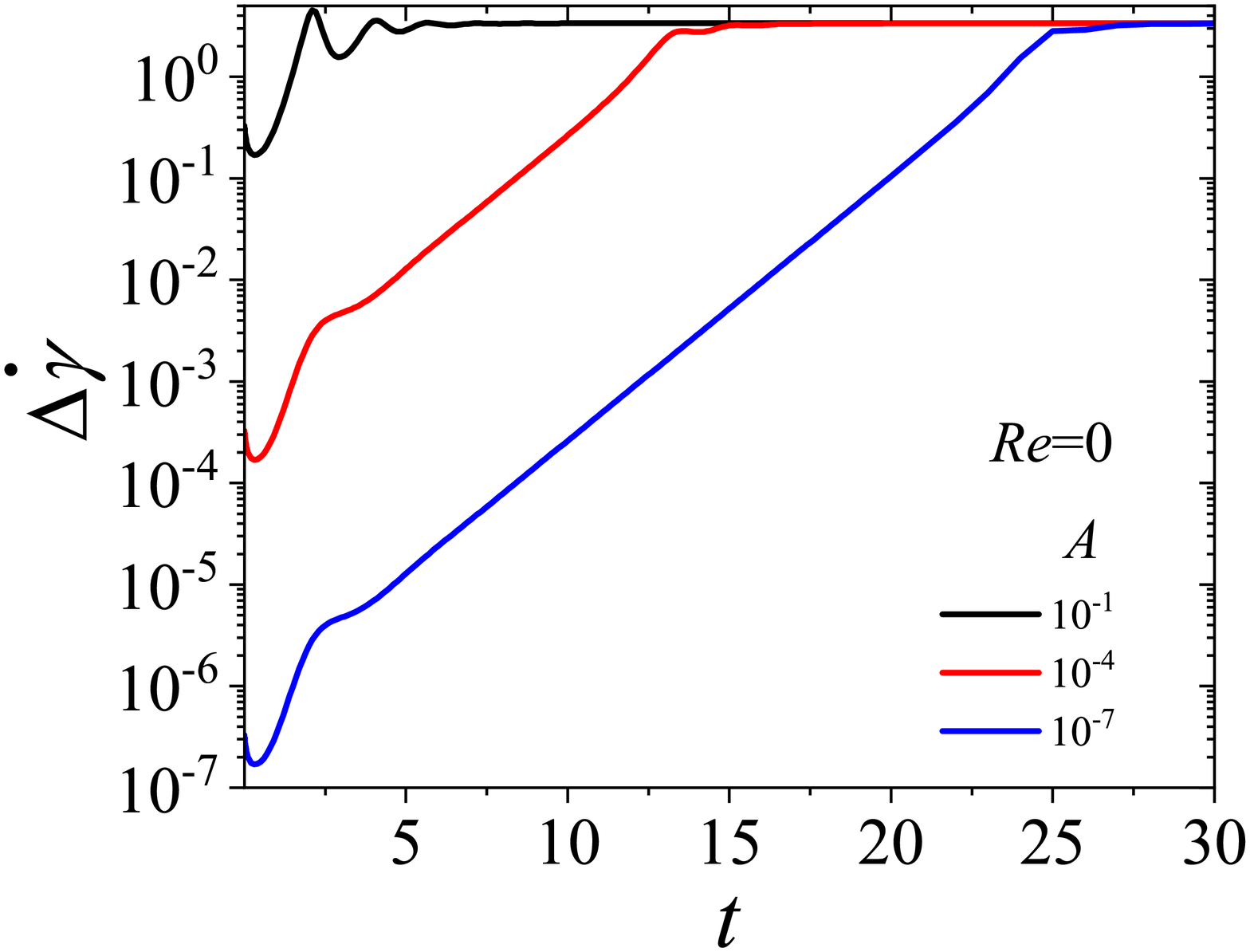}
    \label{js_A_nonmono_dob_Re_zero}
    }
    \caption{Effect of amplitude of perturbation $(A)$ for shear startup of the JS model for $Re=0$. Here, $Wi=12$, and $\eta_s=0.05$ which corresponds to a shear rate in the nonmonotonic region of the constitutive curve. (a) Shear stress evolution is plotted for $A=$ $10^{-1}$, $10^{-4}$, $10^{-7}$ and for a forced homogeneous flow. (b) The variation of degree of banding $(\Delta\dot{\gamma}=\dot{\gamma}_{max}-\dot{\gamma}_{min})$ with time is shown for different values of $A=$ $10^{-1}$, $10^{-4}$, and $10^{-7}$ for $Re=0$. }
    \label{fig:JS_effect_of_A_nonmonotonic}
\end{figure}

We next study the effect of initial amplitude of perturbation in the shear startup flow of the JS model for the case when the shear rate is in the nonmonotonic region of the constitutive curve and steady state shear banding is observed. We solve the JS model at $Wi=12$, $\eta_s=0.05$, and $A=10^{-1}$, $10^{-4}$, and $10^{-7}$ for initial amplitude of perturbations with $Re=0$. Figures \ref{fig:JS_effect_of_A_nonmonotonic} (a) and (b) show the shear stress and the degree of banding as a function of time. The shear stress increases and shows an overshoot before attaining steady state. The evolution of $\Delta\dot{\gamma}$ shows a linear increase before attaining steady state as shown in Fig.\,\ref{js_A_nonmono_dob_Re_zero}. The initial linear increase of $\Delta\dot{\gamma}$ with time on a semi-log plot clearly demonstrates the exponential growth of perturbations, which is expected as shear rate is in the linearly unstable region of constitutive curve [\onlinecite{yerushalmi1970stability}]. Figure \ref{js_dobmax_nonmono} also shows that $\Delta\dot{\gamma}_{max}$ is same as $\Delta\dot{\gamma}_{steady-state}$ for $A=10^{-4}$ and $10^{-7}$, and slightly higher for $A=10^{-1}$. The stress evolution overlap with each other in all these cases, except just before attaining a final steady-state stress. We find that the time of attaining steady state value of degree of banding $(\Delta\dot{\gamma})$ depends on the value of $A$. Higher the value of $A$, lower is the time required for the JS model to become unstable and attain steady state shear banding. These results also demonstrate that perturbations begin to grow exponentially at time $t<1$ (i.e. $t\approx0.3$). Consequently, the onset of instability (as exemplified by the exponential growth at early times) has no correlation with the time during which stress shows any decrease after its overshoot. Also, there is no pronounced transient increase (i.e., $\Delta\dot{\gamma}$ is not greater than $\Delta\dot{\gamma}_{steady-state}$) during the time range in which there is a stress decay after overshoot as shown in Fig.\,\ref{fig:JS_effect_of_A_nonmonotonic}. We also observe that for $A=10^{-1}$, there are some oscillations in the variation of $\Delta\dot{\gamma}$ with time. However, the peak values of $\Delta\dot{\gamma}$ during these oscillations are not significantly higher than the value of $\Delta\dot{\gamma}$ at steady state. Therefore, these oscillations cannot be treated as a signature of transient shear banding. 

The above results show that under creeping-flow assumption and shear rate in the monotonic region of the constitutive curve, JS model shows (i) no transient shear banding even though the linearized perturbation showed growth and decay during the flow, (ii) the maximum value of degree of banding ($\Delta\dot{\gamma}_{max}$) is of the order of $A$ and varies in a linear manner with $A$, and (iii) the time associated with maximum degree of banding depends on $A$. If the shear rate is in the nonmonotonic region of the constitutive curve, then also distinct transient shear banding is not observed and increase in degree of banding begins almost at beginning of the flow and this time is much lesser than the time at which shear stress decreases after its overshoot.

\paragraph{nRP model} \label{scetion_Re_zero_nrp}
\begin{figure}
    \centering
    \subfigure[]{
    \includegraphics[scale=0.25]{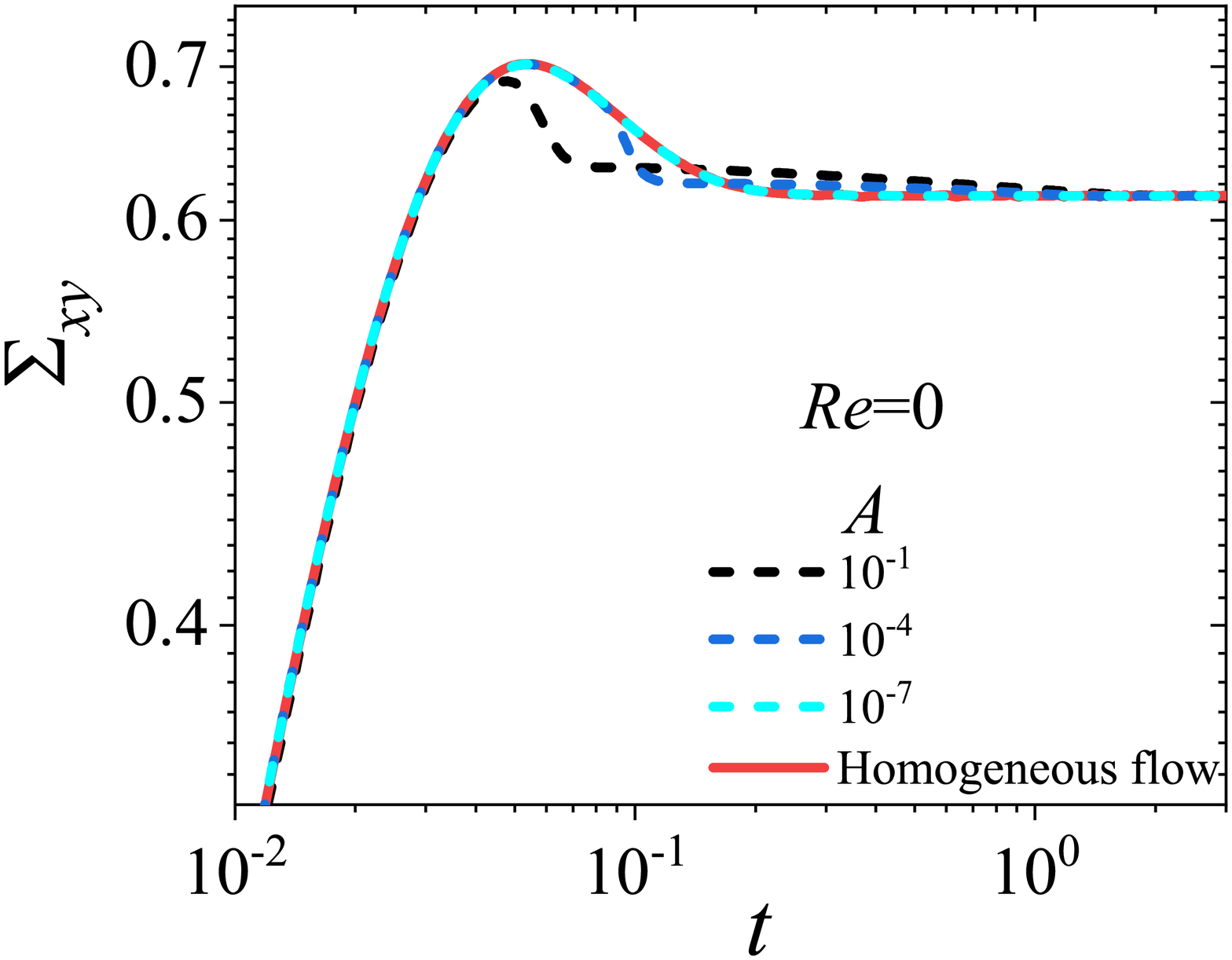}
    \label{nrp_effect_of_A_stress_mono}
    }
    \subfigure[]{
    \includegraphics[scale=0.25]{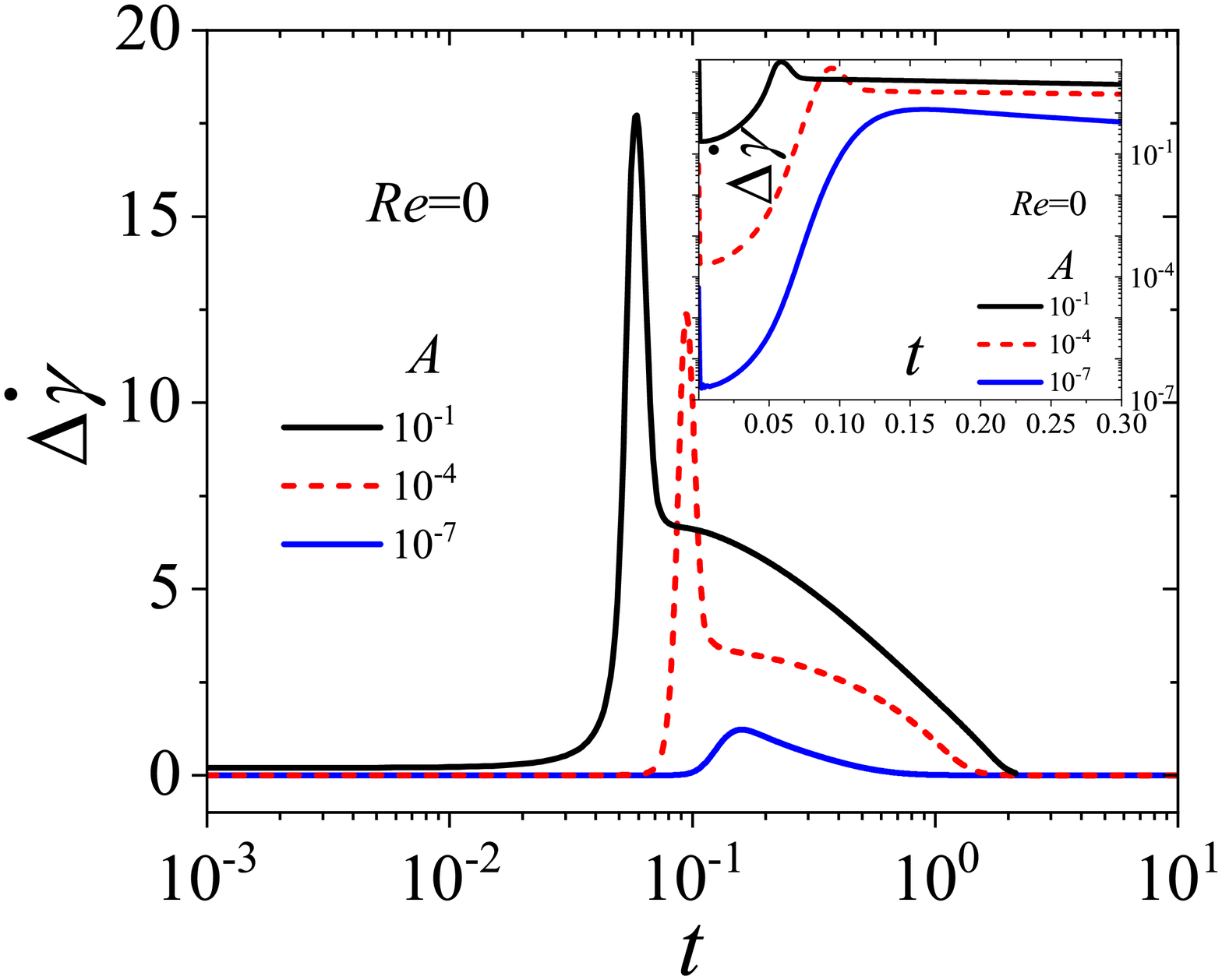}
    \label{nrp_effect_of_A_dob_mono}
    }
      \subfigure[]{
    \includegraphics[scale=0.25]{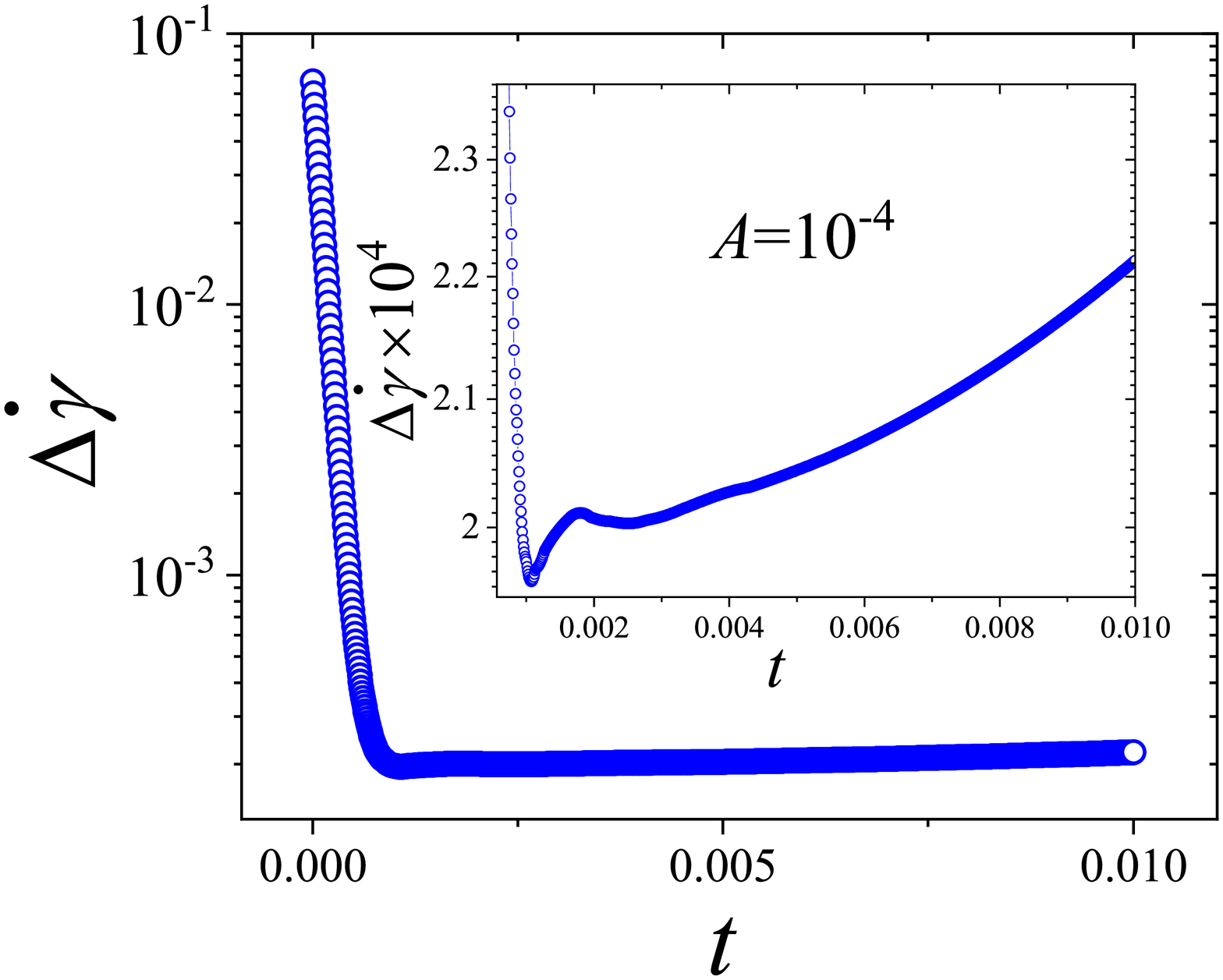}
    \label{nrp_effect_of_A_dob_mono_semilog}
    }
    \caption{Effect of amplitude of perturbation $(A)$ for shear startup of the nRP model for $Re=0$. Here, $Wi=30,\eta_s=10^{-4},\beta=1$ which corresponds to a shear rate in the monotonic region of the constitutive curve. (a) Shear stress evolution is plotted for $A=$ $10^{-1}$, $10^{-4}$, $10^{-7}$ and for a forced homogeneous flow. (b) The variation of degree of banding $(\Delta\dot{\gamma}=\dot{\gamma}_{max}-\dot{\gamma}_{min})$ with time is shown for different values of $A=$ $10^{-1}$, $10^{-4}$, and $10^{-7}$ for $Re=0$. The inset of Fig.\,(b) shows early time $(t=0-0.3)$ for variation of degree of banding on a semi-log plot. (c) The early time $(t=0-0.01)$ variation of degree of banding with time for $\eta_s=10^{-4}$, $A=10^{-4}$ and $Re=0$ on a semi-log plot.  }
    \label{fig:nrp_effect_of_A_Re_zero_mono}
\end{figure}

We next study the effect of initial amplitude of perturbation on the shear start-up of the nRP model under the creeping-flow assumption when the shear rate is in the monotonic region of the constitutive curve $(Wi=30$, $\eta_s=10^{-4}$, and $\beta=1)$. As mentioned in Sec.\,\ref{section_model}, in the absence of inertia, it is not possible to provide initial conditions to the shear rate. Instead, the initial values of the shear rate are fixed by the initial stress perturbation $\sigma_{xy}(t=0)$ via the Cauchy momentum equation (Eq.\,\ref{dimensionless_cauchy}). Because $\dot{\gamma}(t = 0) = \sigma_{xy}(t=0)/\eta_s$, the shear rate perturbations become much larger than the stress perturbations. For $A=10^{-1}$, initial shear rate perturbation is of the order of $10^2$ ($10^4\%$ of base state shear rate for $\eta_s=10^{-4}$). Similarly, for $A=10^{-4}$, initial shear rate perturbation is of the order of $10^{-1}$ ($10\%$ of base state shear rate) and for $A=10^{-7}$, initial shear rate perturbation is of the order of $10^{-4}$ ($0.01\%$ of the base state shear rate).

\begin{figure}
      \centering
    \subfigure[]{
    \includegraphics[scale=0.25]{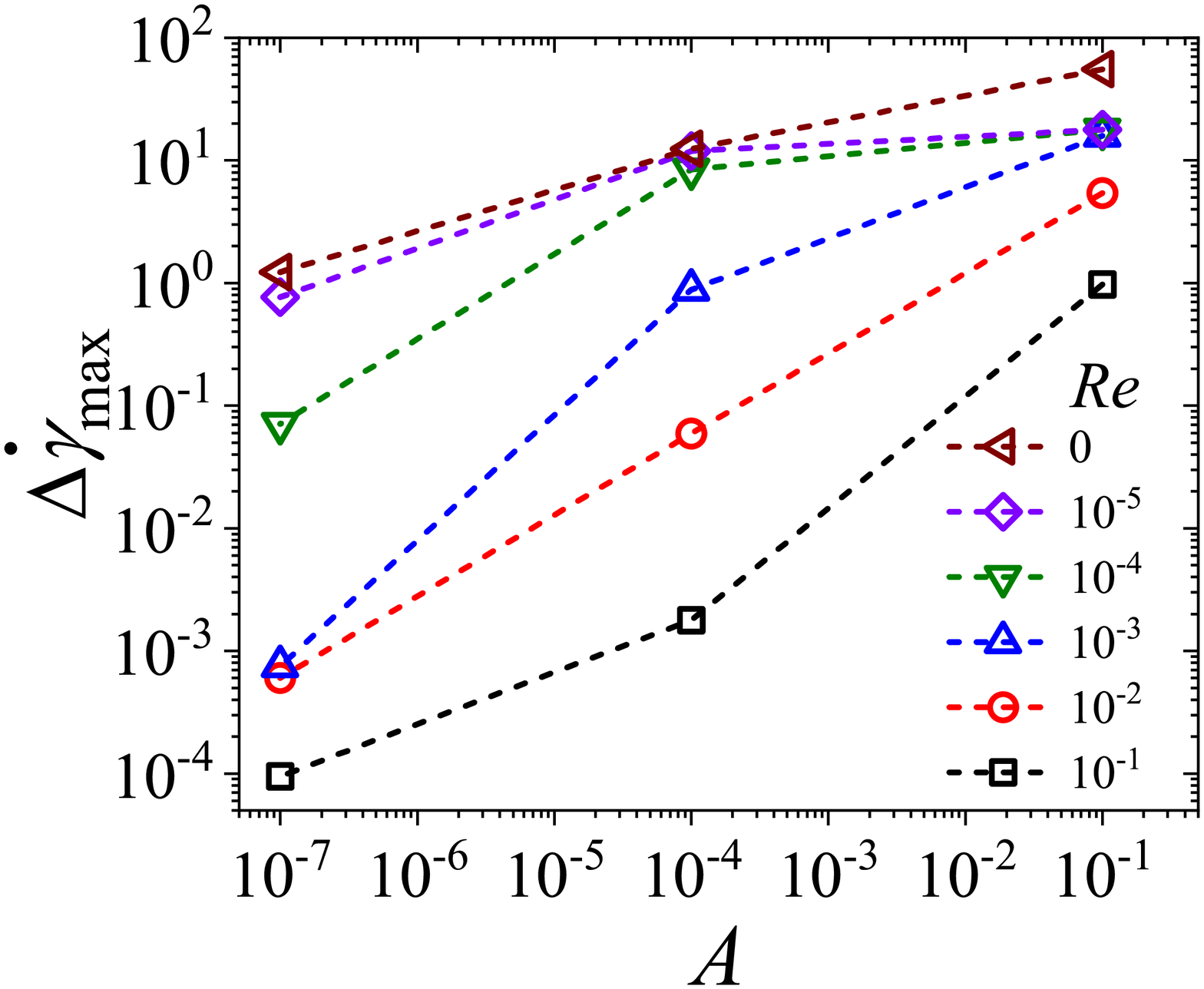}
    \label{nrp_dobmax_A_mono}
    }
    \subfigure[]{
    \includegraphics[scale=0.25]{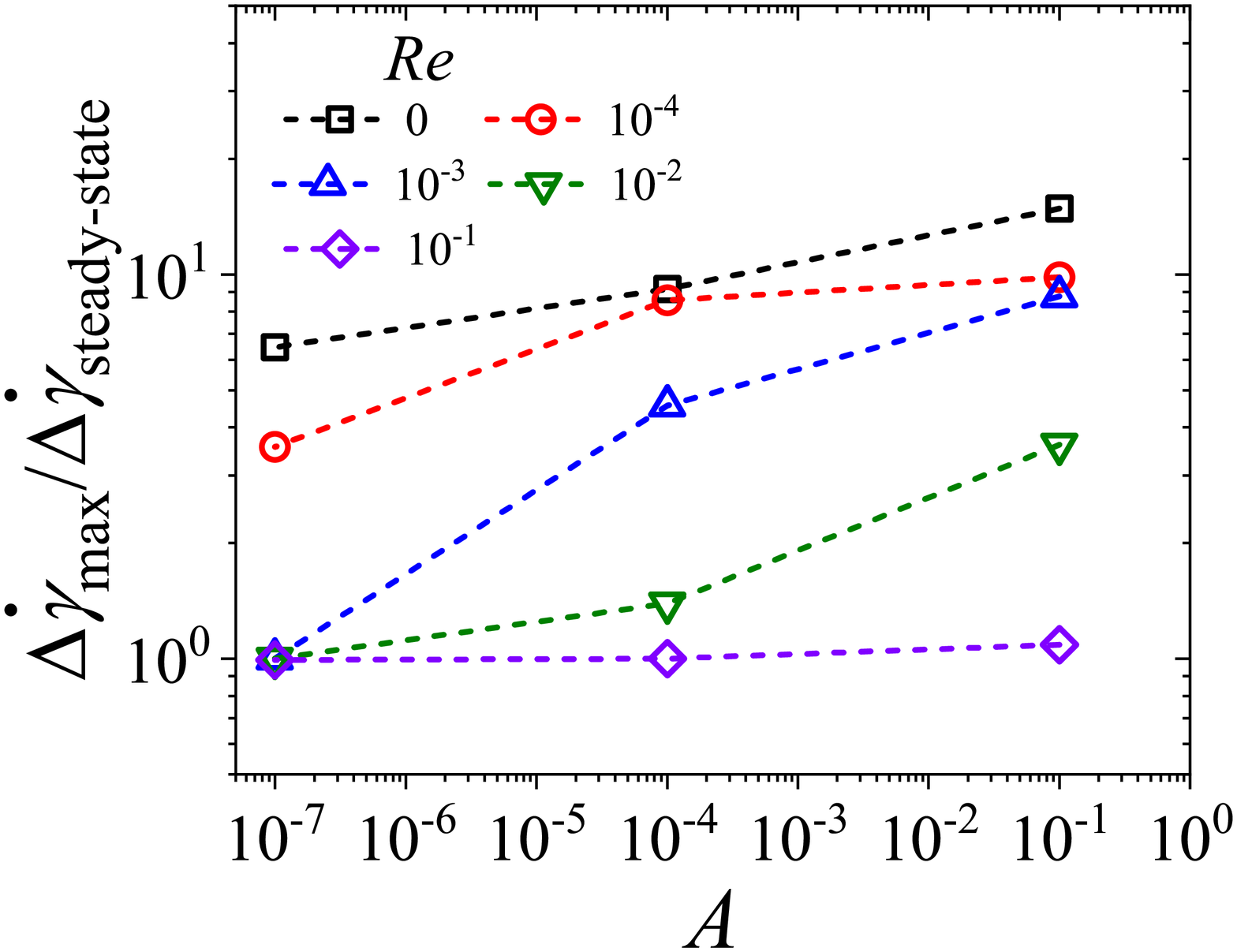}
    \label{nrp_dobmax_A_nonmono}
    }
    \caption{The maximum of degree of banding $(\Delta\dot{\gamma}_{max})$ for $A=$ $10^{-1}$, $10^{-4}$, $10^{-7}$ and $Re=$ $0$, $10^{-2}$, $10^{-1}$ is plotted for shear startup of nRP model if the shear rate is the (a) monotonic $(Wi=30,\eta_s=10^{-4},\beta=1)$ and (b) nonmonotonic $(Wi=30,\eta_s=10^{-4},\beta=0.6)$ region of the constitutive curve. }
    \label{fig:nrp_dobmax_A_mono_nonmono}
\end{figure}

\begin{figure}
\centering
    \subfigure{
    \includegraphics[scale=0.23]{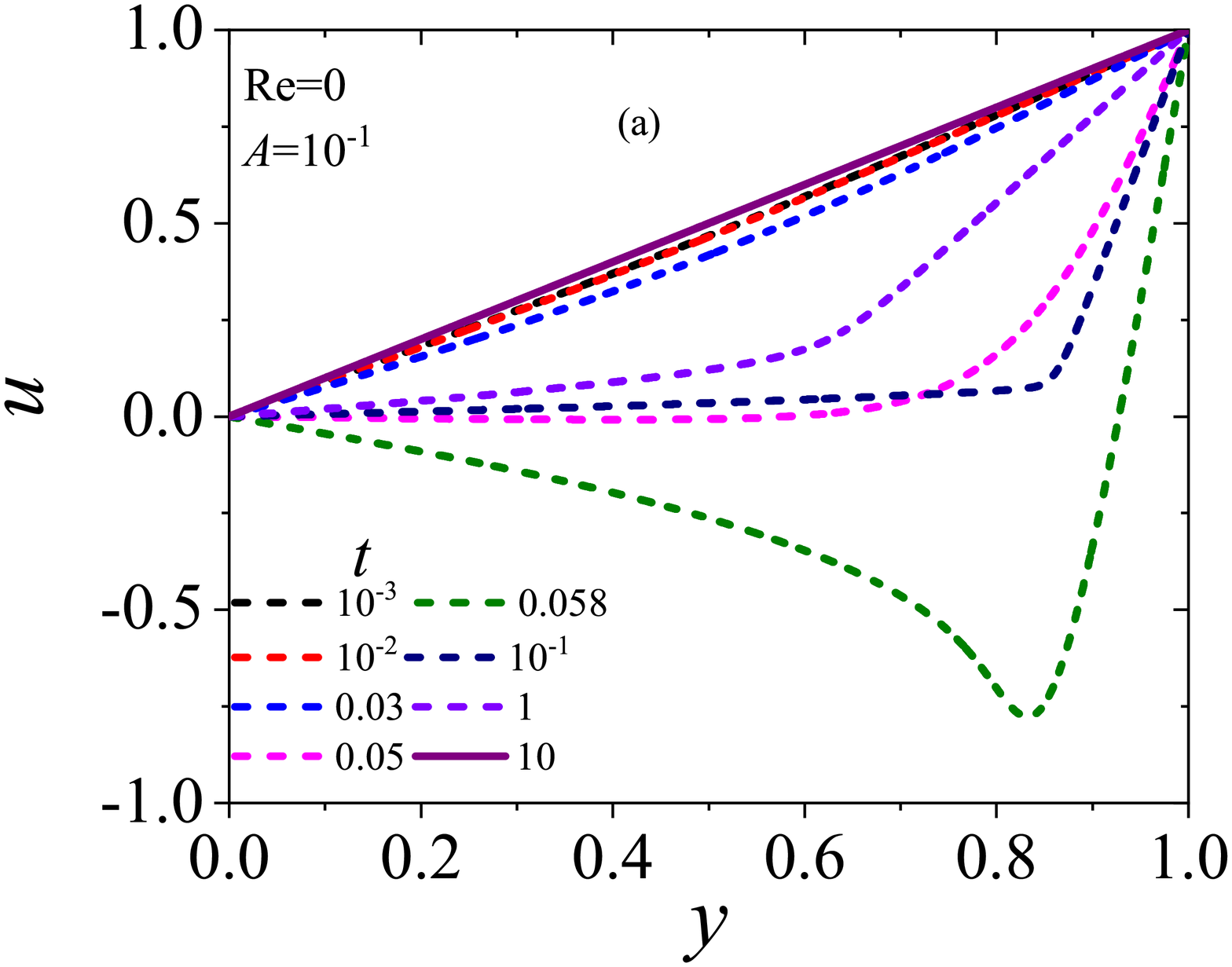}
    \label{nrp_A_1e_1_velocity_Re_0}
    }
    \subfigure{
    \includegraphics[scale=0.23]{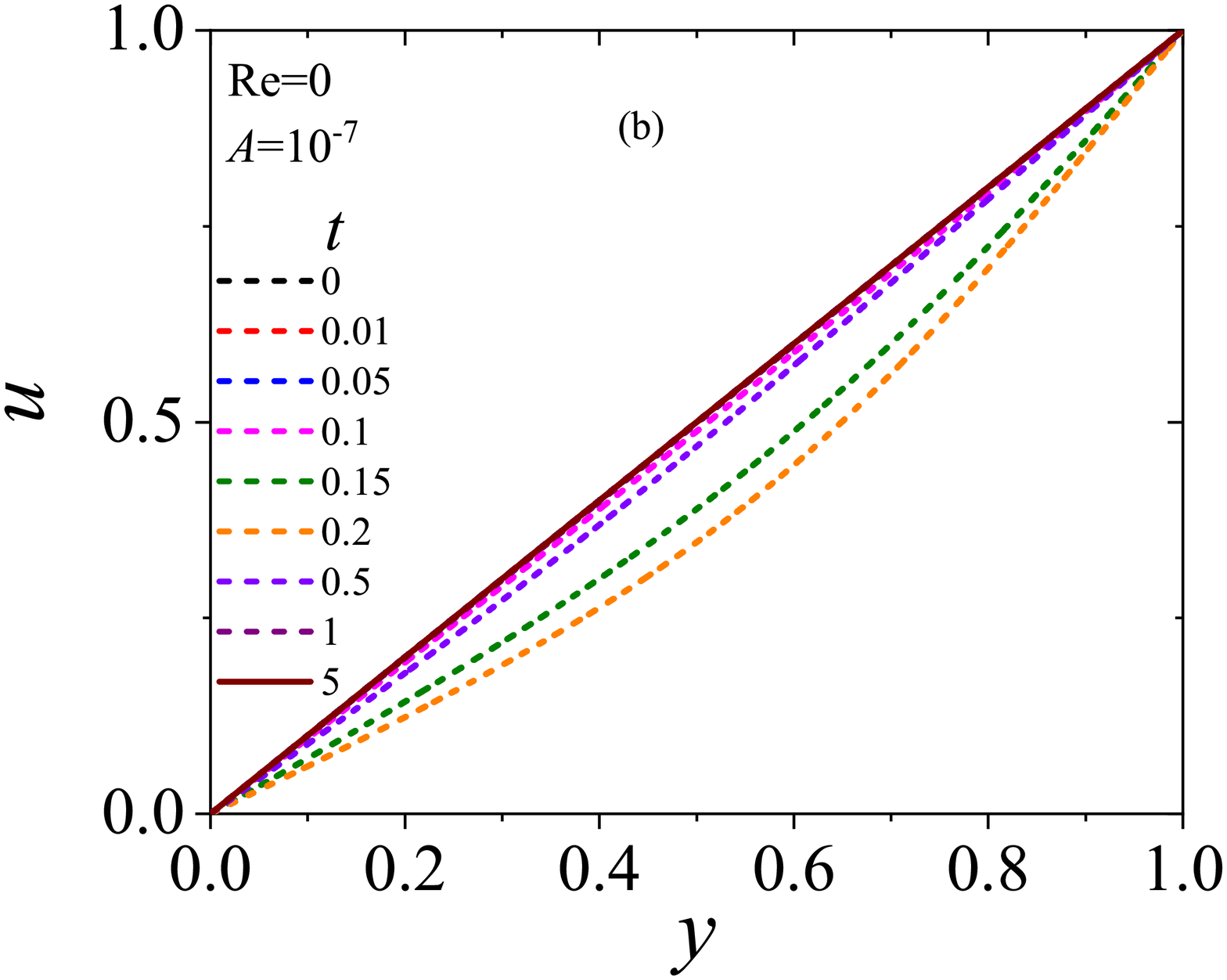}
    \label{nrp_A_1e_7_velocity_Re_0}
    }
    \subfigure{
    \includegraphics[scale=0.23]{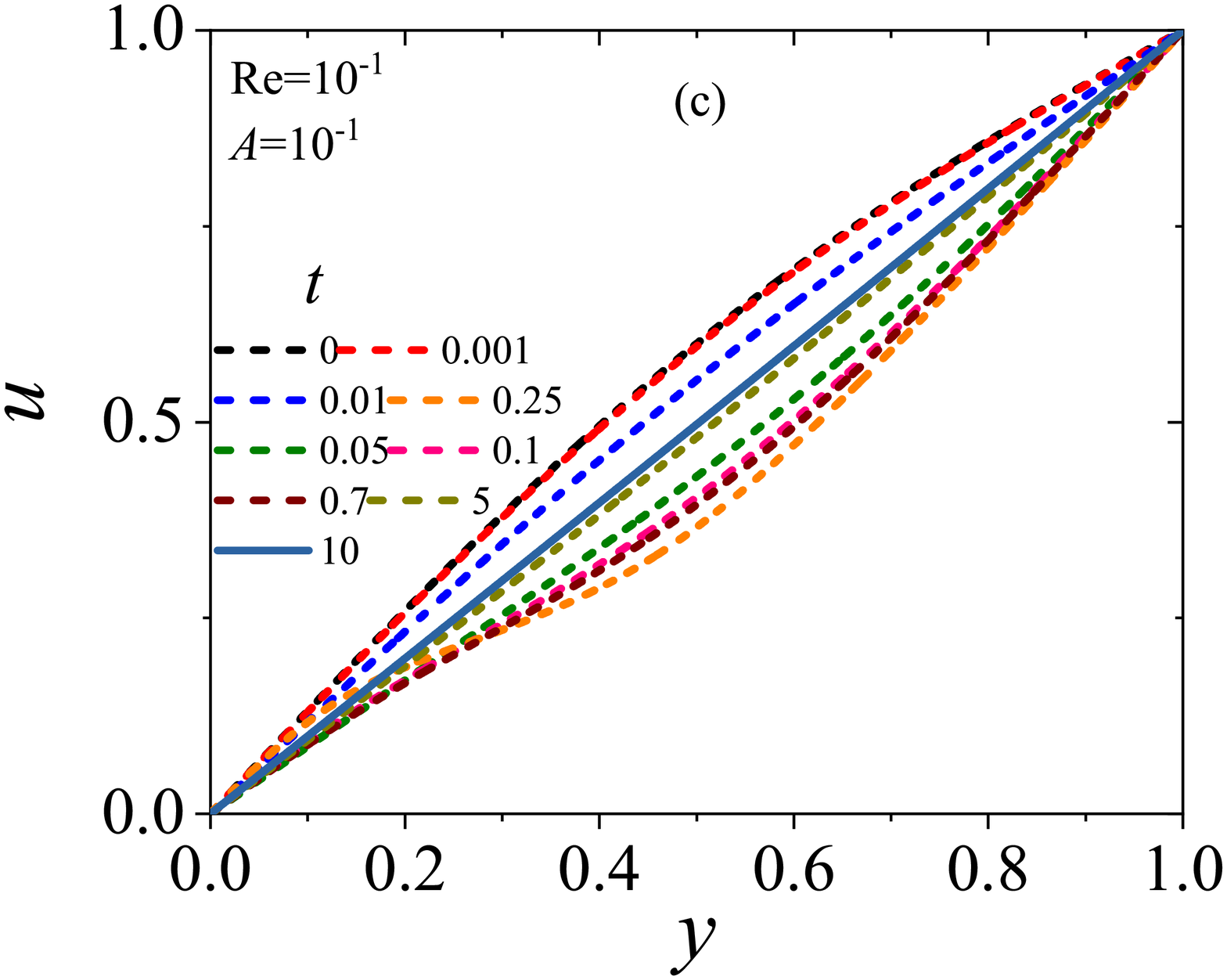}
    \label{nrp_A_1e_1_velocity_Re_1e_1}
    }
    \subfigure{
    \includegraphics[scale=0.23]{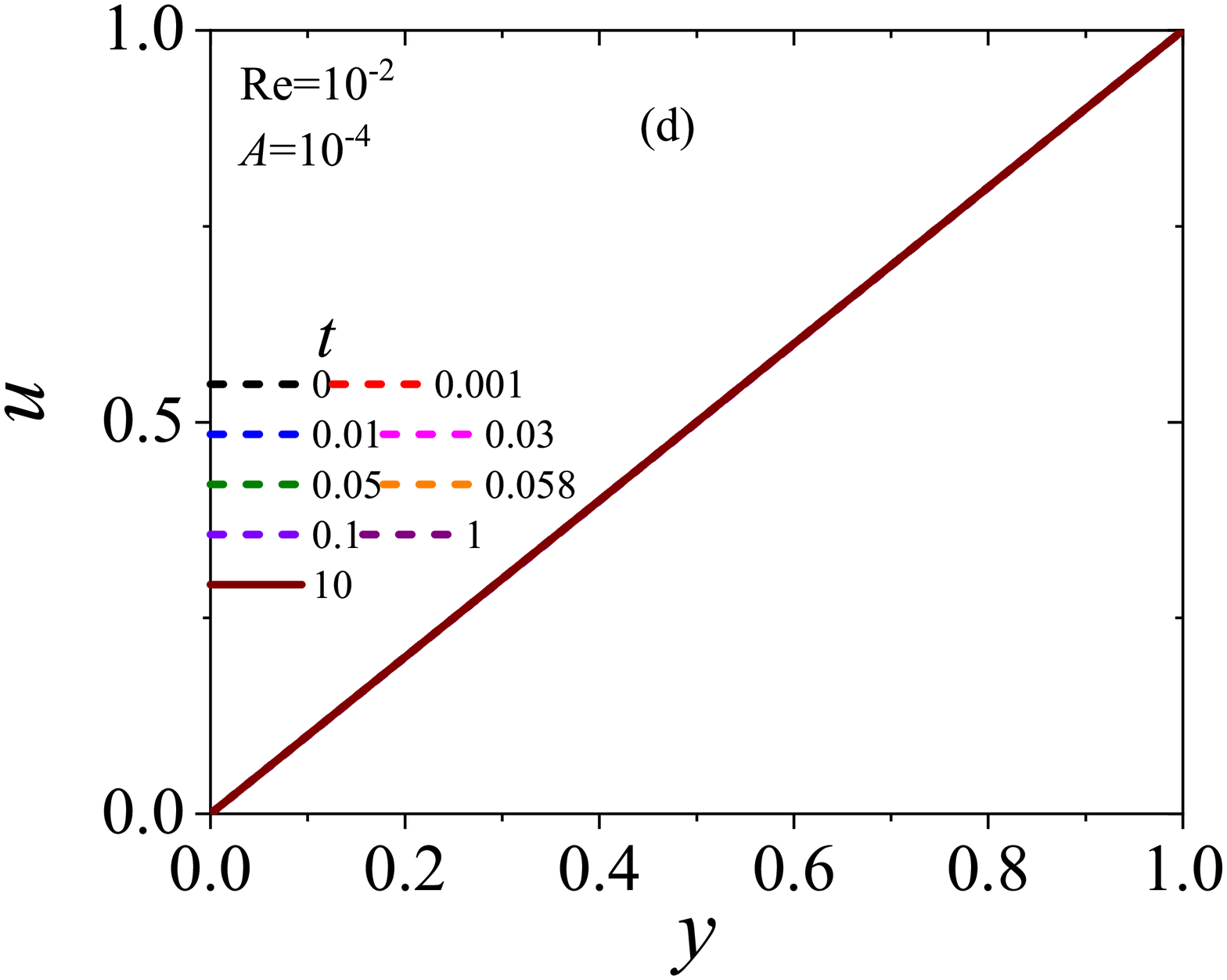}
    \label{nrp_A_1e_4_velocity_Re_1e_2}
    }
    \caption{Velocity profile evolution during the shear startup flow of the nRP model fluid at $Wi=30$, $\eta_s=10^{-4}$, and $\beta=1$ which corresponds to a shear rate in the monotonic region of the constitutive curve. Figure (a) shows result for $Re=0$ and $A=10^{-1}$ and figure (b) shows result for $A=10^{-7}$ and $Re=0$. Figure (c) shows result for $A=10^{-1}$ and $Re=10^{-1}$ and figure (d) shows result for $A=10^{-4}$ and $Re=10^{-1}$.}
      \label{nrp_velocity_profiles_monotonic}
\end{figure}

We find that shear stress evolution does not overlap each other for $A=10^{-1}$, $10^{-4}$, and $10^{-7}$ at $Re=0$ as shown in Fig.\,\ref{nrp_effect_of_A_stress_mono}. The velocity profiles during shear startup flow for $A=10^{-1}$ and $Re=0$ show the presence of transient shear banding with a negative velocity profile as shown in Fig.\,\ref{nrp_A_1e_1_velocity_Re_0}. The corresponding evolution of degree of banding is shown in Fig.\,\ref{nrp_effect_of_A_dob_mono}. The evolution of $\Delta\dot{\gamma}$ as a function of time shows a sharp increase and then finally decays to zero for $A=10^{-1}$, $10^{-4}$, and $10^{-7}$. The inset of Fig.\,\ref{nrp_effect_of_A_dob_mono} shows the early time behavior $(t=0-0.3)$ on a semi-log plot and Fig.\,\ref{nrp_effect_of_A_dob_mono_semilog} shows the evolution only for $A=10^{-4}$ and $t=0-0.01$. These results show that perturbations initially grow linearly followed by a growth which is stronger than linear before final decay to zero. This result shows that the growth of perturbation is exponential which implies that the shear startup of nRP is transiently (and linearly) unstable for $Re=0$. However, the exponential growth of perturbation begins almost at the start of the flow and may not have any correlation with time at which shear stress starts to decrease after its overshoot. Similarly, the value of $\Delta\dot{\gamma}_{max}$ is significantly larger than the input amplitude $A$ as shown in Fig.\,\ref{nrp_dobmax_A_mono}. Therefore, we can treat this as an intrinsic instability of shear startup of the nRP model. 

Interestingly, figure \ref{nrp_effect_of_A_dob_mono} shows that the time of peak of $\Delta\dot{\gamma}$ decreases with increase in value of $A$ and the value of $\Delta\dot{\gamma}_{max}$ also depends on the value of $A$. On increasing the value of $A$, the value of $\Delta\dot{\gamma}_{max}$ also increases but not in a linear manner as shown in Fig.\,\ref{nrp_dobmax_A_mono}. The decrease in $\Delta\dot{\gamma}_{max}$ with $A$ is also clearly visible in the velocity profile evolution for $A=10^{-7}$ as there is only a slight deviation from the linear velocity profile and not any transient shear banding or transient negative velocity profile (Fig.\,\ref{nrp_A_1e_7_velocity_Re_0}). The shear stress, degree of banding and velocity profile obtained using $Wi=30$, $\eta_s=10^{-4}$, $\beta=1$, $A=10^{-1}$, and $Re=0$ in Figs.\,\ref{nrp_effect_of_A_stress_mono}, \ref{nrp_effect_of_A_dob_mono} and \ref{nrp_A_1e_1_velocity_Re_0} has also been appeared in Ref.\,[\onlinecite{moorcroft14}]. We have included these results only for comparison purposes.

\begin{figure}
    \centering
    \subfigure[]{
    \includegraphics[scale=0.25]{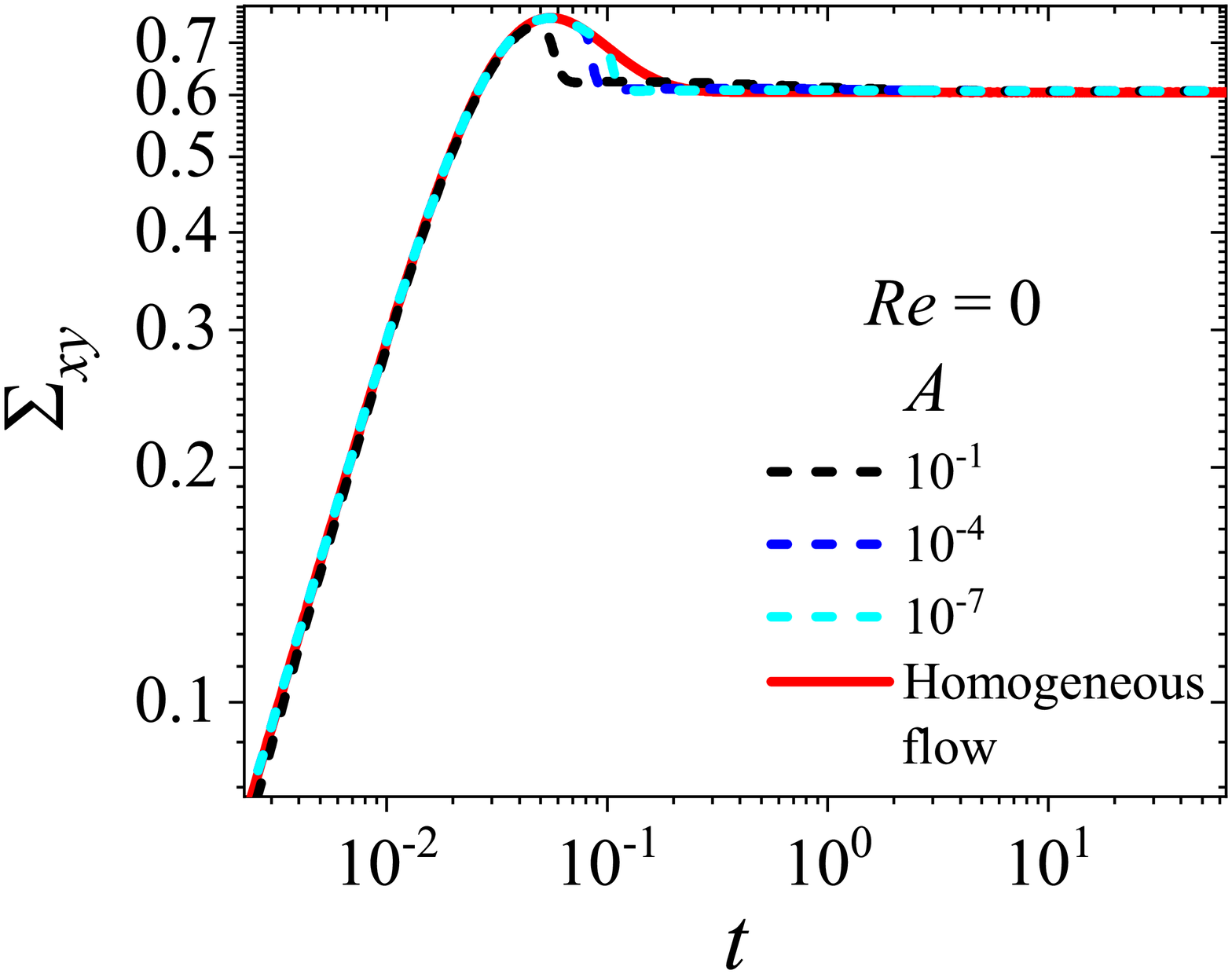}
    \label{nrp_effect_of_A_stress_nonmono}
    }
    \subfigure[]{
    \includegraphics[scale=0.25]{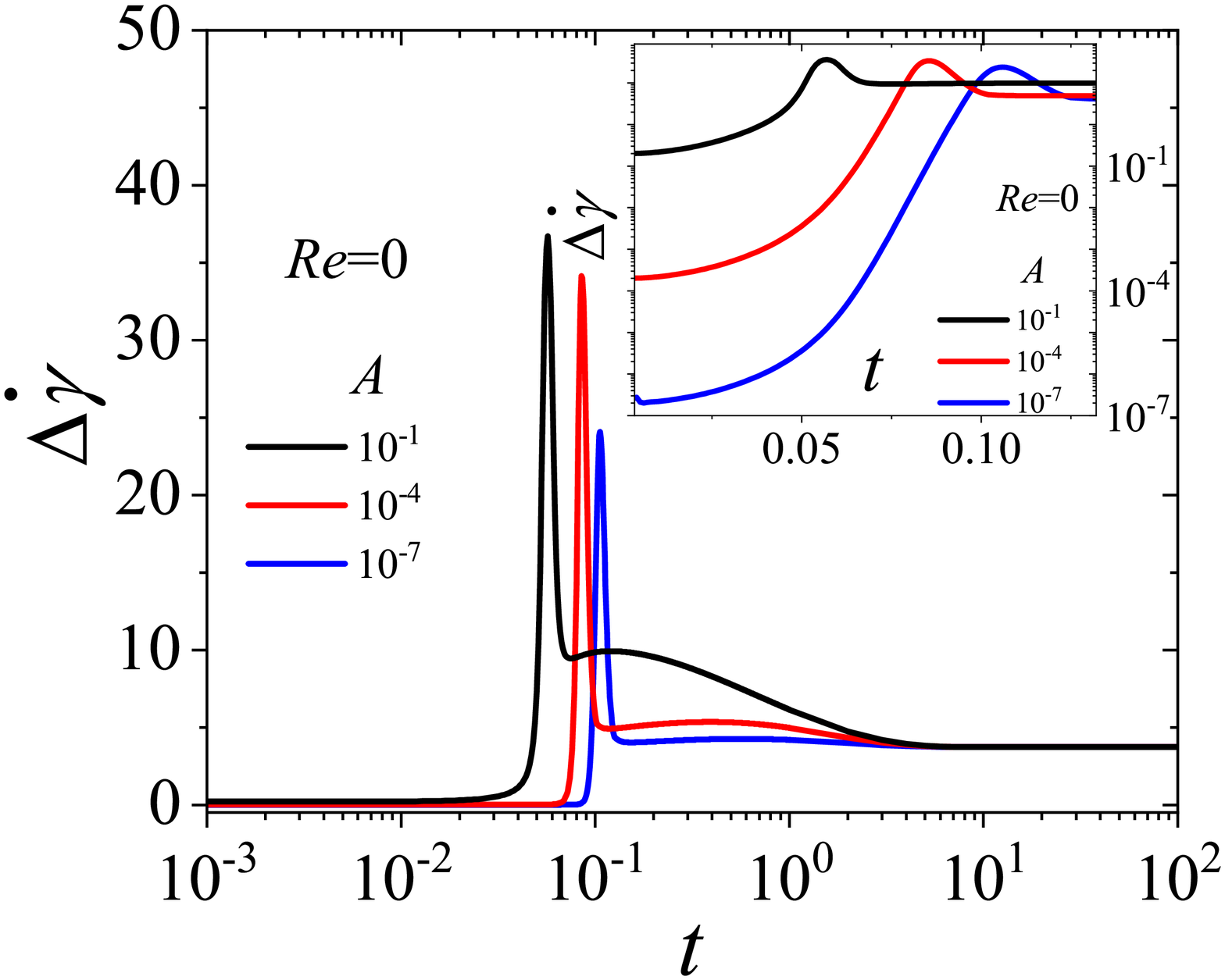}
    \label{nrp_effect_of_A_dob_nonmono}
    }
     \subfigure[]{
    \includegraphics[scale=0.25]{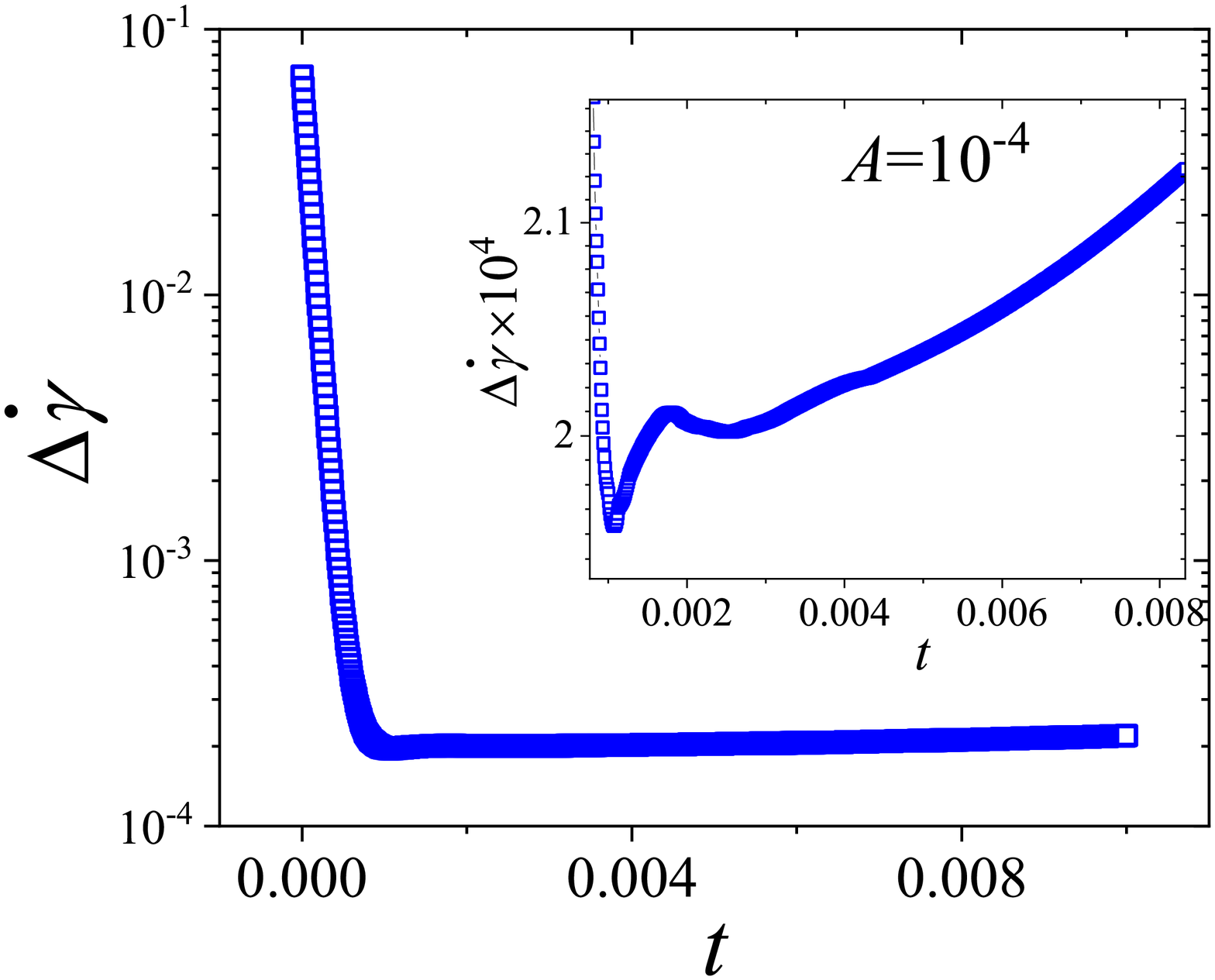}
    \label{nrp_effect_of_A_dob_nonmono_semilog}
    }
    \caption{Effect of amplitude of perturbation $(A)$ for shear startup of the nRP model for $Re=0$. Here, $Wi=30,\eta_s=10^{-4},\beta=0.6$ which corresponds to a shear rate in the nonmonotonic region of the constitutive curve. (a) Shear stress evolution is plotted for $A=$ $10^{-1}$, $10^{-4}$, $10^{-7}$ and for a forced homogeneous flow. (b) The variation of degree of banding $(\Delta\dot{\gamma}=\dot{\gamma}_{max}-\dot{\gamma}_{min})$ with time is shown for different values of $A=$ $10^{-1}$, $10^{-4}$, and $10^{-7}$ for $Re=0$. The inset of Fig.\,(b) shows early time $(t=0-0.13)$ for variation of degree of banding on a semi-log plot. (c) The early time $(t=0-0.01)$ variation of degree of banding with time for $\eta_s=10^{-4}$, $A=10^{-4}$ and $Re=0$ on a semi-log plot.}
    \label{fig:nrp_effect_of_A_Re_zero_nonmono}
\end{figure}

If the shear rate is in the nonmonotonic region of the constitutive curve $(Wi=30$, $\eta_s=10^{-4}$, and $\beta=0.6)$, we find that for $Re=0$, shear stress evolution as a function of time for different values of $A$ does not overlap as also observed in the previous case (the initial shear rate perturbation and shear stress perturbation is same as for previous case as the value of $A$, $\eta_s$, and $Wi$ is same in this case as well).  If the value of $A$ is higher, shear stress attains steady state at lower values of time as shown in Fig.\,\ref{nrp_effect_of_A_stress_nonmono}. The evolution of $\Delta\dot{\gamma}$ with time also confirms that the degree of banding shows a rapid increase to attain a maximum value and then decreases to a constant value if $y-$ axis is linear. If $\Delta\dot{\gamma}$ versus time is plotted on a semi-log scale, then the increase in degree of banding is linear or stronger than linear before attaining a steady state as shown in Figs.\,\ref{nrp_effect_of_A_dob_mono} and \ref{nrp_effect_of_A_dob_mono_semilog}. This result demonstrates the exponential growth of perturbations which is in agreement with criterion of Yerushalmi et. al. [\onlinecite{yerushalmi1970stability}]. However, the time at which degree of banding begins to grow exponentially is much lesser than time at which shear stress starts to decrease after its overshoot. In the literature [\onlinecite{briole2021shear,rassolov2022role,mohagheghi2016elucidating,cao2012shear,zhou2008modeling,hu2008comparison}], the time of decrease in stress after stress overshoot is often linked with time at which steady state shear banding structure begins to form. However, our results show that shear banding structures starts to form at the beginning of the flow, but may not be discernible in experiments, and hence the shear banding structures are not associated with the time of decrease in stress after its overshoot.

More importantly, the evolution of $\Delta\dot{\gamma}$ with time on a semi-log plot for the nRP model shows a linear increase initially followed by a nonlinear increase in both cases of shear rate in the monotonic as well as nonmonotonic region of the constitutive curve (Figs.\,\ref{nrp_effect_of_A_dob_mono} and \ref{nrp_effect_of_A_dob_nonmono}). Because of these findings, it is clear that the linear stability analysis predicts only the exponential growth at very early times, and soon after nonlinear effects take over, and hence the linearized dynamics loses its relevance for times where there is a stress overshoot and decay. The caution regarding the use of linear stability analysis in unsteady flow (shear startup flow) to assess the occurrence of shear banding has also been pointed out by Peterson [\onlinecite{2018PhDT.......101P}].

The transiently high value of $\Delta\dot{\gamma}$ shows the presence of transient shear banding and the constant value of $\Delta\dot{\gamma}$ at steady state show presence of steady state shear banding. Figure \ref{nrp_dobmax_A_nonmono} shows that for $Re=0$, the value of $\Delta\dot{\gamma}_{max}$ is much higher than $\Delta\dot{\gamma}_{steady-state}$, however, $\Delta\dot{\gamma}_{max}$ depends on the value of $A$ but not in a linear fashion. As noted in previous case (i.e., corresponding to results shown in Fig.\,\ref{nrp_effect_of_A_dob_mono}), the time of maximum value of degree of banding depends on the value of $A$. Higher the value of $A$, lower is the time corresponding to maximum value of degree of banding, and higher will be the value of $\Delta\dot{\gamma}_{max}$ as shown in Figs.\,\ref{nrp_effect_of_A_dob_nonmono} and \ref{nrp_dobmax_A_nonmono}. This observation is also similar to the previous case in which shear startup flow is solved for shear rate in the monotonic region of the constitutive curve. Also, transient negative velocity profile is observed in the case of $Re=0$ for $A=10^{-1}$ and $A=10^{-4}$ but no transient negative velocity profile (or a distinct transient shear banding, which is not a part of gradual development to steady state shear banding) has been observed for $A=10^{-7}$ as shown in Fig.\,S7.

In contrast to the JS model results, the nRP model showed that under the creeping-flow assumption (i) there can be transient shear banding in presence of stress overshoot (regardless of whether the steady-state velocity profile is homogeneous or banded). This result is in agreement with the observation reported in our previous study (Fig.\,5(b) of Ref.\,[\onlinecite{sharma2021onset}]) wherein we showed that the linearized perturbations grow rapidly and attain a maximum of the order of $10^4$ before continuous decrease for $\eta_s=10^{-4}$. However, the presence of transient shear banding (degree of banding) is sensitive to the magnitude of $A$, and is observed only if the initial shear rate perturbation is unrealistically high, i.e., $A = 0.1$ for $\eta_s=10^{-4}$ and $Wi=30$ in the case of the nRP model. The effect of amplitude of perturbation on transient shear banding is in agreement with the results obtained in literature for the stretching Rolie-Poly model [\onlinecite{adams2011transient}]. (ii) The time associated with maximum value of degree of banding and the maximum value of the degree of banding depends on the value of $A$. (iii) The variation of $\Delta\dot{\gamma}_{max}$ with $A$ is not linear, and suggests the importance of nonlinearities in the dynamics (Fig.\,\ref{nrp_dobmax_A_nonmono}). (iv) If the magnitude of the initial shear rate perturbation is realistic then there may not be any discernible transient shear banding. (v) In both cases, the increase in degree of banding is exponential initially and occurs at time, which is much lesser than time associated with stress overshoot. 

The contrasting results of the JS and nRP models can be ascribed to the following reasons: (i) In the case of JS model, the magnitude of the initial shear rate perturbation is not unrealistically high because the value of $\eta_s$ was not much smaller compared to unity. (ii) In the case of nRP model, the initial shear rate perturbation becomes unrealistic because of low value of $\eta_s$ if the stress perturbations are $O(1)$.  Therefore, under the creeping-flow assumption, if the initial shear rate perturbation is realistic then there may be no transient shear banding observed in shear startup flow of JS and nRP models.


\subsubsection{$Re\neq0$} \label{section_Re_not_zero}

We now study the shear startup of the JS and nRP models with non-zero inertial effects. We consider $Re=10^{-2}$ and $Re=10^{-1}$ for comparing results of both models. For nonzero inertia, the initial shear rate perturbation can be specified directly in terms of $A$. However, as the value of $\eta_s$ considered differs by orders of magnitude between the two models, the effect of initial amplitude of perturbation in both models is different as discussed below. The effect of inertia on transient shear banding is discussed in Sec.\,\ref{section_Re_effect}. The effect of initial amplitude of perturbation on evolution of stress and degree of banding with time for both the models is plotted in Figs.\,S2, S4, S6 and S8 of the supplementary information.  

\paragraph{JS model} \label{section_Re_not_zero_js}

For the JS model, we find that during shear startup if the shear rate is in the monotonic or nonmonotonic region of the constitutive curve, then the results obtained for different initial amplitude of perturbations for non-zero inertia are similar to those for $Re = 0$. 
Figure~\ref{js_dobmax_mono} shows that the value of $\Delta\dot{\gamma}_{max}$ is of the order of A and increases with A in a linear fashion for $Re\neq~0$ as well as for $Re=0$. 
Similarly, if the shear rate is in the nonmonotonic region of the constitutive curve, then also the variation of $\Delta\dot{\gamma}_{max}$ with $A$ is same for $Re=0$ and $Re\neq0$ as shown in Fig.\,\ref{js_dobmax_nonmono}.

\paragraph{nRP model} \label{section_Re_not_zero_nrp}

For the nRP model, we find that the variation of $\Delta\dot{\gamma}_{max}$ with $A$ is not proportional to the value of $A$ which is in contrast to the results of JS model for $Re\neq0$ as shown in Fig.\,\ref{nrp_dobmax_A_mono}. Interestingly, for $Re=10^{-1}$, $A=10^{-1}$, and $10^{-4}$, the value of $\Delta\dot{\gamma}_{max}$ is proportional to $A$ but it is not the case for $A=10^{-7}$. This result shows that the variation of $\Delta\dot{\gamma}_{max}$ with $A$ is affected by the nonlinearities in the dynamics of shear startup of nRP model. For $Re=1$ and $Re=10$, the variation of  $\Delta\dot{\gamma}_{max}$ is proportional to $A$. This result is discussed in detail in Sec.\,\ref{section_Re_effect}.

Similarly, if the shear rate is in the nonmonotonic region of the constitutive curve, then in presence of inertia, the effect of initial amplitude of perturbation is similar to $Re=0$ case except for $Re=10^{-1}$ as shown in Fig.\,\ref{nrp_dobmax_A_nonmono}. For $Re=10^{-1}$, the value of $\Delta\dot{\gamma}_{max}$ is equal to $\Delta\dot{\gamma}_{steady-state}$ for $A=10^{-4}$ and $10^{-7}$, and $\Delta\dot{\gamma}_{max}$ is slightly higher than $\Delta\dot{\gamma}_{steady-state}$ for $A=10^{-1}$. For $Re<10^{-1}$, the variation of $\Delta\dot{\gamma}_{max}$ depends on $A$ but not in a linear manner, which also shows that the presence of transient shear banding is sensitive to the initial amplitude of perturbation even in the presence of inertia. 

The effect of initial amplitude of perturbation in the case of JS model in the presence of inertia is similar to the non-inertial case where a direct shear rate perturbation amplitude cannot be specified in both the steady state homogeneous and the nonhomogeneous case. Interestingly, the results of nRP model for non-inertial cases are qualitatively similar to results obtained in presence of inertia if $Re<10^{-1}$. We attribute this observation also to smaller value of $\eta_s$ used for the nRP model. This can be rationalized using the Cauchy momentum equation given by:

\begin{equation}\label{dimensionless_cauchy_another}
    \frac{\partial u}{\partial t}=\frac{1}{Re}\frac{\partial\sigma_{xy}}{\partial y}+\frac{\eta_sWi}{Re}\frac{\partial^2 u}{\partial y^2}
\end{equation}
which shows that if $Re$ is $O(10^{-5})$, then first term of R.H.S. of Eq.\,\ref{dimensionless_cauchy_another} becomes much larger in magnitude compared to second term. However in case of JS model, if $Re$ is $O(10^{-5})$, then the order of both terms of R.H.S. of Eq.\,\ref{dimensionless_cauchy_another} will be of the same order as $\eta_s=0.16$. Interestingly, it has indeed been observed in the literature that increase in order of viscous dissipation has a stabilizing effect on flow [\onlinecite{tomar2006instability}]. 

\subsection{Effect of solvent viscosity or flatness of constitutive curve} \label{section_eta_s_effect}

In our previous study (Fig.\,1 of Ref. [\onlinecite{sharma2021onset}]), we showed that decreasing $\eta_s$ leads to more flattening of the constitutive curve for both JS and nRP models. This result is consistent with the results reported in the literature [\onlinecite{adams2011transient,moorcroft14}]. (The flatness of the constitutive curve can also be increased by increasing the number of entanglements [\onlinecite{rassolov2022role}]; however, in this study, we focuse only on the effect of $\eta_s$ by keeping the number of entanglements fixed.) We now discuss the effect of decreasing the solvent viscosity contribution or increase in flatness of constitutive curve. Several results have been reported in the literature [\onlinecite{adams2011transient,moorcroft14}] suggesting that increasing the flatness of constitutive curve increases the chances of transient shear banding in shear startup flow. In this section, we focus on the effect of solvent viscosity only for shear rates in the monotonic region of the constitutive curve. In our previous study [\onlinecite{sharma2021onset}], we showed using the JS model that there is no significant change in the transiently maximum eigenvalue or the maximum value of growth of linearized perturbation with decrease in $\eta_s$. We also noted that in order to study the effect of flatness of the constitutive curve, the value of $\eta_s$ cannot be lowered below $1/9$ in order for the constitutive curve to be monotonic. During the course of the present study, we found that even in the results from nonlinear simulations, there is no significant effect of decreasing the solvent viscosity (from $0.16$ to $0.115$) on degree of banding evolution (data not shown).

\begin{figure}
\centering
    \includegraphics[scale=0.14]{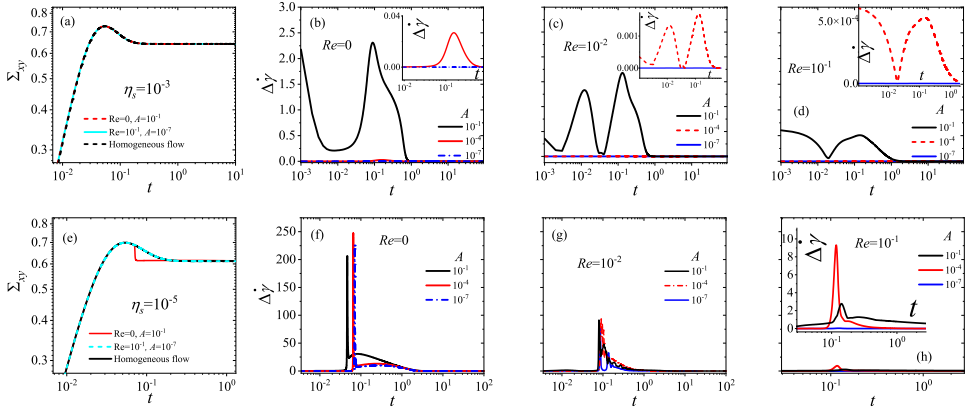}
    \caption{Effect of amplitude of perturbation $(A)$ for $Re=0$, $10^{-2}$, and $10^{-1}$ for shear startup flow of the nRP model for $\eta_s=10^{-3}$ and $\eta_s=10^{-5}$. First row shows results for shear startup flow of nRP model at $Wi=30,\eta_s=10^{-3}$, and $\beta=1$ while second row shows result for shear startup flow of nRP model at $Wi=30,\eta_s=10^{-5}$, and $\beta=1$ and both results are obtained for shear rate in the monotonic region of the constitutive curve. Figure (a) and (e) shows the shear stress evolution as a function of time and figure (b)-(d), figure (f)-(h) shows the degree of banding as a function of time during the shear startup flow. }
      \label{nrp_effect_of_eta_s}
\end{figure}

\begin{figure}
      \centering
    \subfigure[]{
    \includegraphics[scale=0.25]{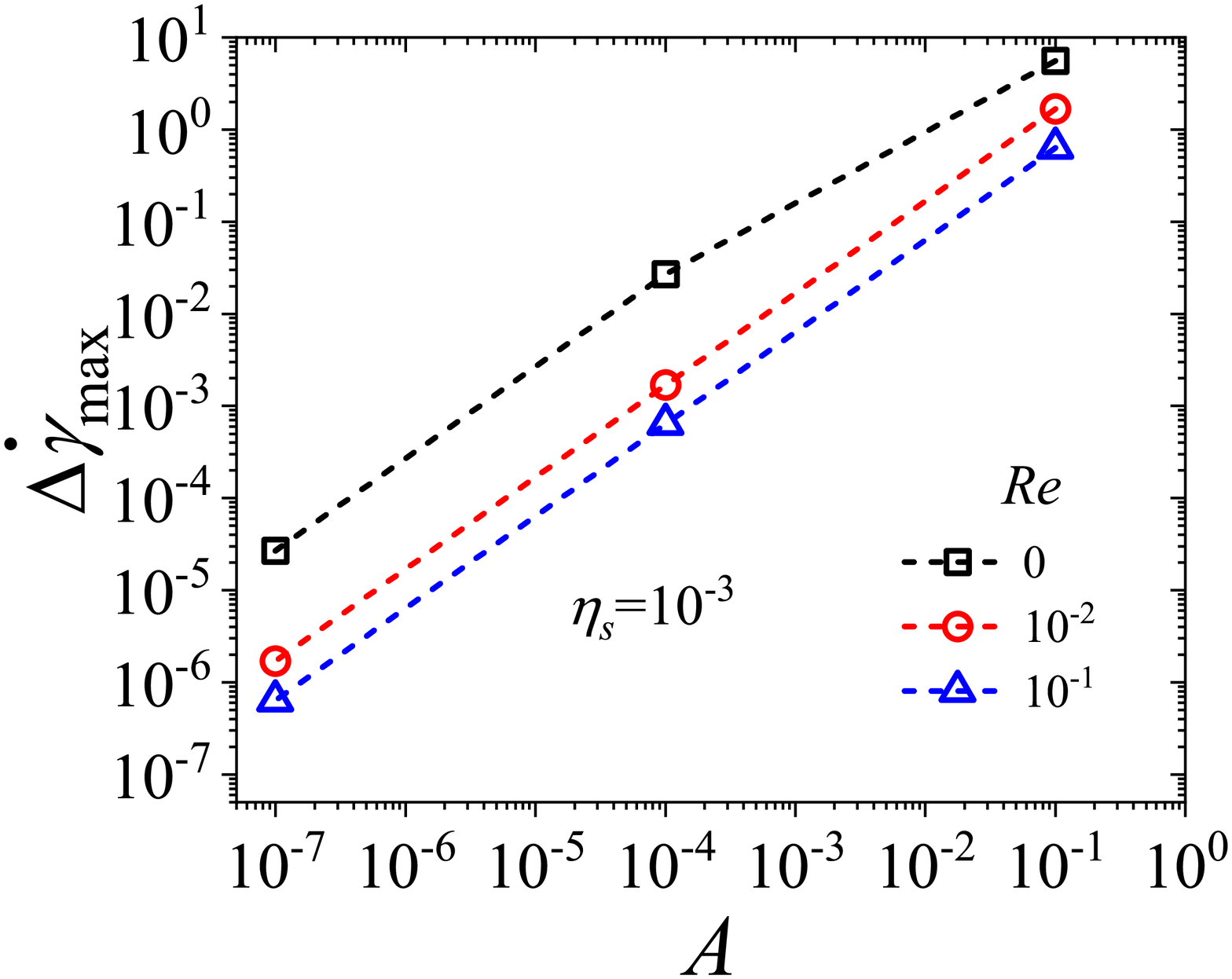}
    \label{nrp_dobmax_mono_N_1e_3}
    }
    \subfigure[]{
    \includegraphics[scale=0.25]{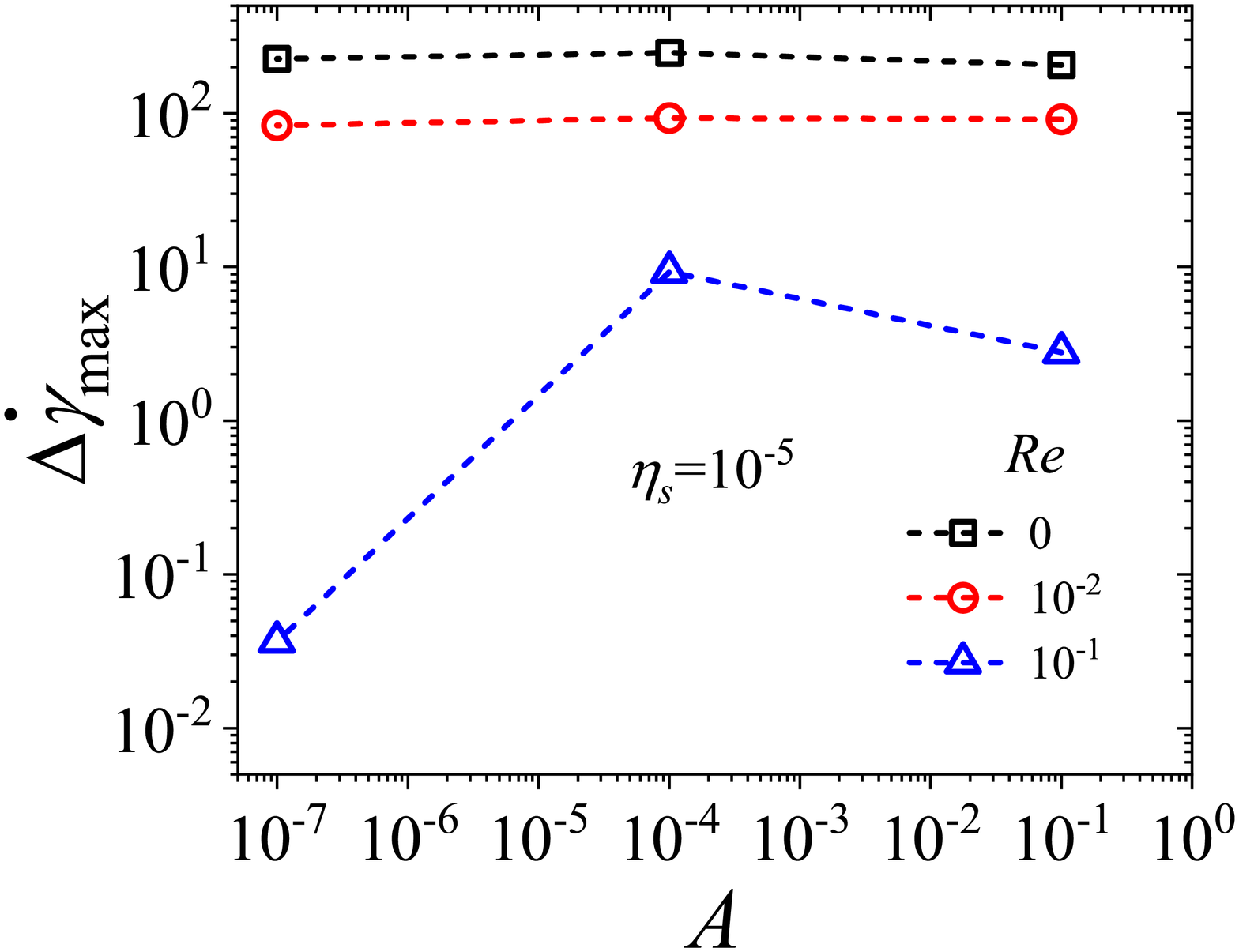}
    \label{nrp_dobmax_nonmono_N_1e_5}
    }
    \caption{The maximum of degree of banding $(\Delta\dot{\gamma}_{max})$ for $A=$ $10^{-1}$, $10^{-4}$, $10^{-7}$ and $Re=$ $0$, $10^{-2}$, $10^{-1}$ is plotted for shear startup of nRP model if the shear rate is the monotonic region of the constitutive curve $(Wi=30, \beta=1)$ for (a)$\eta_s=10^{-3}$ (b)$\eta_s=10^{-5}$. }
    \label{fig:nrp_dobmax_mono_eta_s}
\end{figure}

However, for the nRP model, we showed [\onlinecite{sharma2021onset}] that there is a significant change in the transiently maximum eigenvalue or the maximum value of growth of linearized perturbation with decrease in $\eta_s$. Using nonlinear simulations, we next study the effect of $\eta_s$ on the degree of banding of nRP model for $\eta_s=10^{-3}$ and $\eta_s=10^{-5}$. We study the shear stress and the variation of degree of banding with time for $Re=0$, $10^{-2}$, and $10^{-1}$ with $A=10^{-1}$, $10^{-4}$, and $10^{-7}$. For $\eta_s=10^{-3}$, the variation of $\Delta\dot{\gamma}_{max}$ with $A$ is nonlinear, and it is not of the order of $A$ for $Re=0$ and $Re=10^{-2}$ as shown in Figs.\,\ref{nrp_effect_of_eta_s} (b), (c) and \ref{fig:nrp_dobmax_mono_eta_s}. For $Re=10^{-1}$, the variation of $\Delta\dot{\gamma}_{max}$ with $A$ is linear and it is of the order of $A$. 

\begin{figure}
\centering
    \includegraphics[scale=0.3]{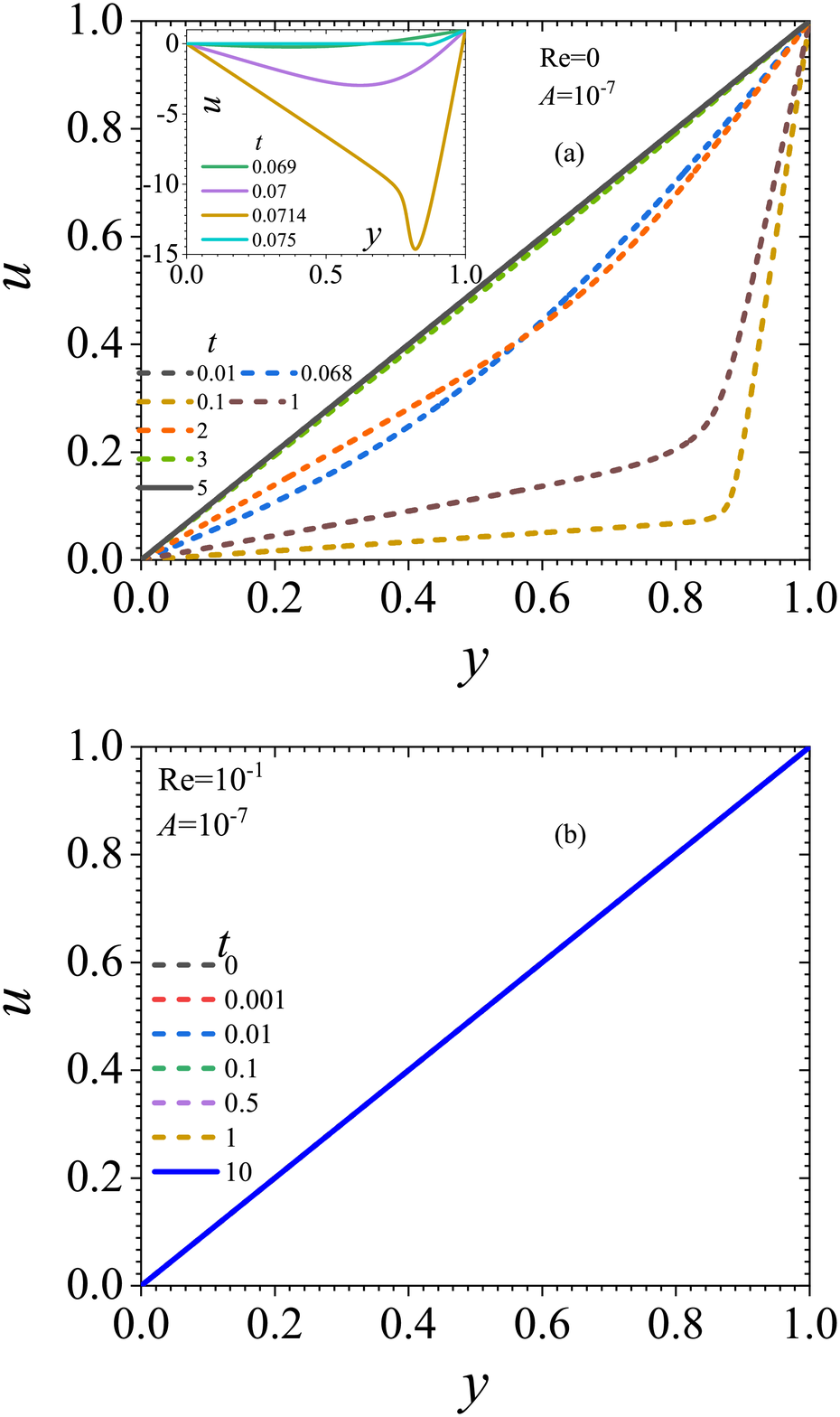}
    \caption{Velocity profile evolution during the shear startup flow of the nRP model at $Wi=30$, $\eta_s=10^{-5}$, and $\beta=1$ which corresponds to a shear rate in the monotonic region of the constitutive curve. Figure (a) shows result for $Re=0$ and $A=10^{-7}$ and figure (b) shows result for $A=10^{-7}$ and $Re=10^{-1}$.}
      \label{nrp_evelocity_N_1e_5}
\end{figure}

\begin{figure}
\centering
    \includegraphics[scale=0.3]{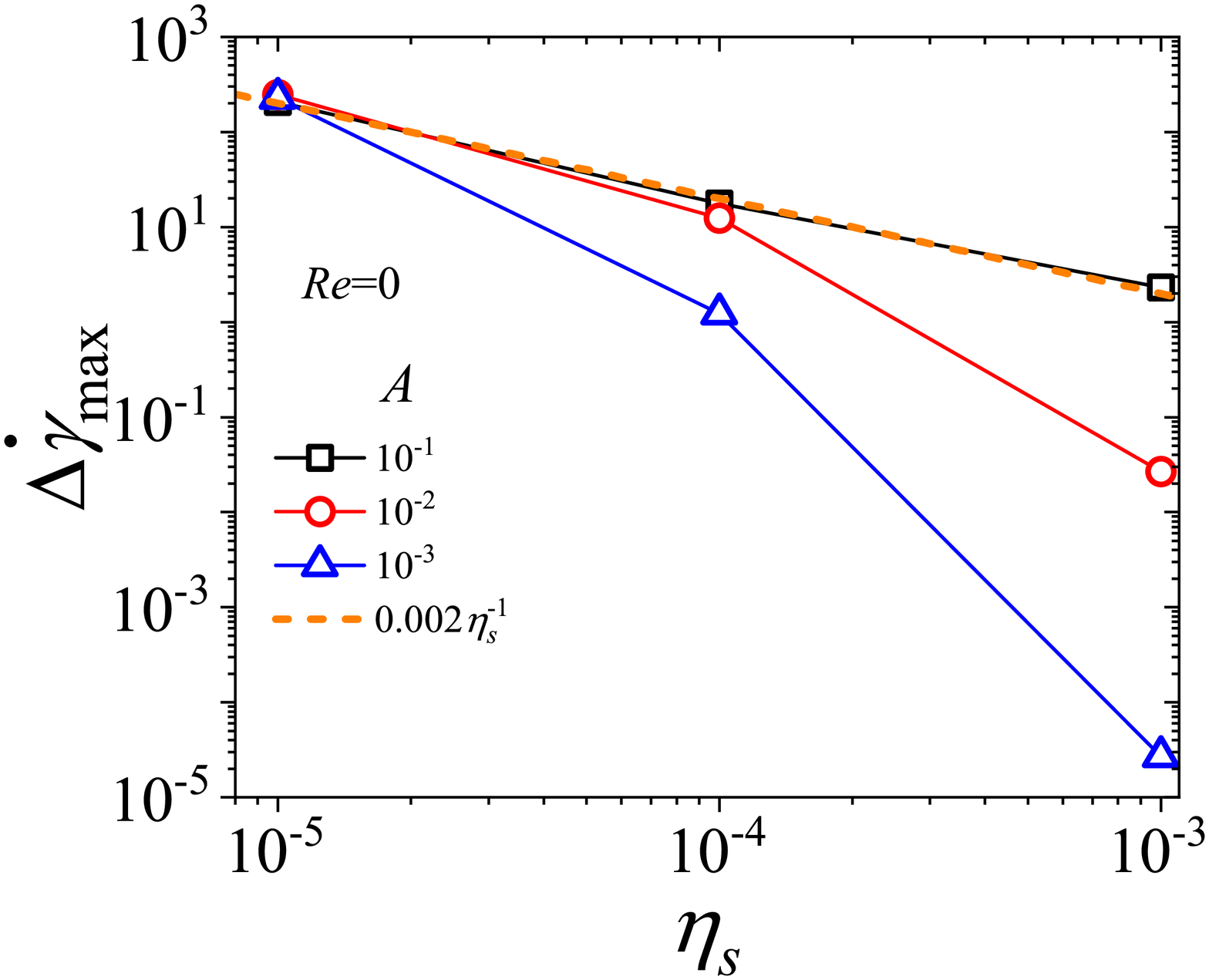}
    \caption{Variation of maximum of degree of banding during the shear startup flow as a function of $\eta_s$ during shear startup flow of the nRP model fluid for $Wi=30$, and $\beta=1$. These results are obtained for $Re=0$.  }
      \label{nrp_dob_effect_of_eta_s}
\end{figure}

For $\eta_s=10^{-5}$, the variation of $\Delta\dot{\gamma}_{max}$ is independent of the order of $A$ for $Re=0$, $10^{-2}$, and $10^{-1}$ as shown in Figs.\,\ref{nrp_effect_of_eta_s} (f)-(h) and \ref{fig:nrp_dobmax_mono_eta_s}. Also, the value of $\Delta\dot{\gamma}_{max}$ is much higher for $\eta_s=10^{-5}$ as compared to $\eta_s=10^{-3}$ for each value of $Re$. (The effect of inertia on transient shear banding is discussed in Sec.\,\ref{section_Re_effect}). The effect of decreasing $\eta_s$ on $\Delta\dot{\gamma}_{max}$ is also summarized in Fig.\,\ref{nrp_dob_effect_of_eta_s}. For $A=10^{-1}$, $\Delta\dot{\gamma}_{max}$ diverges in a power law fashion with exponent $-1$. For $A=10^{-4}$ and $A=10^{-7}$, the divergence of $\Delta\dot{\gamma}_{max}$ does not appear to follow a power law. The divergence of $\Delta\dot{\gamma}_{max}$ with decrease in $\eta_s$ is also similar to divergence of transient growth rate (or eigenvalue) with decrease in $\eta_s$ in our previous study on linearized analysis [\onlinecite{sharma2021onset}]. We also show the corresponding velocity profiles in Fig.\,\ref{nrp_evelocity_N_1e_5} for $Re=0$, $A=10^{-7}$ and $Re=10^{-1}$, $A=10^{-7}$. Figure\,\ref{nrp_evelocity_N_1e_5}~(a) shows that as the velocity profile deviates from the linear profile and shows negative velocity transiently with negative most velocity as -15! This implies that if the top plate velocity is 1, the velocity of fluid between the plates is 15 times the top plate velocity in the opposite direction. This result highlights the unrealistic nature of results obtained due to low value of $\eta_s$ even with $A=10^{-7}$ and is consistent with unbounded growth of linearized perturbations shown in our previous study [\onlinecite{sharma2021onset}]. We discuss how physically realistic results can be obtained below in Section III.

For the nRP model, decrease in solvent viscosity leads to divergence of $\Delta\dot{\gamma}_{max}$. However, the same trend is not observed for the JS model, consistent with results obtained from the linearized analysis in our earlier study  [\onlinecite{sharma2021onset}]. In addition, we also found that there is no effect of variation of $A$ in case of nRP model for $\eta_s=10^{-5}$, which can be attributed to the significantly high $(O(10^3))$ value of transient growth rate. Therefore, on consideration of results of the JS and nRP models, we find that the divergence of $\Delta\dot{\gamma}_{max}$ is not due to increase in flatness of constitutive curve, but rather due to the decrease of $\eta_s$. 


\subsection{Effect of inertia} \label{section_Re_effect}

In our previous study [\onlinecite{sharma2021onset}], we had analyzed the evolution of linearized perturbations in shear startup while the base state is also evolving. We accounted for the effect of inertia by inclusion of inertial effects during the evolution of linearized perturbations, while the base state evolution was inertialess. We found out that if $Re=0$ during the evolution of linearized perturbation, then the transiently maximum eigenvalue diverges with decrease in $\eta_s$. However, on inclusion of inertia during the evolution of linearized perturbations, the transient maximum of eigenvalue saturates to a finite value. The magnitude of saturated eigenvalue decreases with increase in $Re$. In this section, we discuss the results obtained by considering the effect of inertia in full non linear simulations in which perturbations are evolving with the flow. 

\begin{figure}
\centering
    \includegraphics[scale=0.2]{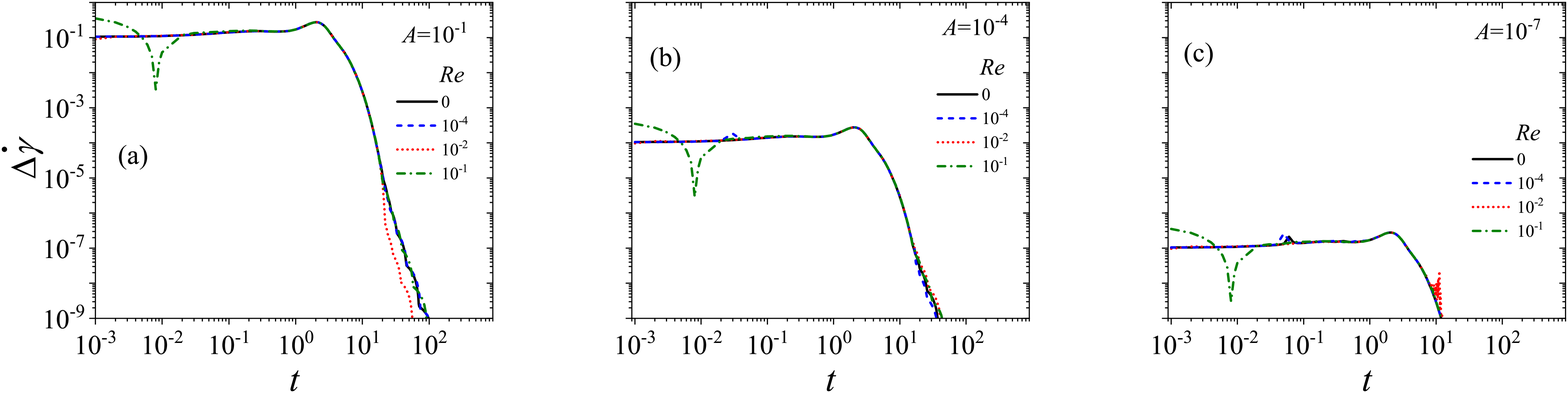}
        \caption{Effect of inertia on the degree of banding in shear startup flow of the JS model for $A=10^{-1}$, $10^{-4}$ and $10^{-7}$, $Wi=12$ and $\eta_s=0.16$. The shear rate is in the monotonic region of the constitutive curve. }
      \label{js_effect_of_inertia_mono}
\end{figure}

\begin{figure}
\centering
    \includegraphics[scale=0.2]{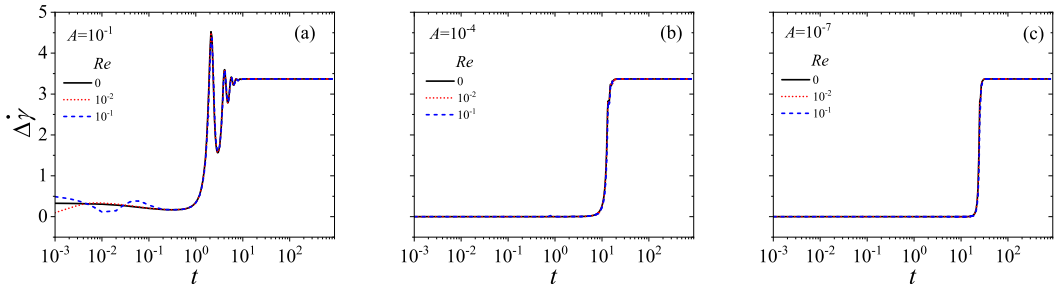}
    \caption{Effect of inertia on the degree of banding in shear startup flow of the JS model for $A=10^{-1}$, $10^{-4}$ and $10^{-7}$, $Wi=12$ and $\eta_s=0.05$. The shear rate is in the nonmonotonic region of the constitutive curve. }
    \label{js_effect_of_inertia_non_mono}
\end{figure}

\paragraph{JS model}  \label{section_Re_effect_js}
We first compare the results obtained for different $Re$ during shear startup of the JS model, with shear rates in monotonic and non-monotonic regions of the constitutive curve. Figure\,\ref{js_effect_of_inertia_mono} shows the effect of inertia on the evolution of the degree of banding as a function of time for $A=10^{-1}$, $A=10^{-4}$, and $A=10^{-7}$ if the shear rate is in the monotonic region of the constitutive curve and Fig.\,\ref{js_effect_of_inertia_non_mono} shows the results when the shear rate is in the nonmonotonic region of the constitutive curve. We have already discussed the evolution of degree of banding for JS model in Figs.\,\ref{js_A_mono_dob_Re_zero} and \ref{js_A_nonmono_dob_Re_zero}. 
As shown in Figs.\,\ref{js_effect_of_inertia_mono} and \ref{js_effect_of_inertia_non_mono}, we find that inertia has no effect on the variation of degree of banding with time for $A=10^{-1}$, $10^{-4}$, and $10^{-7}$. The degree of banding evolution does not overlap initially due to the presence of inertio-elastic waves [\onlinecite{tanner1962note,denn1971elastic}] (that are depicted in the corresponding velocity profiles shown in supplementary information). This result is consistent with our previous study using linear analysis wherein we showed that the magnitude of the transient maximum eigenvalue is small due to the high value of $\eta_s$ and as the transient growth rate is not that high, the inertial effects do not affect the dynamics of the shear startup flow.

\paragraph{nRP model}  \label{section_Re_effect_nrp}

\begin{figure}
\centering
    \includegraphics[scale=0.2]{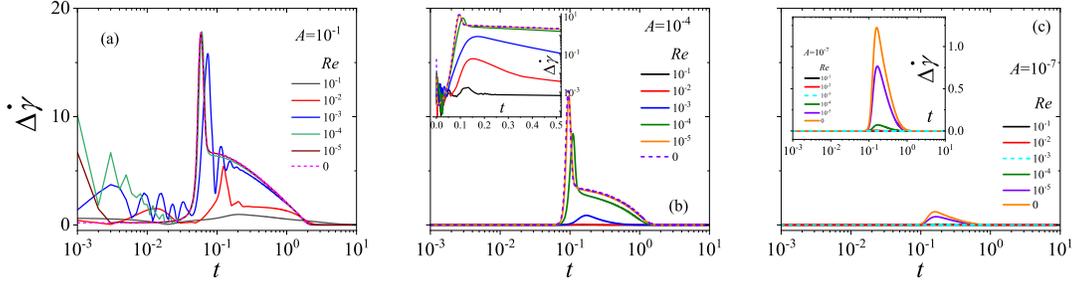}
    \caption{Effect of inertia on the degree of banding in shear startup flow of the nRP model for $A=10^{-1}$, $10^{-4}$ and $10^{-7}$, $Wi=30$, $\eta_s=10^{-4}$ and $\beta=1$. The inset of Fig.\,(b) shows the early time evolution of degree of banding on a semi-log plot. The shear rate is in the monotonic region of the constitutive curve. }
      \label{nrp_effect_of_inertia_mono}
\end{figure}

\begin{figure}
\centering
    \includegraphics[scale=0.2]{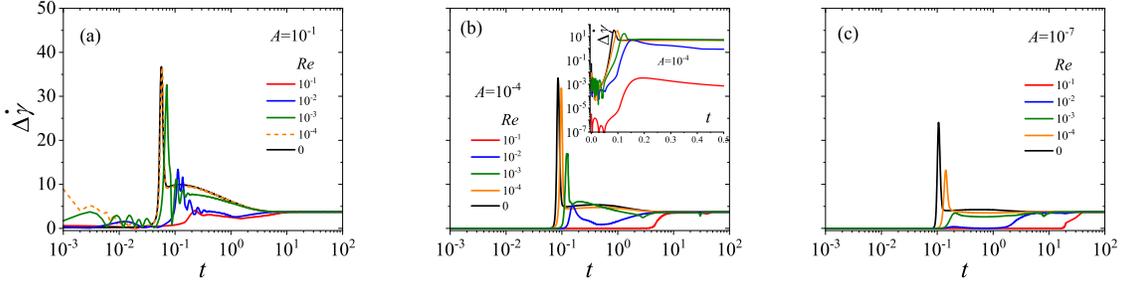}
    \caption{Effect of inertia on the degree of banding in shear startup flow of the nRP model for $A=10^{-1}$, $10^{-4}$ and $10^{-7}$, $Wi=30$, $\eta_s=10^{-4}$ and $\beta=0.6$. The inset of Fig.\,(b) shows the early time evolution of degree of banding on a semi-log plot. The shear rate is in the nonmonotonic region of the constitutive curve. }
        \label{nrp_effect_of_inertia_non_mono}
\end{figure}

We next study the effect of inertia in the shear startup of the nRP model with shear rate in the monotonic and non-monotonic regions of the constitutive curve. Figure\,\ref{nrp_effect_of_inertia_mono} shows the effect of inertia on the degree of banding for $A=10^{-1}$, $A=10^{-4}$ and $A=10^{-7}$ when shear rate is in the monotonic region of the constitutive curve and Fig.\,\ref{nrp_effect_of_inertia_non_mono} shows the results if the shear rate is in the nonmonotonic regions of the constitutive curve. 
Here, we find that the transient maximum of degree of banding or $\Delta\dot{\gamma}_{max}$ decreases significantly with increase in inertia for $A=10^{-1}$, $10^{-4}$ and $10^{-7}$ for shear rates in the monotonic or nonmonotonic regions of the constitutive curve (Figs.\,\ref{nrp_effect_of_inertia_mono} and \ref{nrp_effect_of_inertia_non_mono}). For $A=10^{-1}$, no transient peak is observed for $Re=10^{-1}$  as shown in Fig.\,\ref{nrp_effect_of_inertia_mono}~(a). For $A=10^{-4}$ and $A=10^{-7}$, no peak in the evolution of the degree of banding is observed for $Re\geq10^{-2}$. Similarly, for shear rates in the nonmonotonic region of the constitutive curve, the peak observed in the degree of banding as a function of time decreases with increase in the value of $Re$ for all $A$ as shown in Fig.\,\ref{nrp_effect_of_inertia_non_mono}. For $A=10^{-1}$, there is no peak observed for $Re=10^{-1}$, and for $A=10^{-4}$, no significant peak is observed for $Re\geq10^{-2}$, and for $A=10^{-7}$, no significant peak is observed for $Re\geq10^{-3}$. This result is consistent with the results obtained for low $\eta_s$ in our earlier study (Fig.\,13 of Ref. [\onlinecite{sharma2021onset}]), wherein we showed that with increase in $Re$, the maximum value of transient growth rate for a fixed $\eta_s$ decreases. The effect of inclusion of inertia on initial exponential growth of perturbation during inertialess flow (Figs.\,\ref{nrp_effect_of_A_dob_mono_semilog} and \ref{nrp_effect_of_A_dob_nonmono_semilog}) is also shown for $A=10^{-4}$ in inset of Figs.\,\ref{nrp_effect_of_inertia_mono} (b) and \ref{nrp_effect_of_inertia_non_mono} (b). The inset of Fig.\,\ref{nrp_effect_of_inertia_mono} (b) shows that slope of linear increase of degree of banding decreases with increase in $Re$. For $Re=10^{-1}$, there is no linear increase observed. The inset of Fig.\,\ref{nrp_effect_of_inertia_non_mono} (b) shows that the slope of linear increase decreases with increase in $Re$ for $A=10^{-4}$.

\begin{figure}
\centering
    \includegraphics[scale=0.4]{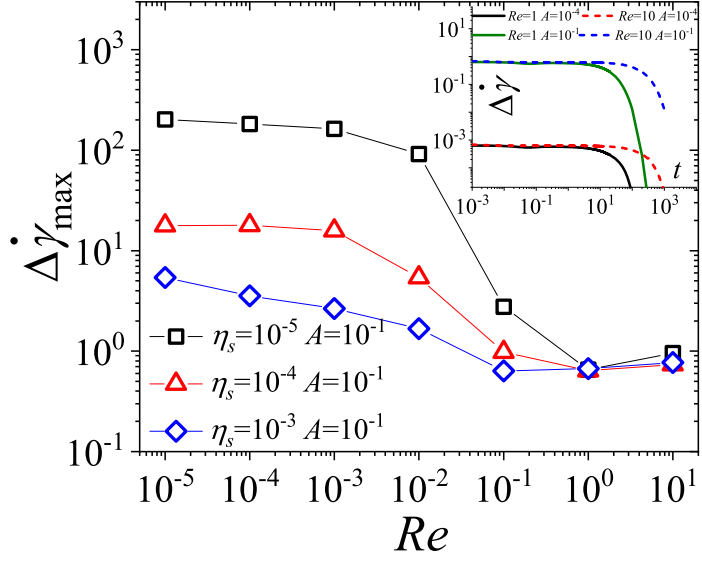}
    \caption{Variation of maximum of degree of banding $(\Delta\dot{\gamma}_{max})$ during the shear startup as a function of $Re$ for the nRP model for $Wi=30$, $\beta=1$. Here, $A=10^{-1}$, and $\eta_s=10^{-3}$, $10^{-4}$, and $10^{-5}$. The inset shows the evolution of degree of banding $(\Delta\dot{\gamma}=\dot{\gamma}_{max}-\dot{\gamma}_{min})$ with time for the nRP model for $\eta_s=10^{-4}$. Here, $Re=1$, $10$ and $A=10^{-1}$, $10^{-4}$.}
        \label{nrp_effect_of_inertia_delta_gamma_dot_max}
\end{figure}

This decrease of $\Delta\dot{\gamma}_{max}$ as a function of $Re$ for $A=10^{-1}$ and $\eta_s=10^{-3}$, $10^{-4}$, and $10^{-5}$ is shown in Fig.\,\ref{nrp_effect_of_inertia_delta_gamma_dot_max}, which summarises the effect of inertia in shear startup flow of the nRP model. It can be seen that with increase in $Re$, $\Delta\dot{\gamma}_{max}$ decreases for $\eta_s=10^{-3}$, $10^{-4}$, and $10^{-5}$. The significant change in $\Delta\dot{\gamma}_{max}$ as a function of $Re$ can be observed for $Re>10^{-2}$ for $\eta_s=10^{-3}$, $10^{-4}$, and $10^{-5}$. 
The inset of Fig.\,\ref{nrp_effect_of_inertia_delta_gamma_dot_max} shows the degree of banding evolution with time for the nRP model for $Re=1$ and $Re=10$ and $A=10^{-1}$ and $A=10^{-4}$ with $\eta_s=10^{-4}$. It shows that $\Delta\dot{\gamma}$ is independent of $Re$ and is of the order $A$ initially before finally decaying down to zero. The constant value of $\Delta\dot{\gamma}_{max}$ and of the order of $A$ for $Re\geq1$ for the nRP model is similar to the JS model results.

In Sec.\,\ref{section_model}, we raised a question about the validity of the creeping-flow assumption in solving the nRP model for $\eta_s\ll1$ when a perturbation is imposed at the beginning of the flow. There are two broad reasons as to why the creeping-flow assumption for $\eta_s\ll1$ might lead to erroneous results in the numerical solution: (i) the transient maximum eigenvalue or transient growth rate diverges with decrease in $\eta_s$ suggesting that transient growth of perturbations in nonlinear simulations can also diverge. However, if the transient growth rate of shear rate perturbation diverges, the creeping-flow assumption cannot hold true. This is because, while neglecting the acceleration term in the Cauchy momentum equation, on account of it being multiplied by $Re$, it is implicitly assumed that for $Re \ll 1$, the acceleration term must remain finite. However, when the perturbation growth diverges, the acceleration terms also exhibit a similar behavior, and the product of $Re$ and $d u/dt$ can no longer be neglected. (ii) Due to the creeping-flow assumption, the initial shear rate perturbation cannot be specified directly and it is of the order of $\frac{\sigma_{xy}}{\eta_s Wi}$ which results in imposing unrealistic magnitudes of initial shear rate perturbation, especially when the initial stress perturbations are kept $O(1)$.

It is important to note that after inclusion of inertial effects, if the solvent viscosity contribution is very low such that $\eta_s\leq O(10^{-4})$, then even with direct imposition of realistic shear rate perturbation, perturbations may eventually reach very high magnitudes. Figure\,\ref{nrp_effect_of_inertia_delta_gamma_dot_max} shows that beyond a critical $Re$, the perturbations do not grow unbounded during shear startup of the nRP model. 

The above results show that if $\eta_s$ is $O(10^{-3}-10^{-1})$, then $\Delta\dot{\gamma}_{max}$ is of the order of $A$ for $Re=0$ and $Re\neq0$. If $\eta_s$ is $O(10^{-4})$, then $\Delta\dot{\gamma}_{max}$ is very sensitive to the value of $A$ under the creeping-flow assumption, and is also affected by inclusion of inertial effects. For $\eta_s=10^{-5}$, the change in $A$ does not affect $\Delta\dot{\gamma}_{max}$ and it is only affected by inclusion of inertial effects. The fact that the velocity profiles during transient evolution can be so sensitive to $Re$, when $\eta_s$ is very small, suggests that the specific details of the experimental conditions will likely play a crucial role in the nature of the observed velocity profiles. 

Recently, Rassolov and Mohammadigoushki [\onlinecite{rassolov2022role}] experimentally obtained velocity profile evolution for wormlike micellar solution during shear startup flow and they found that steady state is banded. During the velocity profile evolution to steady state, the velocity of fluid between two plates attains a negative velocity, however, the magnitude of the negative most velocity reported is at most 25\% of the top plate velocity.  
However, the negative most velocity of fluid between plates obtained for $\eta_s=10^{-4}$ and shear rate in the nonmonotonic region of the constitutive curve can be at most 300\% of the top plate velocity as shown in Fig.\,S7 which is far greater in magnitude as compared to 25\% of top plate negative velocity observed experimentally by Rassolov and Mohammadigoushki [\onlinecite{rassolov2022role}]. Interestingly, it has been observed in literature [\onlinecite{moorcroft14,adams2011transient}] that the negative most velocity in the transient velocity profile during shear startup, if shear rate is in monotonic region of constitutive curve, is lesser than that when the shear rate is in the nonmonotonic region of constitutive curve. Therefore, based on our results we conjecture that either transient negative velocity will not be observed during experiments if shear startup will be performed with shear rate in monotonic region of the constitutive curve or the negative most transient velocity profile will be much lesser than 100\% or 1500\% of the top plate velocity that has been predicted by nRP model in Figs.\,\ref{nrp_A_1e_1_velocity_Re_0} and \ref{nrp_evelocity_N_1e_5}. The comparison with experimental results further highlights the importance of inertial effects during shear startup of nRP model which can help in regularising the results.

\subsection{Comparison of linearized evolution with fully nonlinear simulations}
\begin{figure}
\centering
    \includegraphics[scale=0.3]{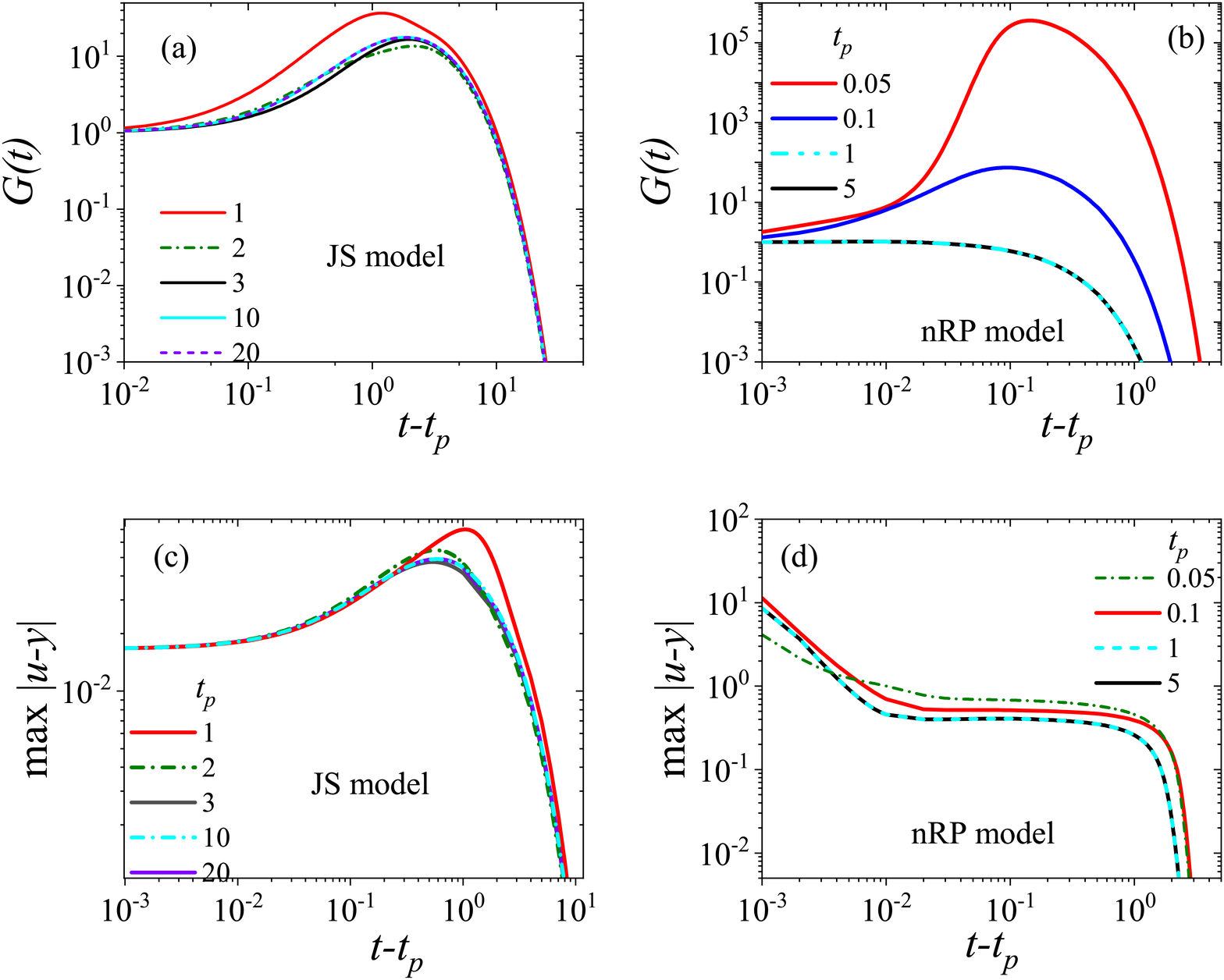}
    \caption{Panels (a) and (b) show the evolution of linearized perturbation ($G(t)$) obtained using the fundamental matrix method for JS and nRP models [\onlinecite{sharma2021onset}]. Panels (c) and (d) show the variation of max|$u-y$| as a function of time for the JS and nRP models for $Re=0$ and $A=10^{-1}$ obtained using nonlinear simulations. Panels (a) and (c) shows results from the JS model for $Wi=12$, $\eta_s=0.16$ and panels (b) and (d) shows results from the nRP model for $Wi=30$, $\eta_s=10^{-4}$, and $\beta=1$.}
        \label{linear_nonlinear}
\end{figure}

In our previous study [\onlinecite{sharma2021onset}], we had also examined the growth of linearized perturbations for different time of imposition of perturbations $(t_p)$ for the JS and nRP models. Therein, it was shown that even if the JS model is perturbed at steady state $(t_p=10$, $50$, and $70)$, the flow exhibits a transient growth and eventual decay of perturbations within the linearized dynamics. On the other hand, for the nRP model, we found that if the flow is perturbed at steady state $(t_p=1$ and $5)$, then the linearized perturbations decay without showing any transient growth. In this study, we compare the magnitude of growth of linearized perturbations with maximum deviation of velocity from a linear profile obtained from nonlinear simulations for the JS and nRP models for $A=10^{-1}$. We study the effect of $t_p$ on linearized growth of perturbations obtained using fundamental matrix method $(G(t))$ if $t_p$ lies in the time range in which stress is decreasing as a function of time after stress overshoot for the JS and nRP model fluids. The value of $G(t)$ at any time $t$ shows the maximum possible growth of linearized perturbation independent of any initial condition. We use $A=10^{-1}$ and $Re=0$ so as to compare the maximum possible linearized growth of perturbation with deviation of linear velocity profile for a very high amplitude perturbation imposed in shear startup that may be practically realisable in an experiment. The evolution of $G(t)$ as a function of $t-t_p$ for the JS and nRP models for different $t_p$ is shown in Figs.\,\ref{linear_nonlinear}(a) and (b). In these figures, the data obtained using $t_p=10$, and $20$ for JS model and for $t_p=1$, and $5$ for the nRP model has appeared before in our previous study [\onlinecite{sharma2021onset}]. We have included this data only for comparison purposes. 

In the case of JS model, we find that for $t=1$, $2$, and $3$, the stress decreases with time after stress overshoot $(Wi=12, \eta_s=0.16$ as shown in Fig.\ref{js_A_mono_stress_Re_zero}). If $t_p=1$, $2$, and $3$, then $G(t)$ shows a slightly higher increase before finally decaying for $t_p=1$ as compared to $t_p=2$, 3, 10 and 20 for which data overlaps onto each other as shown in Fig.\,\ref{linear_nonlinear}(a). The non-linear simulation results show that for $t_p=1$, max|$u-y$| also shows a slight increase before finally decaying down to zero as compared to data obtained for other $t_p$ as shown in Fig.\,\ref{linear_nonlinear}(c). The maximum value of $G(t)$ in Fig.\,\ref{linear_nonlinear}(a) is of the order of 10 wherein the evolution of $G(t)$ is independent of amplitude of initial condition. The peak value of max|$u-y$| is of the order of $10^{-2}$ for the JS model when the initial amplitude of shear rate perturbation is $A=10^{-1}$.

In the case of nRP model, we find that $G(t)$ shows a significantly high growth for $t_p=0.05$ with the maximum value of the order of $10^5$. The growth of $G(t)$ for $t_p=0.1$ is relatively lower than for $t_p=0.05$ as the maximum value is of the order of $10$. The $G(t)$ evolution overlaps and does not show any growth before decay with a maximum value of the order of 1 for $t_p=1$ and 5. On the other hand, the non-linear simulation results show a gradual decrease in the evolution of max|$u-y$| as a function of $t-t_p$. Consequently, the value of max|$u-y$| corresponding to $t-t_p=10^{-1}$ is $10^{-6}$ orders of magnitude lesser than the value of $G(t)$ for $t_p=0.05$ at $t-t_p=10^{-1}$. This result highlights the stabilizing effect of non-linearities of the model that does not allow the divergence of perturbations in the shear startup flow. Therefore, the large growth of perturbations obtained within a linearized analysis does not necessarily result in a corresponding growth of perturbations in the fully nonlinear simulations. Consequently, the actual degree of banding in shear startup flow can only be identified by solving full nonlinear simulations as also pointed out by Peterson [\onlinecite{2018PhDT.......101P}].

\section{Conclusion} \label{section_conclusion}


We presented results from nonlinear simulations of the transient dynamics of shear startup using the JS and nRP models to understand the importance and relevance of transient shear banding during shear startup. We further compared earlier results from linearized dynamics [\onlinecite{sharma2021onset}] with the present nonlinear results in order examine the extent of validity of the former. We find that stress overshoot during shear startup does not necessarily result in transient banding regardless of whether the startup is to shear rates in the monotonic or nonmonotonic regions of the constitutive curve. Our nonlinear simulations reveal that transient shear banding is absent during shear startup of the JS model for the parameter regimes explored.
For the nRP model, however, under the restricted parameter regimes of very low values of solvent to solution viscosity $(\sim O(10^{-4}))$ and in the absence of inertia, we find significant increase in degree of banding before the steady state is reached. We also showed that for the nRP model, the initial exponential growth of degree of banding has no relation with the time at which stress decreases after its overshoot, regardless of whether the startup is to shear rates in the monotonic or nonmonotonic regions of the constitutive curve.

In the inertialess limit, however, the initial shear rate perturbation cannot be prescribed and its magnitude is instead governed by the initial amplitude of the $\sigma_{xy}$ stress perturbation and $\eta_s$. Consequently, the results for shear startup of the nRP model show a nonlinear dependence on initial amplitude of perturbations and a discernible transient shear banding is observed only if $\eta_s < 10^{-3}$ and if the initial shear rate perturbations are unrealistically high for shear rates in the monotonic region of the constitutive curve. For the JS model, we found that the maximum of degree of banding is proportional to the initial amplitude of perturbations, even for a six-fold increase in their order of magnitude, suggesting that there is no intrinsic transient instability in the JS model during shear startup. We also studied shear startup of both JS and nRP models by including inertial effects so that initial shear rate perturbation can be specified directly. The results of JS model again showed a linear dependence for the variation of maximum of degree of banding with initial amplitude of perturbation even in the presence of inertial effects. However, for the nRP model, the maximum of degree of banding showed linear dependence on initial amplitude of perturbation only beyond a threshold level of fluid inertia. This critical magnitude of $Re$ increases with decrease in $\eta_s$. In the absence of inertia, we also showed that maximum of degree of banding diverges with a power law exponent of $-1$ on decreasing $\eta_s$. There is no divergence of maximum of degree of banding on decreasing $\eta_s$ when inertial effects are included. 

The comparison of results obtained using the fundamental matrix method for linearized evolution of perturbations and those from nonlinear simulations showed that nonlinear terms mitigate the divergence of perturbations. The results obtained using different initial amplitude of perturbations showed that the time associated with maximum value of degree of banding and the maximum value of degree of banding depends on the initial amplitude of perturbation. On the basis of these two observations, we conclude that occurrence of transient shear banding, if any, is not governed by a linear instability, and is governed by the following factors: (1) the initial amplitude of perturbation, (2) the solvent to solution viscosity ratio $\eta_s$, and (3) the inertial effects characterized by $Re$. Overall, we show that the results of shear startup of nRP model is very sensitive to initial amplitude of perturbations and magnitude of inertial effects if $\eta_s\ll1$. Therefore, there will not be any transient shear banding in shear startup of JS and nRP model if (i) $\eta_s > 10^{-3}$ (ii) inertial effects are included if $\eta_s\ll1$, because of very high transient growth of perturbations, creeping-flow assumption cannot hold true, and (iii) realistic initial amplitude of perturbation is imposed if $\eta_s\ll1$ and creeping-flow assumption holds true. More importantly, the comparison of linear and nonlinear studies showed that growth of perturbations as indicated by linearized dynamics may  not necessarily signify transient shear banding in the nonlinear simulations. The results of both JS and nRP models show that the ultimate answer of whether there will be any transient shear banding or not can be obtained only using nonlinear simulations.     

\section*{Acknowledgment}
We acknowledge financial support from the Science and Engineering Research Board, Government of India. 

\section*{DATA AVAILABILITY}
The data that supports the findings of this study are available within the article and its supplementary material. Additional data are available from the corresponding author upon reasonable request.

\appendix


\newpage
\section*{Supplementary Information}
\renewcommand\thefigure{\thesection S\arabic{figure}} 
\setcounter{figure}{0}  
\noindent \ul{Shear start up at $Wi=12$, $\eta_s=0.16$ using JS model (monotonic constitutive curve). Effect of inertia and initial amplitude of perturbation}
\begin{figure}[H]
\centering
\includegraphics[scale=0.8]{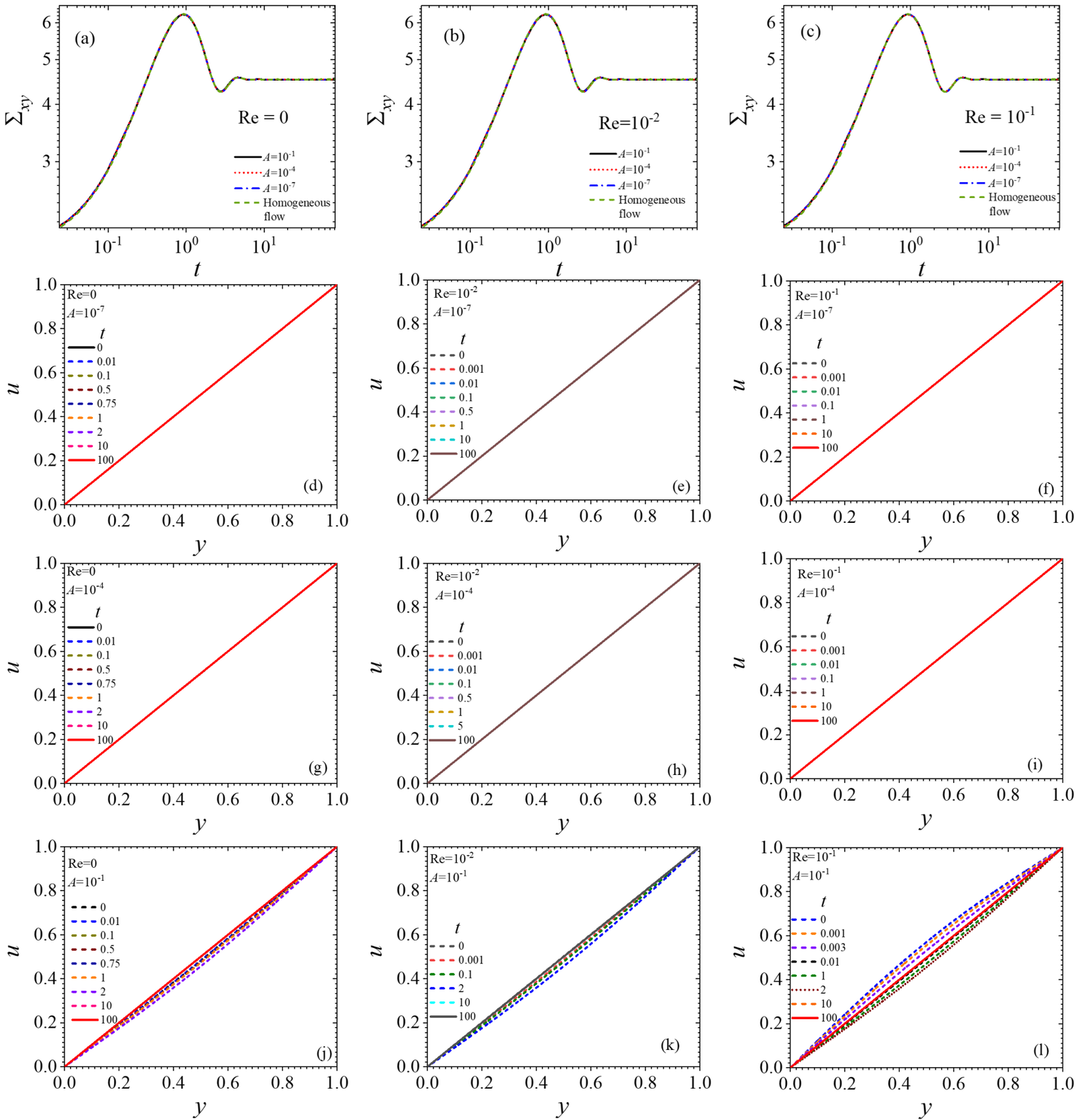}
    \label{js_mono_velocity}
\caption{\footnotesize Effect of inertia $(Re)$ and amplitude of perturbation $(A)$ is shown for shear startup flow of JS model. Shear startup flow is at $Wi=12$, $\eta_s=0.16$ which corresponds to a shear rate in the monotonic region of the constitutive curve. In the first row, shear stress is plotted for $Re=0$, $10^{-2}$, $10^{-1}$. For each value of $Re$, the results are obtained using different values of $A$ which are $10^{-1}$, $10^{-4}$, and $10^{-7}$. Shear stress for a forced homogeneous flow is also plotted in Figs.\,(a), (b), and (c). In the next three rows (Figs.\,(d)-(l)), velocity profile evolution is shown as a function of time at different values of $Re$ and $A$}
\label{fig:S1}
\end{figure}

\begin{figure}[H]
\centering
\includegraphics[scale=0.8]{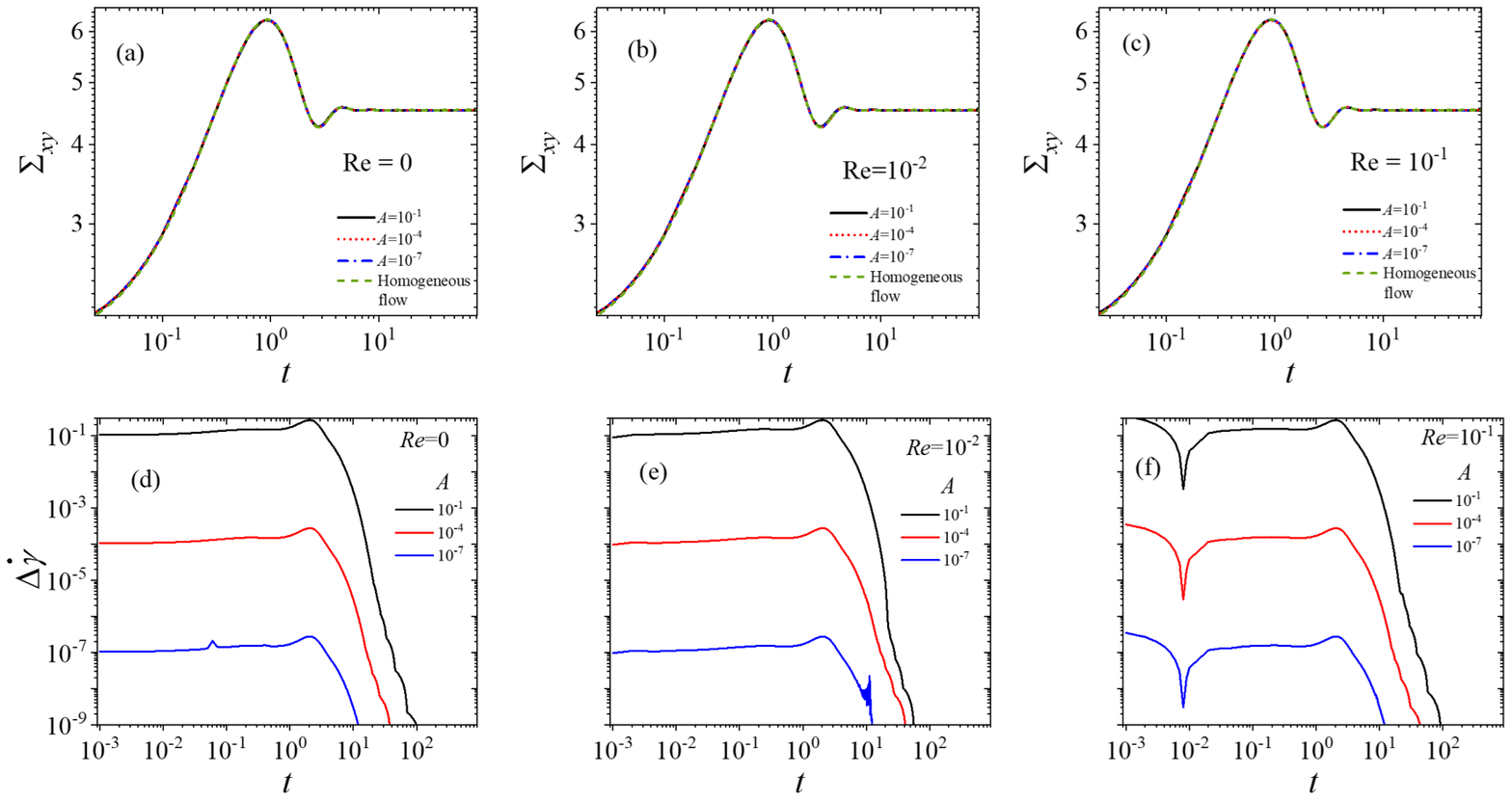}
    \label{js_mono_dob}
\caption{\footnotesize Effect of inertia $(Re)$ and amplitude of perturbation $(A)$ is shown for shear startup flow of JS model. Shear startup flow is at $Wi=12$, $\eta_s=0.16$ which corresponds to a shear rate in the monotonic region of the constitutive curve. In the first row, shear stress is plotted for $Re=0$, $10^{-2}$, $10^{-1}$. For each value of $Re$, the results are obtained using different values of $A$ which are $10^{-1}$, $10^{-4}$, and $10^{-7}$. Shear stress for a forced homogeneous flow is also plotted in Figs.\,(a), (b), and (c). In the second row (Figs.\,(d)-(f)), degree of banding (($\Delta\dot{\gamma}$) evolution is shown as a function of time at different values of $A$ for $Re=0$, $10^{-2}$, $10^{-1}$.}
\label{fig:S2}
\end{figure}
\newpage
\noindent \ul{Shear start up at $Wi=12$, $\eta_s=0.05$ using JS model (nonmonotonic constitutive curve). Effect of inertia and initial amplitude of perturbation}

\begin{figure}[H]
\centering
\includegraphics[scale=0.8]{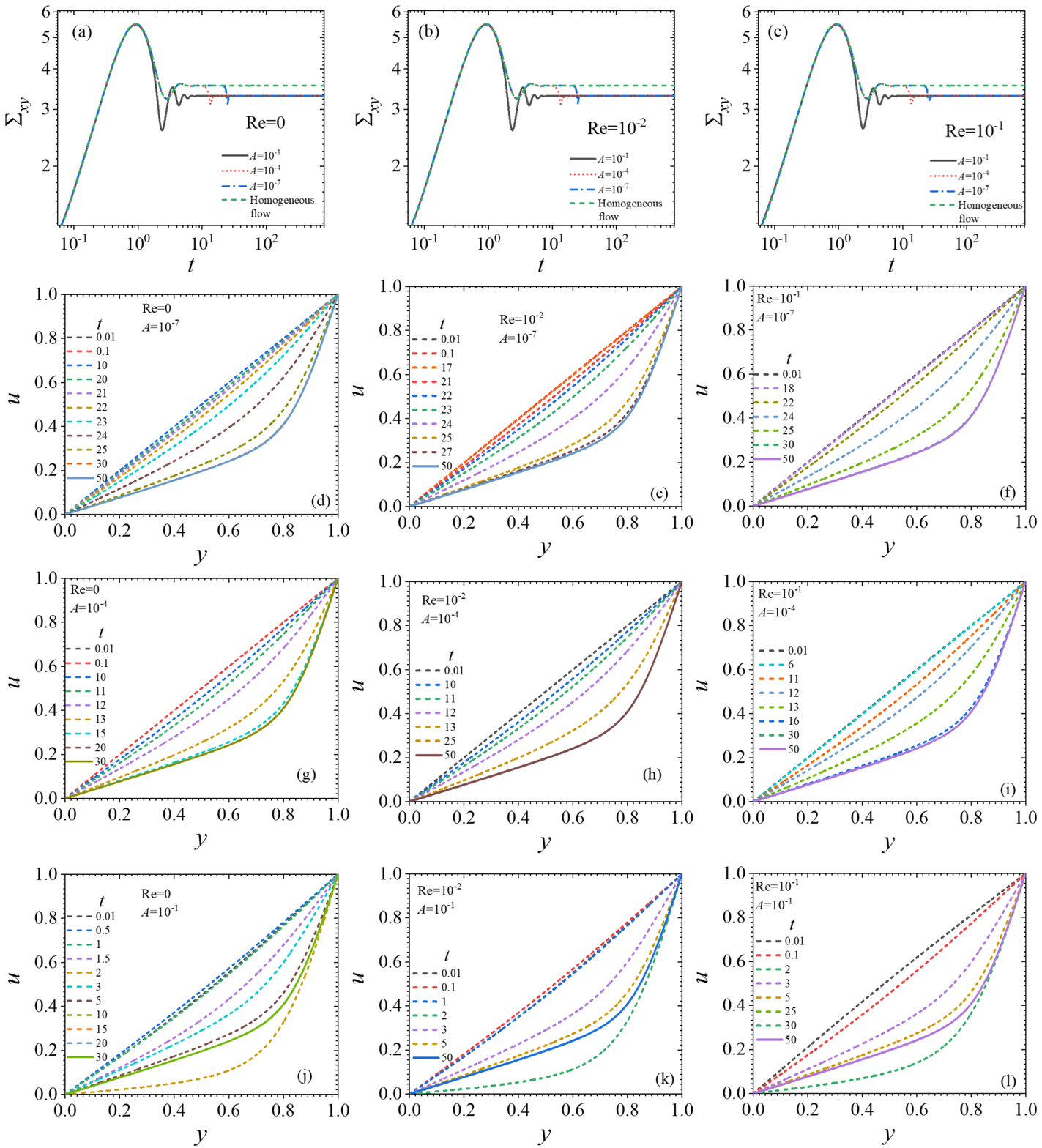}
    \label{js_non_mono_velocity}
\caption{\footnotesize Effect of inertia $(Re)$ and amplitude of perturbation $(A)$ is shown for shear startup flow of JS model. Shear startup flow is at $Wi=12$, $\eta_s=0.05$ which corresponds to a shear rate in the non-monotonic region of the constitutive curve. In the first row, shear stress is plotted for $Re=0$, $10^{-2}$, $10^{-1}$. For each value of $Re$, the results are obtained using different values of $A$ which are  $10^{-1}$, $10^{-4}$, and $10^{-7}$. Shear stress for a forced homogeneous flow is also plotted in Figs.\,(a), (b), and (c). In the next three rows (Figs.\,(d)-(l)), velocity profile evolution is shown as a function of time at different values of $Re$ and $A$.}
\label{fig:S3}
\end{figure}

\begin{figure}[H]
\centering
\includegraphics[scale=0.8]{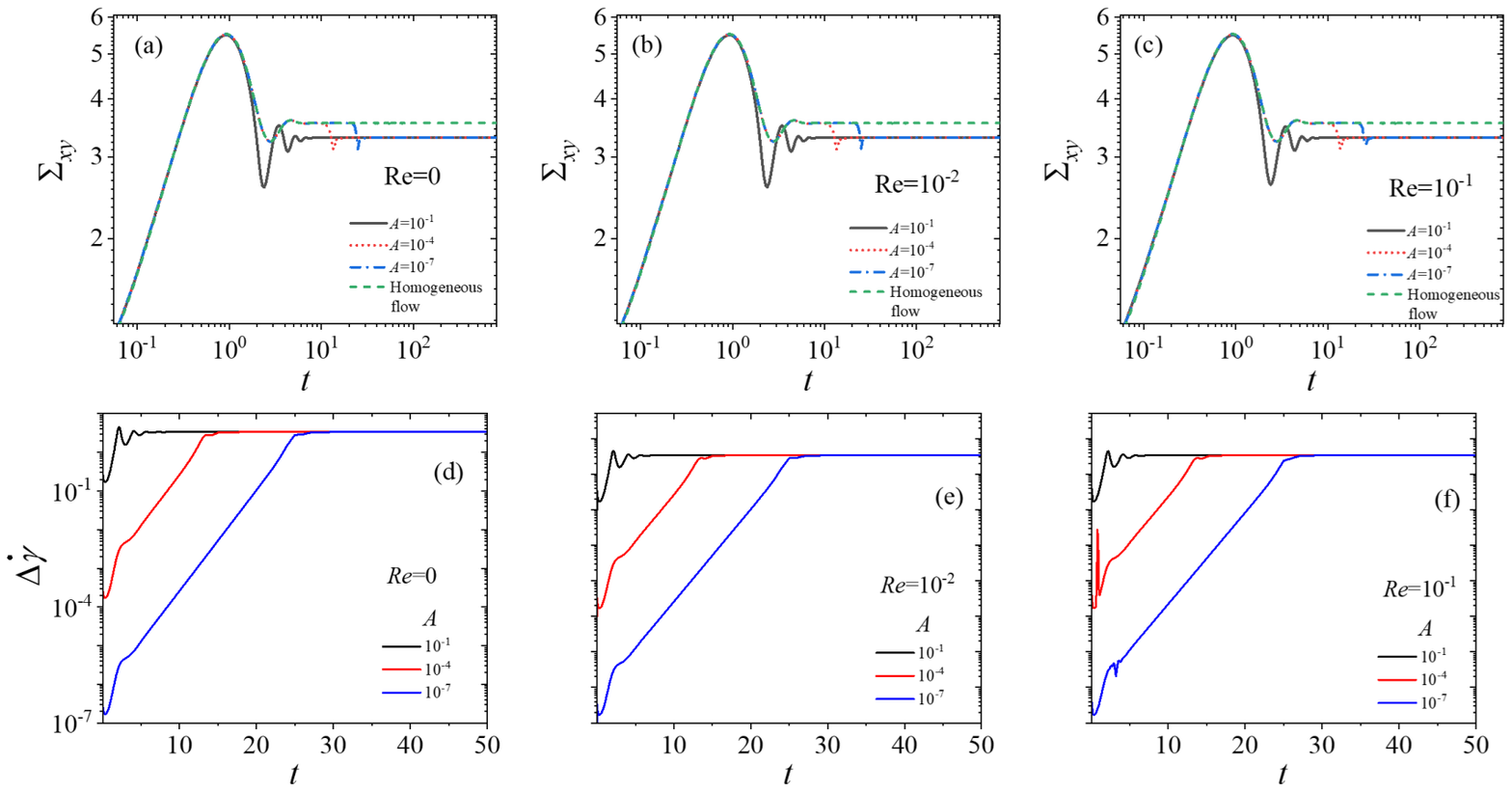}
    \label{js_non_mono_dob}
\caption{\footnotesize Effect of inertia $(Re)$ and amplitude of perturbation $(A)$ is shown for shear startup flow of JS model. Shear startup flow is at $Wi=12$, $\eta_s=0.05$ which corresponds to a shear rate in the non-monotonic region of the constitutive curve. In the first row, shear stress is plotted for $Re=0$, $10^{-2}$, $10^{-1}$. For each value of $Re$, the results are obtained using different values of $A$ which are $10^{-1}$, $10^{-4}$, and $10^{-7}$. Shear stress for a forced homogeneous flow is also plotted in Figs.\,(a), (b), and (c). In the second row (Figs.\,(d)-(f)), degree of banding ($\Delta\dot{\gamma}$) evolution is shown as a function of time at different values of $A$ for for $Re=0$, $10^{-2}$, $10^{-1}$. }
\label{fig:S4}
\end{figure}

\newpage

\noindent \ul{Shear start up at $Wi=30$, $\eta_s=10^{-4}$, $\beta=1$ using nRP model (monotonic constitutive curve). Effect of inertia and initial amplitude of perturbation}
\begin{figure}[H]
\centering
\includegraphics[scale=0.8]{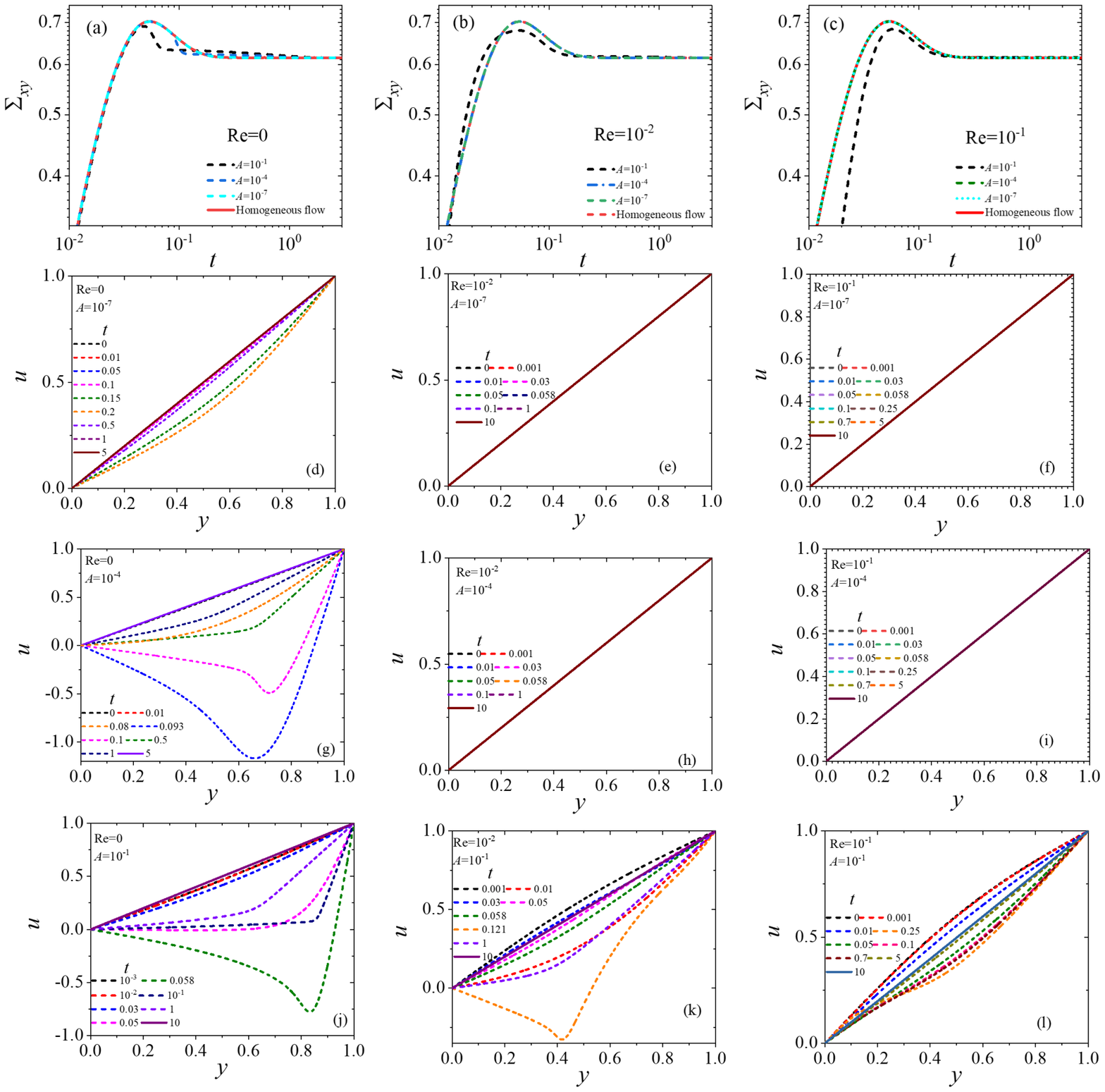}
    \label{nrp_mono_velocity}
\caption{\footnotesize Effect of inertia $(Re)$ and amplitude of perturbation $(A)$ is shown for shear startup flow of nRP model. Shear startup flow is at $Wi=30,$ $\eta_s=10^{-4}$, $\beta=1$   which corresponds to a shear rate in the monotonic region of the constitutive curve. In the first row, shear stress is plotted for 
$Re=0$, $10^{-2}$, $10^{-1}$. For each value of $Re$, the results are obtained using different values of $A$ which are $10^{-1}$, $10^{-4}$, and $10^{-7}$. Shear stress for a forced homogeneous flow is also plotted in Figs.\,(a), (b), and (c). In the next three rows (Figs.\,(d)-(l)), velocity profile evolution is shown as a function of time at different values of $Re$ and $A$.}
\label{fig:S5}
\end{figure}

\begin{figure}[H]
\centering
\includegraphics[scale=0.8]{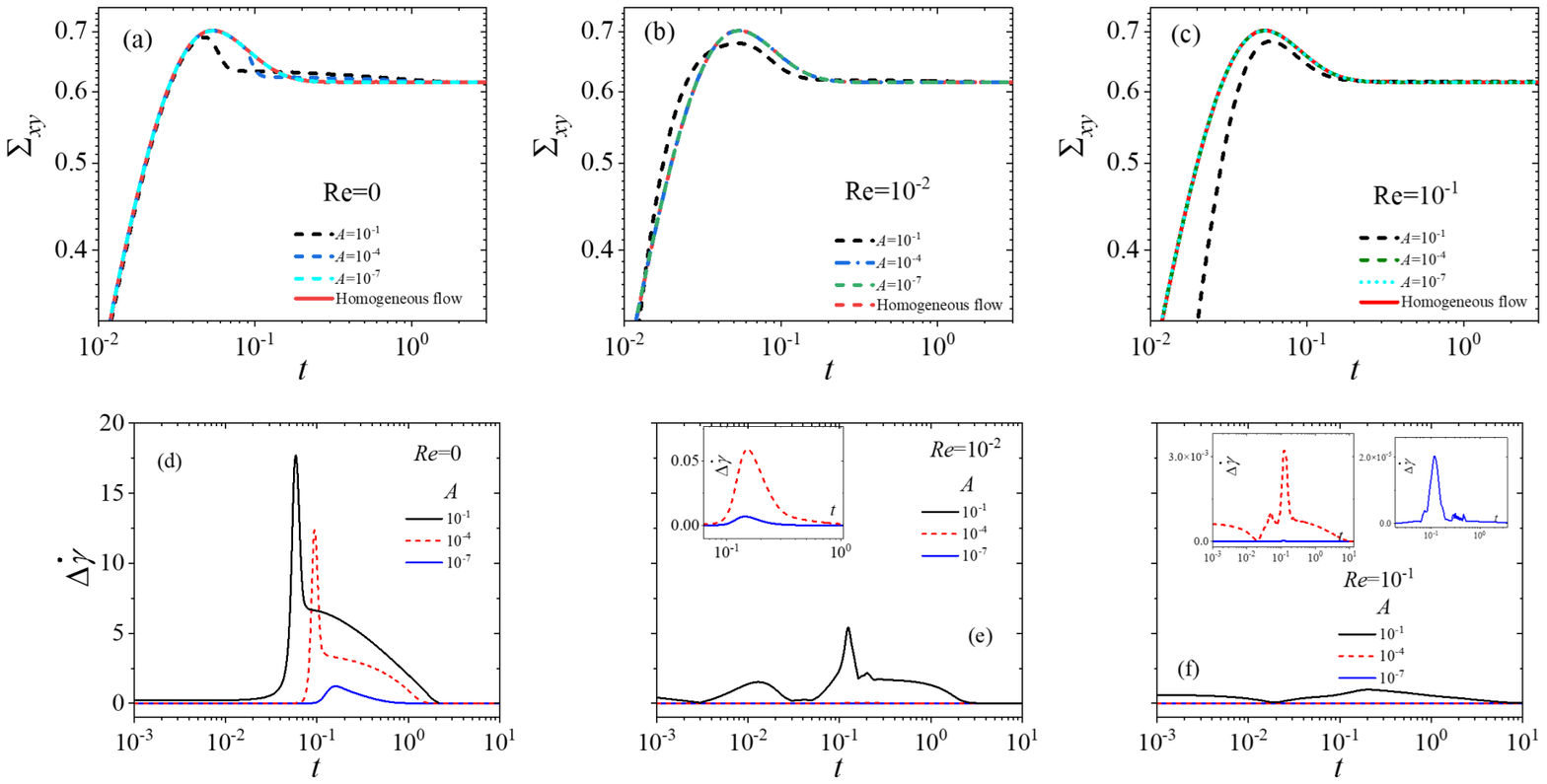}
    \label{nrp_mono_dob}
\caption{\footnotesize Effect of inertia $(Re)$ and amplitude of perturbation $(A)$ is shown for shear startup flow of nRP model. Shear startup flow is at $Wi=30,$ $\eta_s=10^{-4}$, $\beta=1$ which corresponds to a shear rate in the monotonic region of the constitutive curve. In the first row, shear stress is plotted for 
$Re=0$, $10^{-2}$, $10^{-1}$. For each value of $Re$, the results are obtained using different values of $A$ which are $10^{-1}$, $10^{-4}$, and $10^{-7}$. Shear stress for a forced homogeneous flow is also plotted in Figs.\,(a), (b), and (c). In the second row (Figs.\,(d)-(f)), degree of banding ($\Delta\dot{\gamma}$ in) evolution is shown as a function of time at different values of $A$ for $Re=0$, $10^{-2}$, $10^{-1}$.}
\label{fig:S6}
\end{figure}

\newpage
 \noindent \ul{Shear start up at $Wi=30$, $\eta_s=10^{-4}$, $\beta=0.6$ using nRP model (nonmonotonic constitutive curve). Effect of inertia and initial amplitude of perturbation}
\begin{figure}[H]
\centering
\includegraphics[scale=0.8]{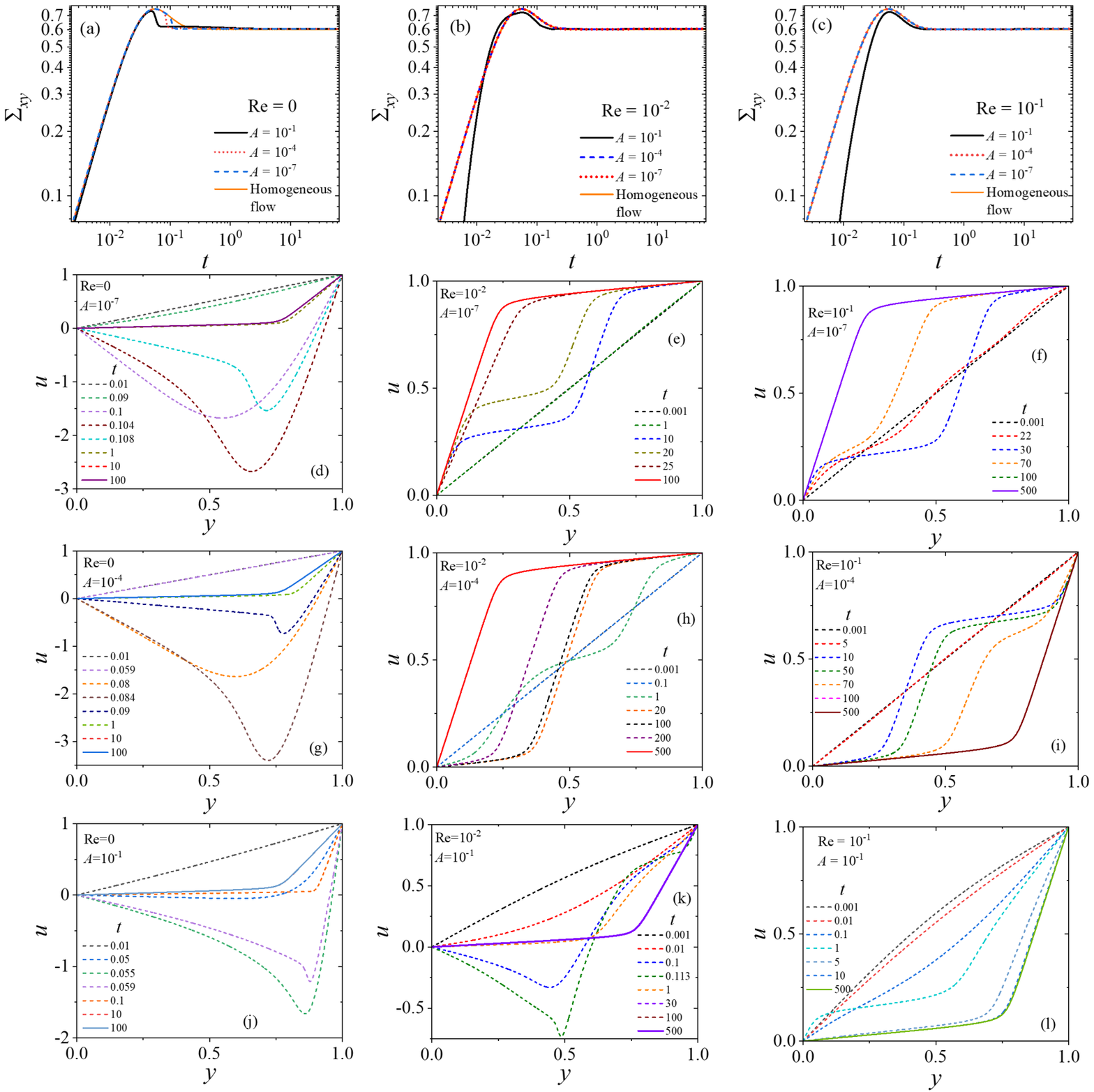}
    \label{nrp_non_mono_velocity}
\caption{\footnotesize Effect of inertia $(Re)$ and amplitude of perturbation $(A)$ is shown for shear startup flow of nRP model. Shear startup flow is at $Wi=30,$ $\eta_s=10^{-4}$, $\beta=0.6$ which corresponds to a shear rate in the non-monotonic region of the constitutive curve. In the first row, shear stress is plotted for $Re=0$, $10^{-2}$, $10^{-1}$. For each value of Re, the results are obtained using different values of $A$ which are $10^{-1}$, $10^{-4}$, and $10^{-7}$. Shear stress for a forced homogeneous flow is also plotted in Figs.\,(a), (b), and (c). In the next three rows (Figs.\,(d)-(l)), velocity profile evolution is shown as a function of time at different values of $Re$ and $A$.}
\label{fig:S7}
\end{figure}

\begin{figure}[H]
\centering
\includegraphics[scale=0.8]{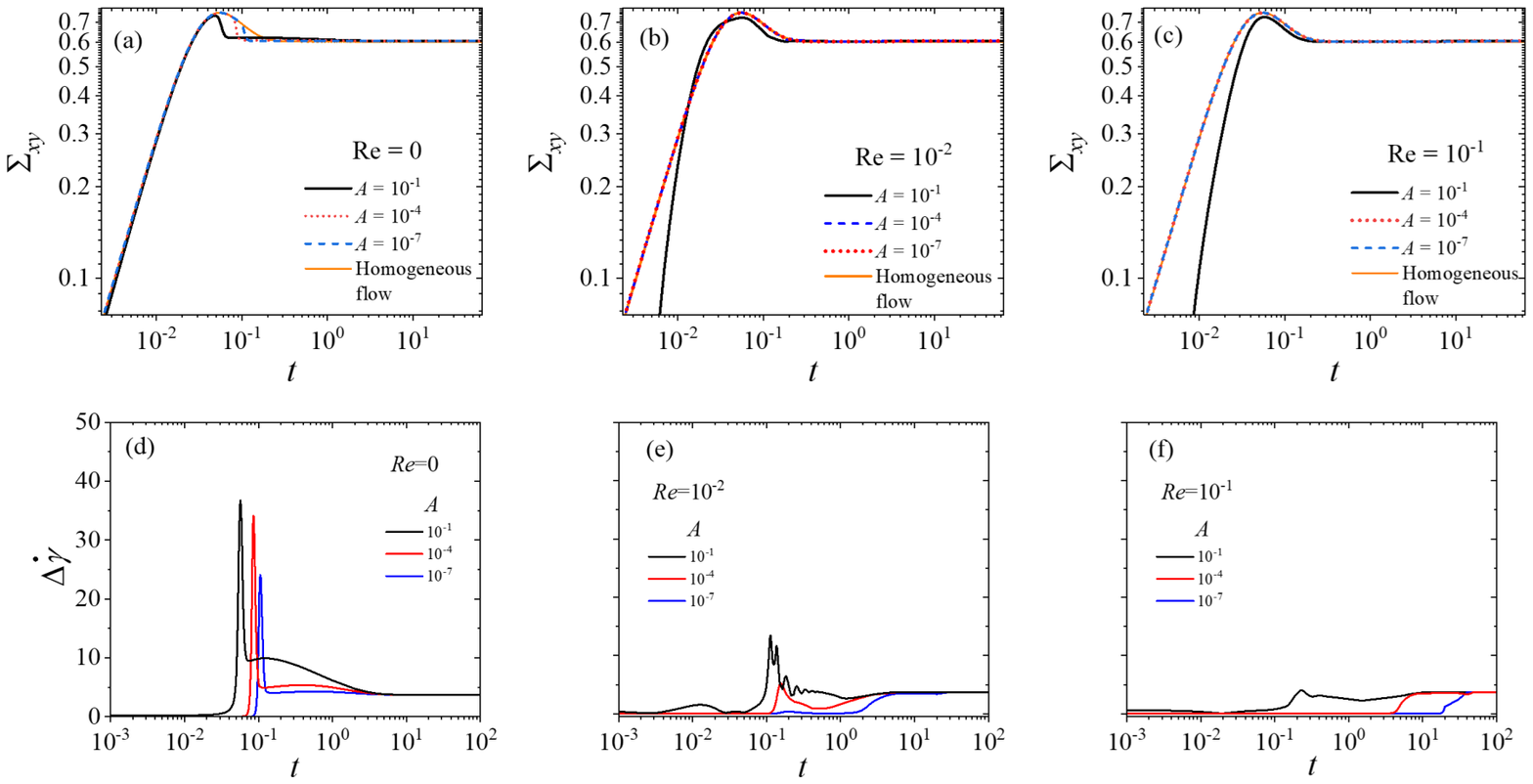}
    \label{nrp_non_mono_dob}
\caption{\footnotesize Effect of inertia $(Re)$ and amplitude of perturbation $(A)$ is shown for shear startup flow of nRP model. Shear startup flow is at $Wi=30,$ $\eta_s=10^{-4}$, $\beta=0.6$ which corresponds to a shear rate in the non-monotonic region of the constitutive curve. In the first row, shear stress is plotted for $Re=0$, $10^{-2}$, $10^{-1}$. For each value of $Re$, the results are obtained using different values of $A$ which are $10^{-1}$, $10^{-4}$, and $10^{-7}$. Shear stress for a forced homogeneous flow is also plotted in Figs.\,(a), (b), and (c). In the second row (Figs.\,(d)-(f)), degree of banding $\Delta\dot{\gamma}$ evolution is shown as a function of time at different values of $A$ for $Re=0$, $10^{-2}$, $10^{-1}$.}
\label{fig:S8}
\end{figure}

\bibliography{mybibfile.bib}
\end{document}